\documentclass[english,12pt]{article}
\usepackage[T1]{fontenc}
\usepackage[headheight=0pt, headsep=0pt,margin=1.25in]{geometry}
\usepackage{float}
\usepackage{babel}
\usepackage{epsfig}
\usepackage[space]{grffile}
\usepackage{etoolbox}
\usepackage{dcolumn}
\usepackage{pdflscape}
\usepackage{caption}
\usepackage{setspace}
\usepackage{subcaption}
\usepackage{booktabs}
\usepackage{amsmath}
\usepackage{xfrac}
\usepackage{fp}
\usepackage{pgfplotstable}
\usepackage{dsfont}
\usepackage{graphicx}
\usepackage{overpic}
\usepackage{textgreek}
\usepackage{natbib}
\usepackage{xurl}
\usepackage[colorlinks=true, linkcolor=black, citecolor=black, urlcolor=black]{hyperref}
\usepackage[author=Dan Herbst]{pdfcomment}
\usepackage[title]{appendix}
\usepackage[us]{datetime}
\usepackage{mypackage}


\makeatletter

\def \bpath {.}
\def \outpath {\bpath/output}

\def\input@path{{\outpath/}{\outpath/tables/}}
\graphicspath{{\outpath/figures/}{\bpath/output/figures_manual/}}

\definecolor{wbarcolor}{RGB}{178,34,34}
\def \wbar{\makebox(0,0){\large\color{wbarcolor}{\fontfamily{phv}\selectfont $\overline{\text{w}}(\text{\texttheta})$}}}
\def \wbarb{\makebox(0,0){\large\color{wbarcolor}{\fontfamily{phv}\selectfont $\overline{\text{w}}_B(\text{\texttheta})$}}}

\definecolor{avcolor}{RGB}{0,139,69}
\def \av#1{\makebox(0,0){\large\color{avcolor}{\fontfamily{phv}\selectfont AV$_{#1}$({\texttheta})}}}

\definecolor{mvcolor}{RGB}{24,116,205}
\def \mv#1{\makebox(0,0){\large\color{mvcolor}{\fontfamily{phv}\selectfont MV$_{#1}$({\texttheta})}}}

\def \N {3030}
\def \Ncontrol {302}
\def \meancorvalue {7.003}

\def \meantakeup {0.486}

\def \meanscore {19.453}

\def \meannumscore {21.982}

\def \maxscore {52}
\def \bwllossbunching {3.93}
\def \bwrlossbunching {3.38}
\def \plossbunching {0.803}
\def \corvaluePA {7.279}

\def \corvaluesePA {0.179}

\def \corvalueA {7.279}

\def \corvalueseA {0.179}

\def \corvaluePB {7.656}

\def \corvalueTB {4.619}

\def \corvaluesePB {0.116}

\def \corvalueB {7.110}

\def \corvalueseB {0.108}

\def \hourlyB {0.210}

\def \rangeBlo {1.20}
\def \rangeBhi {3.00}
\def \corvaluePC {7.871}

\def \corvalueTC {6.041}

\def \corvaluesePC {0.130}

\def \corvalueC {6.980}

\def \corvalueseC {0.093}

\def \rangeClo {3.60}
\def \rangeChi {9.60}
\def \corvaluePD {7.840}

\def \corvalueTD {6.558}

\def \corvaluesePD {0.227}

\def \corvalueD {6.843}

\def \corvalueseD {0.104}

\def \hourlyD {0.738}

\def \rangeDlo {10.80}

\def \bSc {1.202}
\def \seSc {0.056}
\def \bAScPcorvalue {0.093}
\def \seAScPcorvalue {0.100}
\def \bAScHcorvalue {0.666}
\def \seAScHcorvalue {0.121}

\def \bTEccorvalue {-0.612}
\def \seTEccorvalue {0.215}

\def \bTEcXcorvalue {-0.606}
\def \seTEcXcorvalue {0.208}

\def \bTEcXXcorvalue {-0.591}
\def \seTEcXXcorvalue {0.207}

\def \bScXXX {1.248}
\def \seScXXX {0.058}

\def \bTEcXXXcorvalue {-0.475}
\def \seTEcXXXcorvalue {0.193}

\def \obsconditionaldecline {423}
\def \obsconditionalaccept {515}
\def \obsunconditionaldecline {679}
\def \obsunconditionalaccept {560}
\def \gainconditionaldecline {115}
\def \gainconditionalaccept {97}

\def \meanbreakevenaccept {270.52}
\def \meanbreakevendecline {273.59}
\def \feereduction {0.224}
\def \thetaav {0.549}
\def \uberalpha {408.409}
\def \uberbeta {211.376}
\def \ubereq {0.717}
\def \ubereff {0.762}
\def \uberdwl {3.631}

\def \popDWLPcorvalueXXfullsemi {0.0282}
\def \popDWLPcorvalueXXsemi {0.0401}

\def \popthetaeqHcorvalueXXsemi {0.4898}

\def \popthetaeffHcorvalueXXsemi {0.5546}

\def \popDWLHcorvalueXXfullsemi {0.0390}
\def \popDWLHcorvalueXXsemi {0.0461}

\def \popDWLPsecorvalueXXfullsemi {0.0286}
\def \popDWLPsecorvalueXXsemi {0.0368}

\def \popthetaeqHsecorvalueXXsemi {0.0153}

\def \popthetaeffHsecorvalueXXsemi {0.0188}

\def \popDWLHsecorvalueXXfullsemi {0.0199}
\def \popDWLHsecorvalueXXsemi {0.0152}

\def \poptaxstarcorvalueXXsemi {0.1830}
\def \poptaxstarsecorvalueXXsemi {0.0461}

\def \hoursperweek {14.73}

\pgfkeys{/pgf/number format/.cd,
    fixed,
    fixed zerofill,
    precision=2,
    set thousands separator={{{,}}},
    set decimal separator={.},
  }

\newcolumntype{M}[1]{>{\centering\arraybackslash}m{#1}}
\newcolumntype{L}[1]{>{\raggedright\arraybackslash}p{#1}}
\newcolumntype{Q}{l}
\newcolumntype{d}{D{.}{.}{-1}}

\def\sym#1{\ifmmode^{#1}\else\(^{#1}\)\fi}
\newcommand\starnote{Heteroskedasticity-robust standard errors are reported in parentheses. *** = significant at 1\% level, ** = significant at 5\% level, * = significant at 10\% level.}

\DeclareMathOperator*{\argmax}{argmax}
\DeclareMathOperator*{\argmin}{argmin}

\begin{document}

\author{Daniel Herbst\thanks{\fontsize{9.5}{11}\selectfont Department of Economics, University of Arizona, McClelland Hall 401QQ, Tucson, AZ 85721 (email: dherbst@arizona.edu, website: \href{http://www.danjherbst.com}{\color{blue}{www.danjherbst.com}}). I am thankful to David Arnold, Asher Baraban, Jesse Bruhn, Maria Cardi, Sydnee Caldwell, Will Dobbie, Jamie Fogel, Felipe Goncalves, Nathan Hendren, Hide Ichimura, Andreas Kostøl, Ilyana Kuziemko, John List, Derek Neal, Juan Pantano, Stephen Ross, and Evan Taylor for their helpful comments in the development of this paper. I am also thankful to seminar participants at Brown University, Carnegie Mellon University, Northwestern University, the University of Arizona, the University of Chicago Harris School of Public Policy, the Advances in Field Experiments Conference, the ASSA Annual Meeting, the ASU Empirical Microeconomics Conference, the Bank of Italy CEPR Labor Workshop, the CESifo Labor Economics Conference, the NBER Labor Workshop, the SOLE/EALE Labor Conference, and the Yale Cowles Conference on Labor and Public Economics for their thoughtful insights. Financial support was provided by the University of Arizona Eller College of Management.}\\
University of Arizona}

\title{\vspace{-1cm}Asymmetric Information in Wage Contracts: \\ Experimental Evidence and Welfare Implications}
\date{October 5, 2026}
\maketitle
\vspace{-.6cm}

\begingroup
\hyphenpenalty=10000
\exhyphenpenalty=10000
\begin{abstract}
  Compared to self-employment or performance-based pay, fixed wages reduce earnings uncertainty but are prone to market distortions through moral hazard and adverse selection. Using a model of wage contracts under asymmetric information, I show that these distortions map to marginal treatment response functions in a marginal treatment effects (MTE) framework. I apply this framework to a field experiment in which data-entry workers choose between a randomized fixed wage and a piece rate. I find evidence of both moral hazard and adverse selection: Fixed wages reduce worker productivity by an estimated \fp[2]{-100*\bTEccorvalue/\meancorvalue} percent relative to the mean. Meanwhile, a 10 percent increase in the wage offer attracts a marginal worker whose productivity is higher by \fp[2]{20*\bAScHcorvalue/\meancorvalue} percent of the mean. Using semi-parametric MTE estimation, I calculate the welfare loss associated with asymmetric information and the marginal values of public funds (MVPFs) for policies aimed at recovering this loss. I find that a tax on piece rates can mitigate adverse selection into fixed wages, generating tax revenue at social costs as low as \$0.88 per dollar.

\end{abstract}

{\raggedright\noindent \textbf{Keywords:} compensation structure, wage insurance, performance pay, adverse~selection, moral hazard, information asymmetries, marginal treatment effects.\par}\medskip

\endgroup


\onehalfspacing

\vspace{-1.5cm}
\newpage
\section{Introduction}\label{sec:intro}

Fixed wages offer workers a guaranteed return on each hour of work, even when output is unpredictable. In financial terms, they operate like equity contracts in which the employer buys a claim on the worker's next hour of labor product. But these contracts are also vulnerable to moral hazard and adverse selection---a fixed wage might induce less effort or attract less productive workers than freelance hiring or piece-rate pay. Such information asymmetry problems can distort equilibrium wages, leaving workers over-reliant on output-based compensation. The potential for these distortions is perhaps highest in short-term gig labor markets, where a growing number of workers are compensated by the number of miles driven, pages written, or tasks completed \citep{garin2020new,koustas2018consumption}.

The potential effects of moral hazard and adverse selection on wage contracts have important policy implications.
For example, an optimal piece-rate tax must balance the welfare benefits of inducing the marginal worker into fixed wages with the distortionary costs of encouraging that worker to shirk. Moral hazard and adverse selection are inherent to this trade-off, but identifying their effects on equilibrium wages is challenging for several reasons. First, both forces can lead to lower observed productivity among fixed-wage workers compared to self-employed or piece-rate workers, making them hard to distinguish without exogenous variation in wages.
Second, measuring these forces requires estimates of labor supply and selection at the market level, which are difficult to obtain from samples of existing employees or job applicants facing unobserved outside options \citep{dube2020monopsony,caldwell2023gender}.
Finally, adversely selected contracts may be too unprofitable to exist in equilibrium.
Methods that extrapolate from local price variation often overlook these ``missing contracts,'' excluding the selection margins most relevant to measuring welfare loss \citep{einav2010beyond}.

In this paper, I experimentally estimate the equilibrium and welfare effects of moral hazard and adverse selection in a large online labor market for data-entry work. To identify these forces, I conduct a field experiment with two stages of randomization: First, I offer workers a choice between a randomized fixed hourly wage and a standardized piece rate. Then, after workers choose a payment option but before they begin their assigned task, I increase hourly wages for a randomized subset of those who accepted hourly offers, bringing them to parity with the highest offered wage. Using the initial wage offer as an instrument for accepting an hourly contract allows me to identify moral hazard as the treatment effect of fixed-wage compensation on worker output. Meanwhile, comparing output across workers on the same contract who faced different ex-ante offers identifies adverse selection into fixed wages.

To map my experimental estimands into a welfare framework, I show how both moral hazard and adverse selection can be expressed using marginal treatment response (MTR) functions under fixed-wage and piece-rate counterfactuals---the objects identified in a marginal treatment effects (MTE) analysis \citep{bjorklund1987estimation,heckman1999local,heckman2005structural,heckman2007econometric}. Embedding these MTR functions into a model of asymmetric information allows me to flexibly characterize equilibrium and efficient allocations of fixed-wage labor. Much like canonical models of insurance markets \citep{efc,akerlof}, equilibrium is determined by two curves: a worker's hourly reservation wage---the minimum payment they will accept for an hour of labor---and the average value of output among workers with lower reservation wages than their own. Employers cannot profitably pay a given worker's reservation wage if that wage exceeds the average output value of lower-reservation-wage workers. Relative to a full-information equilibrium, this profit condition leads to an underprovision of hourly work---some piece-rate workers would like to forfeit a portion of their expected earnings in exchange for a fixed-wage contract, but the threat of adverse selection prevents employers from offering such contracts at those workers' reservation wages. My empirical design allows me to estimate the welfare loss implied by these distortions using semi-parametric MTE methods.

Results from my experiment provide evidence of both moral hazard and adverse selection into fixed-wage contracts. Estimated treatment effects imply that working under hourly pay reduces workers' output value by \fp{-100*\bTEccorvalue/\meancorvalue} percent relative to the mean (SE=\fp{100*\seTEccorvalue/\meancorvalue}). At the same time, a ten-percent increase in the hourly wage offer attracts a marginal worker whose productivity is higher by \fp[2]{20*\bAScHcorvalue/\meancorvalue} percent of the mean (SE=\fp{20*\seAScHcorvalue/\meancorvalue}), raising the average productivity of fixed-wage workers by \fp{10*\bAScHcorvalue/\meancorvalue} percent of the mean (SE=\fp{10*\seAScHcorvalue/\meancorvalue}). This finding appears driven by selection on \emph{unobserved} determinants of productivity; results persist after conditioning on demographics and work experience, suggesting that pre-employment screening cannot fully resolve asymmetric information problems in this labor market.

Using local polynomial MTE estimation, I find that \fp[0]{100*\popthetaeffHcorvalueXXsemi} percent (SE=\fp[2]{100*\popthetaeffHsecorvalueXXsemi}) of workers in my sample would pay a premium for fixed-wage contracts over piece-rate pay. This share reflects the efficient allocation that would exist if employers were fully informed of workers' reservation wages, inclusive of any moral hazard effects. By contrast, only \fp[0]{100*\popthetaeqHcorvalueXXsemi} percent (SE=\fp[2]{100*\popthetaeqHsecorvalueXXsemi}) of workers would find fixed-wage positions in a competitive equilibrium.
This underprovision of fixed-wage positions reflects adverse selection on potential output under fixed wages, identified as the MTR function under treatment. Comparing this function to estimates of potential output under the piece rate suggests a nonlinear pattern of selection on moral hazard: At low reservation wages, those with marginally stronger preferences for fixed wages have a slightly higher propensity to shirk, consistent with patterns of selection in insurance markets \citep{einav2013selection}. At higher reservation wages, however, shirkers are more likely to prefer piece rates, perhaps because the workers most tolerant of variable pay are also less fearful of poor performance reviews or other repercussions from shirking.

Estimated MTR functions imply that the welfare loss arising from the underprovision of fixed-wage contracts is \$\fp{\popDWLHcorvalueXXsemi} (SE=\$\fp[3]{\popDWLHsecorvalueXXsemi}) per hour of labor. Scaling this per-hour loss by roughly 35 billion annual paid worker-hours in online clerical and data-entry gig work implies aggregate annual welfare losses on the order of \$\fp[1]{35*(\popDWLHcorvalueXXsemi)} billion globally.\footnote{I calculate aggregate annual welfare loss by multiplying my loss-per-hour estimate of \$\fp{\popDWLHcorvalueXXsemi} by 35 billion worker hours---the estimated number of paid worker hours spent on online clerical and data-entry tasks each year \citep{worldbank2024workingwithoutborders,ilo2021digitalplatforms}. Details are provided in Footnote~\ref{fn:welfarescale}.}
Using an MVPF analysis \citep{hendren2020unified}, I show how governments can efficiently recover these losses with policies that balance the welfare benefits of promoting fixed wages with the fiscal costs of increased shirking. I find that a tax on piece-rate earnings can yield a marginal cost of public funds as low as 0.88, meaning each dollar of tax revenue carries a net social cost of just \$0.88---more efficient than a distortionless tax.

These estimates are specific to the data-entry workers in my experimental setting; as in much of the applied literature on worker incentives, both selection and treatment effects may not extrapolate across labor markets. In the final section, I discuss external validity and show how my framework can be applied to other settings. To illustrate, I calibrate my model to the market for rideshare drivers using experimental estimates from \citet{angrist2021uber}.
I find that adverse selection into less ride-dependent compensation results in an approximate welfare loss of $\$\fp[2]{\feereduction*\uberdwl/\hoursperweek}$ per worker-hour.
This exercise illustrates how my methods can be applied to a variety of labor markets, especially those dominated by tips, commissions, and bonuses.

This study relates to several strands of labor research. Theory shows how asymmetric information can generate shirking, sorting, and inefficient wage setting or labor allocation \citep{mirrlees1971,miyazaki1977rat,Holmstrom1979,GrossmanHart1983,jovanovic1982favorable,greenwald1986adverse,lazear1986salaries,MacLeodMalcomson1989,levine1991just,kugler2004firing,moen2005performance,Shimer2005Assignment, stantcheva2014optimal}.\footnote{Related work links information asymmetries to ``efficiency wages''---wages paid above the market-clearing rate---\citep{weiss1980job,krueger1988efficiency,weiss2014efficiency,yellen1984efficiency,malcomson1981unemployment,katz1986efficiency}.} Empirical studies document incentive and sorting effects across compensation schemes in windshield repair \citep{lazear2000}, teaching \citep{brown2020inducing,leaver2021recruitment,johnston2020preferences}, rideshare \citep{angrist2021uber,caldwell2023gender}, sales \citep{butschek2022motivating}, mining \citep{shearer1996piece}, health care \citep{kantarevic2016physician,gaynor2004}, and agriculture \citep{foster1996information,foster2022comparative}.\footnote{Recent evidence also points to selection effects on other aspects of labor contracts like remote work \citep{emanuel2023working} and paid time off \citep{koenig2026regulating}.} Experimental evidence on selection and treatment effects of payment schemes includes \citet{flory2015competitive}, \citet{dohmen2011performance}, \citet{glewwe2010teacher}, \citet{dellavigna2018motivates}, \citet{shearer2004piece}, and \citet{guiterasjack2018}.

This paper also contributes to research on insurance and insurance-like contracts. The model draws directly on \citet{efc} and \citet{herbst2024opportunity}, and the empirical mapping to marginal treatment effects complements \citet{kowalski2023reconciling}. I extend this line of work by using MTE methods to quantify welfare loss from adverse selection and moral hazard in fixed-wage labor contracts.

Methodologically, my experimental design builds upon \citet{karlanzinman}, who identify selection and incentive effects for microfinance loans by combining variation in offered terms with variation in the terms ultimately received. They find strong evidence of moral hazard and weaker evidence of adverse selection. In a related experiment, \citet{bryan2015referrals} estimate how referral bonuses induce selection into consumer loans.\footnote{Like my study, \citet{bryan2015referrals} identify selection on potential outcomes (repayment propensity) with and without treatment (peer enforcement bonuses). But unlike my marginal treatment effects approach,
  their design randomly removes the treatment condition for a subset of those who select into it.} They estimate large treatment effects of social pressure on repayment, but find little evidence of selection on potential repayment or resistance to social pressure. In another two-stage experiment, \citet{beaman2023selection} randomize village access to agricultural credit and subsequent cash grants to identify selection into borrowing on heterogeneous potential returns to capital.

This study makes three contributions to the existing literature. First, I provide new evidence on how moral hazard and adverse selection can lead to an underprovision of fixed-wage jobs. While previous work has documented average incentive and sorting effects for single wage contracts, my analysis identifies the marginal treatment and selection schedules across a range of wage offers, including those not observed in equilibrium. These estimates form the primitives of an insurance-based model of wage contracts, allowing me to flexibly recover equilibrium, welfare loss, and policy counterfactuals in markets for fixed-wage labor---objects that have not been identified in previous research on worker compensation structure.

Second, this paper investigates the public costs and benefits of policies that promote fixed wages for short-term gig workers. Specifically, I estimate the marginal costs of public funds (MCPFs) from piece-rate taxes on online data-entry. The components of these MCPFs map directly to my experimental estimands, allowing me to flexibly compare the welfare benefits of inducing more fixed wages with the fiscal costs of increased worker shirking. Aside from its direct relevance to the regulation of online labor platforms, this analysis provides a lens through which future work might examine trade-offs associated with policies around tips, commissions, or bonuses in other labor markets.

Finally, this paper provides a methodological framework to flexibly estimate welfare loss from information asymmetries in markets for insurance or insurance-like contracts. Building on methods from \citet{karlanzinman} and \citet{bryan2015referrals}, my approach separately identifies treatment and selection over a range of experimental contract offers. Importantly, these offers include contracts that may not be profitable to a real-world firm, avoiding the ``under-the-lamppost'' problem inherent to many empirical studies of information asymmetries \citep{einav2010beyond}. Moreover, because I observe outcomes for both accepters and decliners of these contracts, I can use MTE methods to identify marginal selection on potential outcomes in both insured and uninsured states. The resulting MTR functions map directly to the components of a ``cost-curve'' insurance model \citep{efc,herbst2024opportunity}, allowing me to semi-parametrically identify welfare loss from asymmetric information. Applying this flexible estimation approach to a wide range of experimental contract offers improves upon traditional methods that rely on local price changes and linear extrapolation \citep{efc,hackmann2012health,hackmann2015mandate,finkelstein2019subsidizing}.

The rest of this paper proceeds as follows: In Section~\ref{sec:experiment}, I describe my experiment and underlying empirical strategy. In Section~\ref{sec:results}, I discuss the baseline results of the experiment. Section~\ref{sec:model} presents a model of hourly wage contracts under asymmetric information, and Section~\ref{sec:estmodel} estimates that model using MTE methods. Section~\ref{sec:policy} uses estimated MTR functions and associated value curves to calculate MVPFs for fixed-wage subsidies and piece-rate taxes. In Section~\ref{sec:discuss}, I discuss the broader implications of my findings and calibrate my model to the market for rideshare drivers. Section~\ref{sec:conclusion} concludes.

\section{Experimental Design}\label{sec:experiment}
In this section, I describe my experimental design and empirical strategy. The goal of my experiment is twofold: First, I want to identify the incentive effects of fixed-wage contracts on worker performance (moral hazard). Second, I want to identify how workers with different unobserved productive potentials self-select into these contracts (adverse selection). Separately identifying these forces poses an empirical challenge---differences in realized output between workers who opted into a given wage offer reflect both the ex-ante productivity differences between those self-selected groups and the causal effect of the different contracts they chose.

To overcome this challenge, my experiment offers data-entry workers a choice between a randomized fixed hourly wage and a standardized piece rate. Comparing realized output between individuals who faced different hourly wage offers but work under the same contract identifies adverse selection---both groups ultimately face the same treatment condition but made decisions under different ex-ante choice sets. At the same time, using wage offers as an instrument for take-up of the hourly contract allows me to separately identify treatment effects of hourly wages among those who accept the offer. I first illustrate the intuition behind this design with a stylized two-contract setting before extrapolating to settings with multiple wage offers.

\subsection{Example using a Single Wage Offer}\label{sec:single}

Consider a potential outcomes framework in which a worker $i$ chooses one of two mutually exclusive contracts---a piece rate and an hourly wage. Let $Y_{1i}$ denote $i$'s output if they work under the hourly wage, and let $Y_{0i}$ denote their output if they work under the piece rate. Given these potential outcomes, worker $i$'s observed output, $Y_i$, is given by $Y_i = D_iY_{1i} + (1-D_i)Y_{0i}$, where $D_i$ is a binary indicator for whether $i$ chooses the hourly wage. Differencing realized outputs between hourly workers ($D_i=1$) and piece-rate workers ($D_i=0$) would yield the sum of two components---average treatment-on-the-treated and average selection on untreated outcomes:
\begin{align}
  E[Y_i |D_i&=1] - E\left[Y_i |D_i=0\right]\nonumber\\
                                                     &=\underbrace{E\left[Y_{1i}-Y_{0i} |D_i=1\right]}_{\text{\phantom{U}Average Treatment on the Treated }\phantom{n}} +& \underbrace{E\left[Y_{0i} |D_i=1\right] - E\left[Y_{0i}|D_i=0\right]}_{\text{Average Selection on $Y_0$ }}.\label{eq:overall}\end{align}
The treatment-on-the-treated component corresponds to the average effect of hourly pay ($Y_{1i}-Y_{0i}$) among those who accept the hourly wage offer ($D_i=1$), while the selection component corresponds to the average difference in potential piece-rate outcomes ($Y_{0i}$) between those choosing hourly pay ($D_i=1$) and those choosing the piece rate ($D_i=0$). These components are difficult to separate because piece-rate outcomes among hourly workers ($Y_{0i}|D_i=1$) are not observed.

To separate these forces of treatment and selection, my experimental design randomizes the menu of compensation options faced by workers. In this simple two-contract setting, that means randomly assigning each worker to one of two offer conditions, $W_i\in\left\{0,1\right\}$. Only workers assigned to $W_i=1$ are offered the choice between the piece rate ($D_i=0$) and the hourly wage ($D_i=1$), while workers assigned to $W_i=0$ are paid the piece rate with no alternative.

Assume the offer condition, $W_i$, can only affect $Y_i$ through the choice of contract, $D_i$. Let $D^*_i$ denote worker $i$'s potential take-up of the hourly wage if given the option, so that observed take-up is given by $D_i=W_iD^*_i$. Comparing worker output across the two treatment-offer groups and scaling by the hourly-wage take-up rate yields the classic treatment-on-the-treated estimator from \citet{wald1940}:
\begin{eqnarray}
  \label{eq:wald}
            \underbrace{E\left[Y_{1i}-Y_{0i} |D^*_i=1\right]}_{\text{Average Treatment on the Treated}} & = & \frac{E\left[Y_i |W_i=1\right]  - E\left[Y_i |W_i=0\right]}{\pi},
\end{eqnarray}
where  $\pi\equiv \Pr\left(D_i = 1 | W_i=1\right)$, the share of hourly contracts accepted among those offered a choice ($W_i=1$).

Turning to the selection, I can identify the second component in Equation~\eqref{eq:overall}, average selection on $Y_0$, by comparing output between piece-rate workers in the control group ($W_i=0$) and piece-rate workers in the hourly-offer group ($W_i=1$) who declined the hourly wage offer:
\begin{eqnarray}
\underbrace{\left(E\left[Y_{0i} |D^*_i=1\right] - E\left[Y_{0i} |D^*_i=0\right]\right)}_{\text{Average Selection on $Y_0$ }} & = & \frac{E\left[Y_i |W_i=0\right]  - E\left[Y_i |D_i=0,W_i=1\right]}{\pi},\label{eq:selection}
\end{eqnarray}
where equality follows from randomized assignment.\footnote{Randomized assignment implies $E\left[Y_{0i} |W_i=1\right]=E\left[Y_{0i} |W_i=0\right]=E\left[Y_{i} |W_i=0\right]$, so \hbox{$E\left[Y_{0i} |D_i=1,W_i=1\right] = \frac{E\left[Y_i |W_i=0\right] - (1-\pi)E\left[Y_{i} |D_i=0,W_i=1\right]}{\pi}$.} Equation~\eqref{eq:selection} can also be derived by subtracting the Wald estimator~\eqref{eq:wald} from the difference in hourly versus piece-rate outcomes in the treatment-offer group,  $E\left[Y_i |D_i=1,W_i=1\right]  - E\left[Y_i |D_i=0,W_i=1\right]$.}

Figure~\ref{fig:design_1}, Panel A illustrates the intuition from Equation~\eqref{eq:selection}. The control group, by construction, is subject to the standardized piece rate, while the treatment-offer group is offered an hourly wage as an alternative. Selection is identified by comparing the control group (top left box) to those in the treatment group who chose to remain on the piece rate (top right box). This selection-on-unobservables estimator captures the average difference in potential untreated outcomes for ``compliers'' versus ``never-takers'' \citep{black2022simple,kowalski2023examine,mogstad2018using,huber2013simple}.

\subsection{Multiple Wage Offers and Second-Stage Randomization}\label{sec:estwage}
The example above provides a simplified illustration of my experimental design with a binary treatment assignment, $W_i\in\left\{0,1\right\}$. In practice, however, my experiment features several treatment groups facing different hourly wage offers alongside a standardized piece-rate offer.\footnote{The standardized piece-rate option is designed to represent a ``no-insurance'' outside option, under which workers sell their labor product at its market price.
  See footnote~\ref{fn:pfix}.} Including multiple wage offers with incomplete take-up allows me to identify selection on potential outcomes under both the piece rate (the untreated state, $Y_0$) and hourly wages (the treated state, $Y_1$) across multiple margins.

To see how, consider an example experiment with three offer conditions, $W_i\in \left\{0,L,H\right\}$. As in the previous example, control workers assigned to $W_i=0$ are offered the piece rate with no alternative. But now the remaining workers are randomly separated into two groups---workers assigned to $W_i=L$ are offered the choice between the piece rate and a low hourly wage, while workers assigned to $W_i=H$ are offered the choice between the piece rate and a high hourly wage.

As before, I can identify selection on $Y_0$ by comparing outcomes between piece-rate workers who faced different ex-ante hourly offers---none ($W_i=0$), low ($W_i=L$), or high ($W_i=H$). Because all three of these groups face the same ex-post payment terms but different ex-ante wage offers, comparisons between them isolate worker selection on potential output under the piece rate, $Y_0$. But unlike the single-wage-offer example, this multiple-offer design allows me to also identify selection on $Y_1$, potential output under hourly pay, by comparing hourly workers who \emph{accepted} their respective wage offers. Importantly, however, this identification result requires wage offers to only affect workers' output through their choice of hourly versus piece-rate contracts. If accepters in both high- and low-offer groups work under the wages they were offered, this exclusion restriction could be violated through wage effects---higher pay might induce greater effort through increased motivation or satisfaction, biasing my estimates. To separate these wage effects from the incentive effects of hourly contract structure, the second stage of my experiment increases wages for a random subset of workers accepting the low hourly wage offer, equalizing their pay with the high hourly wage offer just before the task begins. This surprise top-up creates random variation in \emph{effective} wages among those accepting a given \emph{offered} wage, allowing me to separate potential wage effects from both moral hazard and adverse selection.

\begin{figure}[h!]\vspace{-.5cm}
    \centering \caption{Experimental Design}
    \label{fig:design_1}
  \begin{subfigure}{\textwidth}
    \vspace{-.1cm}
    \centering
    \includegraphics[scale=0.35,trim=3cm 0cm 3cm 0cm,clip]{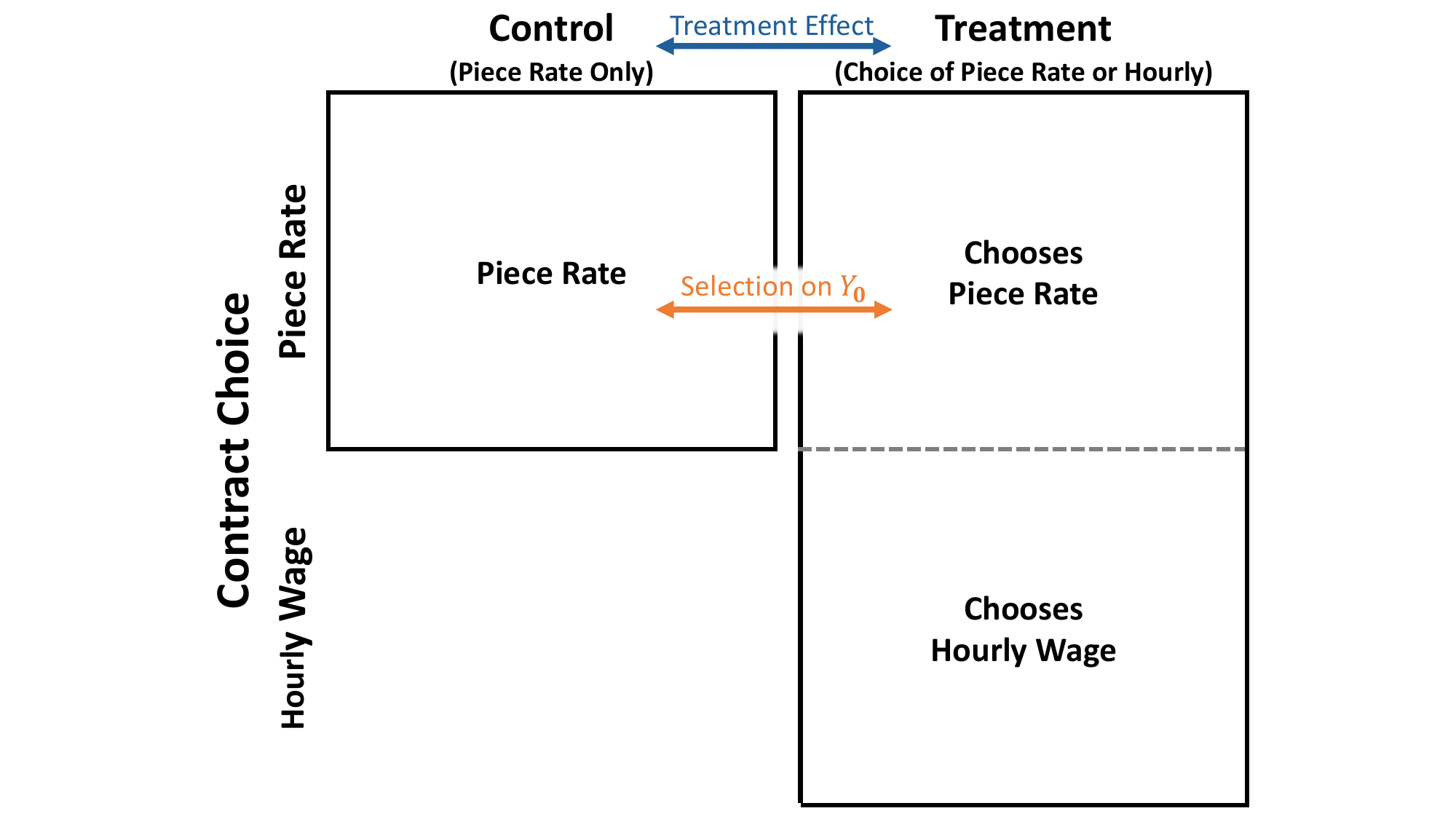}
    \caption{Single Treatment Example}
  \end{subfigure}
  \\\vspace{.2cm}
  \begin{subfigure}{\textwidth}
    \includegraphics[scale=0.4,trim=0cm 1cm 0cm .5cm,clip]{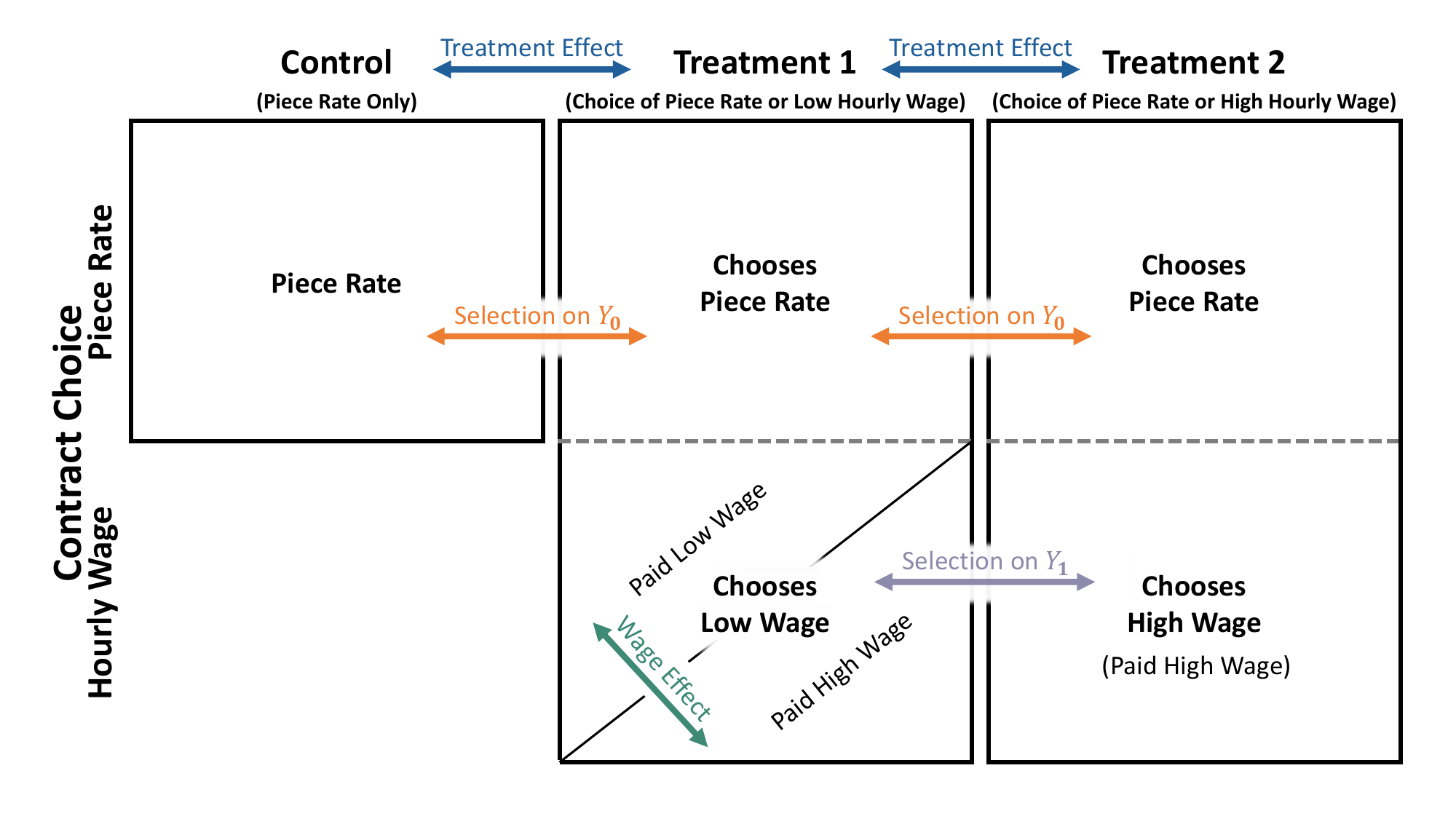}
    \centering \caption{Multiple Treatment Example}\label{fig:design_2}
\vspace{.32cm}
  \end{subfigure}
  \begin{figurenotes}This figure provides a graphical illustration of my experimental design, simplified to include only two (Panel A) or three (Panel B) wage-offer conditions. In each panel, columns denote experimental groups with different offer conditions. The control group is only compensated through the piece-rate contract, and treatment groups are offered a choice between the piece-rate contract and different hourly wages. Treatment groups are separated into those who accept the piece-rate contract (upper boxes) and those who accept their respective hourly wage offers (lower boxes). The diagonal split in the bottom box of Treatment 1, Panel B represents the second stage of randomization, in which some workers accepting the low hourly wage are promised the higher wage before they begin the task. Arrows denote comparisons that, after scaling by take-up, identify treatment effects (blue), selection on output under the piece rate, $Y_0$ (orange), selection on output under hourly wages, $Y_1$ (purple), and wage effects (teal). See Section~\ref{sec:experiment} for details.
  \end{figurenotes}
  \vspace{-.5cm}
\end{figure}

Figure~\ref{fig:design_1}, Panel B illustrates the intuition behind the multiple wage-offer design. The top row of boxes represents individuals in each of the three experimental groups, $W_i\in \left\{0,L,H\right\}$, who remain on the piece rate. Because all three of these groups face the same ex-post payment terms but different ex-ante wage offers, comparisons between them isolate worker selection on productivity under the piece rate, $Y_0$. The bottom two boxes represent workers who accepted low ($W_i=L$) and high ($W_i=H$) hourly wage offers, respectively. In the second stage of the experiment, a random subset of those accepting the low hourly wage is promised additional top-up compensation before beginning the task. This surprise top-up equalizes their effective wage with that of the high-offer group, allowing me to separate wage effects (diagonal arrow) from selection on productivity under hourly wages, $Y_1$ (horizontal arrow). I formally derive these identification results in Appendix~\ref{sec:expdetails}.

\subsection{Experimental Setting and Implementation}\label{sec:setting}

The design and recruitment details for this experiment were pre-registered on the AEA RCT Registry under ID AEARCTR-13138 \citep{registry}. Appendix~\ref{sec:settingdetails} provides complete details on the experimental protocol.

Participants in my experiment were recruited on Prolific, an online platform that allows clients to hire online workers for short-term tasks.\footnote{\citet{douglas2023data} find that the Prolific platform compares favorably to Amazon Mechanical Turk (``MTurk'') and other platforms across several dimensions of data quality.}  The experimental job posting offered workers a \$1.00 reward plus \$0.03 for each correctly typed sentence transcribed over five minutes. Similar ``micro tasks'' of a few minutes or less are commonly requested on Prolific and other online platforms for the purpose of training artificial intelligence (AI) algorithms \citep{hornuf2022hourly}. Participants were not informed that they were part of an experiment until after they performed the task.

Upon accepting the experimental job, workers were randomized into one of eighteen experimental groups. Each group was offered a different menu of compensation options in exchange for completing the task. In the first treatment group, workers were offered a choice between a fixed \$0.10 payment (\$1.20 per hour) and a piece rate of \$0.03 per correctly typed sentence.\footnote{A piece rate of \$0.03 per correctly typed sentence was chosen to roughly align with observed market rates for online image-to-text transcription services. In 2024, Transcription US advertised \$2 per 350-word page for handwritten typing \citep{TranscriptionUS_TypingRates2024}, while Upwork reported a median hourly rate of \$13 for data-entry specialists and a typing benchmark of 50--60 words per minute \citep{Upwork_DataEntryRates2024,Upwork_DataEntryGuide2024}. Applied to the average sentence in my experiment, these figures imply rates of \$0.042 and \$0.026--\$0.031 per sentence, respectively.} In the second treatment group, workers were offered a choice between a fixed \$0.15 payment (\$1.80 per hour) and the same \$0.03 piece rate. Additional treatment groups follow the same structure, with each condition offering the \$0.03 piece rate but increasing the flat wage offer by multiples of \$0.05, up to a maximum of \$1.75 (\$21.00 per hour). A control group was offered the \$0.03 piece rate for each correctly typed sentence, with no alternative option. Each of these options was offered on top of the \$1.00 reward advertised in the job posting, which all workers received for agreeing to the task. Experimental conditions are listed in Table~\ref{tab:groups}.

The experiment took place in ten waves of three hundred job postings. Waves were launched at a broad range of days and times to ensure the sample was not restricted to workers with a particular schedule. Data on task performance were collected at the conclusion of each wave. The primary outcome of interest is hourly output value, defined as
\begin{equation}
  \label{eq:corvalue}
  \text{Output Value}\equiv \frac{\text{Correct Sentences} \times \$0.03}{\sfrac{1}{12} \text{ Hours}}.
\end{equation}
Output value is linked to self-reported background information from workers' Prolific profiles. Specifically, I observe each worker's gender, ethnicity, age, employment status, and whether they are a student. I also observe the prior number of tasks they have successfully completed through the Prolific platform. Because the goal of my experiment is to identify selection on \emph{private} information, conditioning on these potentially screenable characteristics is important to simulate a sample of workers who would be observably equivalent to a hypothetical employer.

Importantly, clients on the Prolific platform have the ability to reject or approve a given worker's assignment. Rejected assignments do not earn rewards and lower workers' approval ratings. The reputational damage from rejected assignments is a salient concern among workers on Prolific and similar platforms.\footnote{See, for example, the Reddit forum r/ProlificAc (\url{https://www.reddit.com/r/ProlificAc/comments/11yt72u/}).}
As in most labor markets, this threat of rejection creates an incentive for online workers like those in my experiment to maintain a minimum standard of performance, even if they are paid a fixed wage.

\paragraph{Setting Advantages} The online labor platform and data-entry task used in my experiment offer several distinct advantages over alternative settings or tasks. First, online labor markets and data-transcription services represent a large and growing part of the broader gig economy. Over 160 million workers are registered on online labor platforms \citep{KassiLehdonvirta2018,KassiEtAl2021ssrn}, which have drawn increased attention from both policymakers and academics \citep{stanton2025benefits,Horton2025,caplin2024economic,pallais2016referential,pallais2014inefficient,Horton2017,HortonRandZeckhauser2011}. Meanwhile, the market for AI ``data-labeling'' tasks like the one in my experiment has been valued at up to \$19 billion and is expected to grow rapidly \citep{grandview2024datalabeling, mordor2025datalabeling}. Second, I can control the complete menus of wage contracts and observe workers' output regardless of their contract choice. This means I can observe worker selection between a single predetermined outside option (the \$0.03 piece rate) and a wide range of experimental wage offers, including those that would be unprofitable to a real-world employer. These margins of selection are critical determinants of welfare loss but impossible to observe in most settings.\footnote{For example, an experiment that recruits participants through randomized posted wages can only measure selection relative to workers' existing outside options, which likely include competing offers from other fixed-wage employers \citep{dube2020monopsony}. In this case, experimental estimates would likely overstate the elasticity of fixed-wage labor supply and fail to capture selection into below-market offers. By contrast, my design ensures that participants' choice sets only include experimental offers, which they observe after accepting a generous upfront fee recruiting them to the task.}
Third, I can recruit a broad sample of workers in the targeted labor market, not just those opting into a particular employer or wage contract. By contrast, an experiment conducted at an existing workplace would likely exclude workers who prefer self-employment or freelance work, eliminating the very margin of selection I seek to identify. Finally, the protocol for this experiment can be easily replicated: Researchers can validate or extend my experimental intervention in comparable populations of online workers at minimal cost.

\section{Baseline Experimental Results}\label{sec:results}
This section describes baseline reduced-form results from the experiment.
A total of \fp[0]{\N} workers participated in the experiment: \fp[0]{\Ncontrol} workers were assigned the control condition with no fixed-wage offer, and the remaining \fp[0]{\N-\Ncontrol} workers were each assigned to one of the 17 wage-offer conditions.\footnote{One worker exited the task before observing their experimental wage offer and was dropped from the experimental sample. All other workers remained in the sample, even if they failed to enter sentences or click the submit button after the five-minute timer expired. If a worker failed to click the submit button within thirty minutes of accepting the task, the Prolific task scheduler automatically re-assigned their treatment condition to a new worker, even while the unsubmitted task remained in my sample. These re-assigned treatments result in an observation count ($N=\fp[0]{\N}$) that exceeds my pre-registered sample size of 3,000.} Sample sizes for each experimental group are listed in Table~\ref{tab:groups}.\footnote{Appendix Table~\ref{tab:balance} reports balance tests of joint significance between baseline demographics and group assignment.}

Table~\ref{tab:summ} reports summary statistics for the experimental sample. Excluding those assigned to the no-offer control group, \fp[0]{100*\meantakeup} percent of workers accepted their experimental fixed-wage offers. Across all experimental conditions, workers completed an average of \fp{\meannumscore} sentences within five minutes, of which \fp{\meanscore} were typed correctly, resulting in a mean hourly output value of \$\fp[2]{\meancorvalue}.

\begin{figure}[p]
     \centering{}\protect\caption{Hourly Wage Take-Up}\label{fig:bar_hourly}
\resizebox{\columnwidth}{!}{\includegraphics[trim=.4cm .5cm 0cm 0cm,clip]{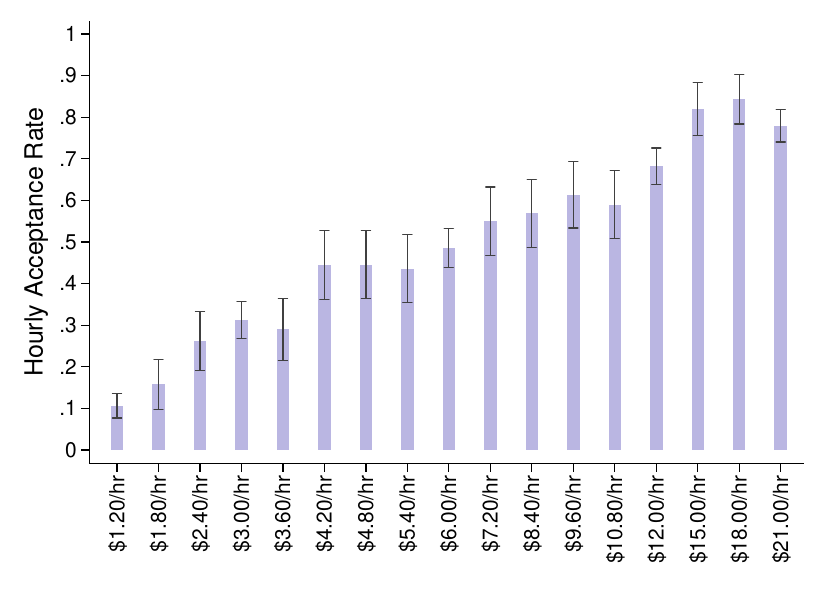}}
    \begin{figurenotes}This figure reports hourly-wage acceptance rates by treatment group. The y-axis measures the share of workers in each group who declined the \$0.03 piece rate in favor of the hourly wage offer displayed on the x-axis. Note that both piece rate and hourly wage options are offered as supplements to a base wage of \$12.00 per hour.  Bands indicate 90\% confidence intervals.
    \end{figurenotes}
  \end{figure}

The bar chart in Figure~\ref{fig:bar_hourly} shows the share of workers in each treatment group who accepted their hourly wage offer instead of the \$0.03 piece rate. Unsurprisingly, the relative supply of hourly workers increases with the offered wage. On average, wage offers of \$\fp[0]{round(\rangeBhi,0)} or lower were accepted at a rate of \fp[0]{100*\hourlyB} percent, while wage offers above \$\fp[0]{round(\rangeChi,0)} were accepted at a rate of \fp[0]{100*\hourlyD} percent. To fit these results to a continuous supply curve, Table~\ref{tab:Sc} reports estimated coefficients from a logistic regression of a binary indicator for hourly acceptance against log wage offer, excluding the control group. Column 1 reports estimates from a univariate specification, while Columns 2 through 4 successively add controls for task timing, employment, and demographics. In each specification, I find a statistically significant effect of log wage offer on hourly take-up, with estimates ranging from \fp{\bSc} (SE=\fp{\seSc}) to \fp{\bScXXX} (SE=\fp{\seScXXX}) depending on the inclusion of controls.

\begin{figure}[p]
    \centering{}\protect\caption{Worker Output Value by Treatment Offer and Acceptance Status}\label{fig:bar_value}
    \resizebox{\columnwidth}{!}{\includegraphics[trim=.3cm .4cm 0cm .5cm,clip]{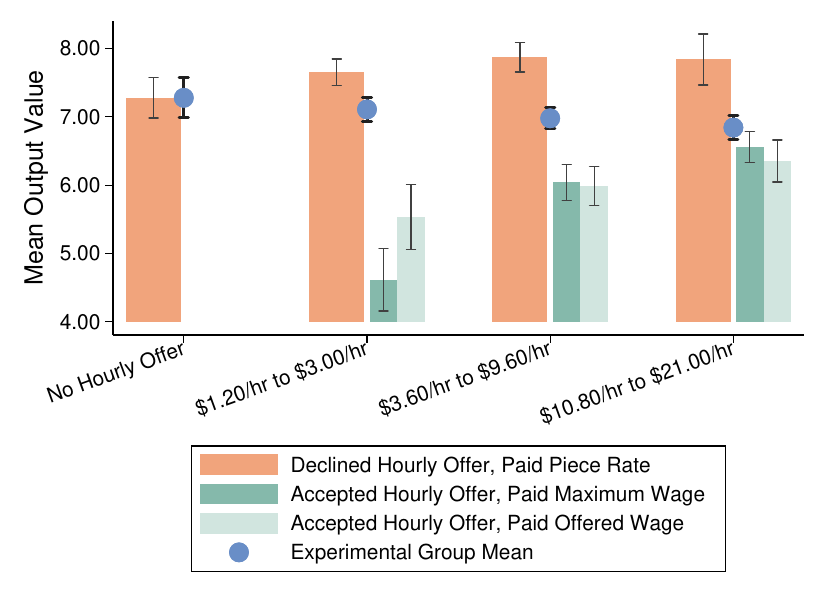}}\\
    \begin{figurenotes}This figure shows mean worker output value by wage-offer groups and compensation choice. ``Output value'' is defined as the number of correctly typed sentences per hour multiplied by \$0.03. Control and treatment groups are labeled on the x-axis. Blue circles measure mean output values among all individuals in each group. Orange bars measure mean output values among those who were paid the \$0.03 piece rate. Dark green bars measure mean output values among hourly workers paid the \$21.00 per hour maximum, either through a randomized top-up or by accepting that offer directly. Light green bars measure mean output values among hourly workers paid their original offers below the maximum. Note that both piece rate and hourly wage options are offered as supplements to a base wage of \$12.00 per hour. Bands indicate 90\% confidence intervals.
    \end{figurenotes}
    \vspace{-.5cm}
  \end{figure}

The increase in take-up rates observed in Figure~\ref{fig:bar_hourly} generates different groups of accepters (hourly workers) and decliners (piece-rate workers) for each hourly offer. In Figure~\ref{fig:bar_value}, I examine how average output value for these groups varies with the generosity of the wage offer.\footnote{Output value is measured using correctly completed sentences. Appendix Tables~\ref{tab:TEcnum} and~\ref{tab:AScnum} report the corresponding treatment and selection estimates when the outcome variable is instead defined using completed sentences, regardless of correctness.} For readability, experimental conditions are aggregated into four categories---those in the control group who received no hourly offers, those receiving wage offers between \$\fp[2]{\rangeBlo} and  \$\fp[2]{\rangeBhi}, those receiving wage offers between \$\fp[2]{\rangeClo} and \$\fp[2]{\rangeChi}, and those receiving wage offers of \$\fp[2]{\rangeDlo} or higher. Blue circles measure average output values among all individuals in each of these categories. Orange bars measure average output values among workers on the piece rate. Dark green bars measure average output values among hourly workers paid the \$21.00 per hour maximum, either through a randomized top-up or by accepting that offer directly. Light green bars measure average output values among hourly workers paid their original offers below the maximum. Following Section~\ref{sec:experiment}, I compare these different groups across wage-offer categories to identify treatment and selection into hourly pay, providing evidence for both moral hazard and adverse selection.

\paragraph{Treatment Effect of Hourly Wages} In the absence of wage effects, comparing means across each aggregated group in Figure~\ref{fig:bar_value} identifies the intent-to-treat effect of hourly wages. The blue circles decline with the generosity of the wage offer, suggesting hourly wages reduce worker output, consistent with moral hazard. Those in the control group, who received no hourly wage offer, produce \$\fp[2]{\corvalueA} (SE=\$\fp[2]{\corvalueseA}) of output value. Those receiving wage offers between \$\fp[2]{\rangeBlo} and  \$\fp[2]{\rangeBhi} produce \$\fp[2]{\corvalueB} (SE=\$\fp[2]{\corvalueseB}) of output value, those receiving offers between \$\fp[2]{\rangeClo} and \$\fp[2]{\rangeChi} produce \$\fp[2]{\corvalueC} (SE=\$\fp[2]{\corvalueseC}) of output value, and those receiving offers of \$\fp[2]{\rangeDlo} and above produce \$\fp[2]{\corvalueD} (SE=\$\fp[2]{\corvalueseD}) of output value.

To obtain point estimates for the treatment effect of hourly wages on workers' output value, I disaggregate the groups from Figure~\ref{fig:bar_value} into a continuous measure of wage offers, which I use as an instrumental variable for hourly contract take-up. To remove the potential influence of wage effects, I first regress output value against log effective hourly wages (i.e., inclusive of top-ups) and a full set of experimental group dummies for the sample of hourly workers who were eligible to receive random top-ups. I then residualize hourly workers' output values by subtracting the demeaned wage effect implied by the coefficient on log effective hourly wages. I use this residualized measure of output value as the dependent variable in a two-stage least-squares (2SLS) regression where I instrument for hourly wage take-up with log wage offers and an indicator variable for being in the no-offer control group.\footnote{Appendix Table~\ref{tab:NTEc} reports estimates from 2SLS regressions using unresidualized output values. Results from these specifications are nearly identical to baseline results that adjust for potential wage effects.}

Table~\ref{tab:TEc} reports 2SLS estimates of the treatment effects of hourly wages on output value. As in Table~\ref{tab:Sc}, Column 1 reports estimates from a univariate specification, while Columns 2 through 4 successively add controls for task timing, employment, and demographics. Across all four specifications, hourly contracts reduce worker productivity; the estimates are significant at the 1 percent level in Columns 1 through 3 and at the 5 percent level in Column 4. Absent controls, accepting an hourly contract reduces a worker's output value by \$\fp{-1*\bTEccorvalue} (SE=\$\fp{\seTEccorvalue}) or \fp{-100*\bTEccorvalue/\meancorvalue} percent of the sample mean. Adding controls for task specifics changes this estimate to \$\fp{-1*\bTEcXcorvalue} (SE=\$\fp{\seTEcXcorvalue}), while adding employment and demographic controls reduces it to  \$\fp{-1*\bTEcXXcorvalue} (SE=\$\fp{\seTEcXXcorvalue}) and  \$\fp{-1*\bTEcXXXcorvalue} (SE=\$\fp{\seTEcXXXcorvalue}), respectively.
These negative and significant treatment effect estimates are consistent with moral hazard---on average, workers' output is lower under hourly wages, when they do not bear the financial cost of decreased effort, than under piece rates, when their earnings are more closely tied to output.

\paragraph{Selection into Hourly Wages} While comparisons of pooled means across wage-offer groups identify treatment effects, comparisons of means for accepters and decliners across wage-offer groups identify selection on potential output under hourly wages and piece rates, respectively. In Figure~\ref{fig:bar_value}, the orange bars generally increase with the wage offer, meaning those who decline the most generous hourly wages in favor of the piece rate have relatively high output. Those declining offers between \$\fp[2]{\rangeBlo} and  \$\fp[2]{\rangeBhi} produce \$\fp[2]{\corvaluePB} (SE=\$\fp[2]{\corvaluesePB}) of average output value, those declining offers between \$\fp[2]{\rangeClo} and \$\fp[2]{\rangeChi} produce \$\fp[2]{\corvaluePC} (SE=\$\fp[2]{\corvaluesePC}) of average output value, and those declining offers of \$\fp[2]{\rangeDlo} and above produce \$\fp[2]{\corvaluePD} (SE=\$\fp[2]{\corvaluesePD}) of average output value. All three of these averages exceed the \$\fp[2]{\corvaluePA} (SE=\$\fp[2]{\corvaluesePA}) of average output value produced by exclusively piece-rate workers in the no-offer control group. These patterns are consistent with adverse selection on $Y_0$, potential output under the piece rate.

The green bars also rise with the wage offer, meaning those who accept the lowest hourly offers over the piece rate have relatively low output. Restricting attention to hourly workers who were paid the same effective rate of \$21.00 per hour (dark green bars), I find that those accepting offers between \$\fp[2]{\rangeBlo} and  \$\fp[2]{\rangeBhi} produce \$\fp[2]{\corvalueTB} of average output value, those accepting offers between \$\fp[2]{\rangeClo} and \$\fp[2]{\rangeChi} produce \$\fp[2]{\corvalueTC} of average output value, and those accepting offers of \$\fp[2]{\rangeDlo} and above produce \$\fp[2]{\corvalueTD} of average output value. These patterns are consistent with adverse selection on $Y_1$, potential output under the hourly wage. Note that the average output values among hourly workers paid their original offers below the maximum (light green bars) exhibit a similar pattern to those paid the maximum, suggesting that wage effects are not important in this setting.

Disaggregating the groups from Figure~\ref{fig:bar_value} into respective experimental wage offers, I use ordinary least squares (OLS) estimation to fit the selection patterns seen in Figure~\ref{fig:bar_value} to the following linear model:
\begin{equation}
  \label{eq:ols}
  Y_i= \alpha D_i + \beta_0(1-D_i)\times W_i + \beta_1D_i\times W_i + \gamma D_i \times W^P_i + \boldsymbol{\xi}'X_i + \epsilon_i,
\end{equation}
where $W_i$ is worker $i$'s log hourly wage offer, $D_i$ is a binary indicator for whether they accept the hourly wage, $W^P_i$ is the log wage that hourly workers are actually paid (equal to zero for piece-rate workers), and $X_i$ represents a vector of covariates, a constant term, and a dummy variable for being in the control group.\footnote{Rather than include a common $W_i$ term and only one interaction term, Equation~\eqref{eq:ols} includes full interactions of wage offers with acceptance status, $(1-D_i)\times W_i$ and  $D_i\times W_i$. While the two models are equivalent, the parameterization of $\beta_0$ and $\beta_1$ in the fully interacted specification is easier to interpret. Note that $(1-D_i)\times W_i^P$ is excluded because $W_i^P=0$ for all $D_i=0$ (piece-rate workers are ineligible for top-ups).}

Table~\ref{tab:ASc} reports OLS estimates of coefficients from Equation~\eqref{eq:ols}. In Column~1, the estimated coefficient on ``Declined $\times$ Log Hourly Wage Offer'' implies that increasing wage offers by one log point corresponds to a \$\fp[2]{\bAScPcorvalue} (SE=\$\fp[2]{\seAScPcorvalue}) increase in output value among those who remain on the piece rate, while the coefficient on ``Accepted $\times$ Log Hourly Wage Offer'' implies that productivity among hourly workers increases by \$\fp[2]{\bAScHcorvalue} (SE=\$\fp[2]{\seAScHcorvalue}) per log point. In all four columns, the latter coefficient is similar in magnitude and significant at the 1 percent level, suggesting worker selection on ex-ante productivity is not captured by observable characteristics.
The former remains positive in all four columns but reaches conventional significance only in Column~4, where it is significant at the 10 percent level.
Note that these coefficients correspond to changes in the average productivity of the respective worker pools; the productivity of \emph{marginal} workers---those who switch to the fixed-wage contract in response to a higher offer---would need to increase by roughly twice these magnitudes to generate these estimated coefficients. For example, a coefficient of  \fp[2]{\bAScHcorvalue} on ``Accepted $\times$ Log Hourly Wage Offer'' implies that a one-log-point higher wage offer attracts a \$\fp[2]{2*\bAScHcorvalue}-more productive marginal worker, an increase of \fp[1]{200*\bAScHcorvalue/\meancorvalue} percent over the mean. See Appendix~\ref{sec:expdetails} for details.

Finally, note that the estimated coefficient on ``Accepted $\times$ Log Effective Hourly Wage'' in Table~\ref{tab:ASc} is small and statistically insignificant in all four columns. These estimates imply that randomized wage top-ups failed to elicit a meaningful productivity response, once again suggesting that wage effects are not important in this setting.

\begin{figure}[p]
    \centering{}\protect\caption{OLS Estimates of Selection on Output Value by Wage Offer}\label{fig:coef}
    \resizebox{\columnwidth}{!}{\includegraphics[trim=0cm .4cm 0cm 0cm,clip]{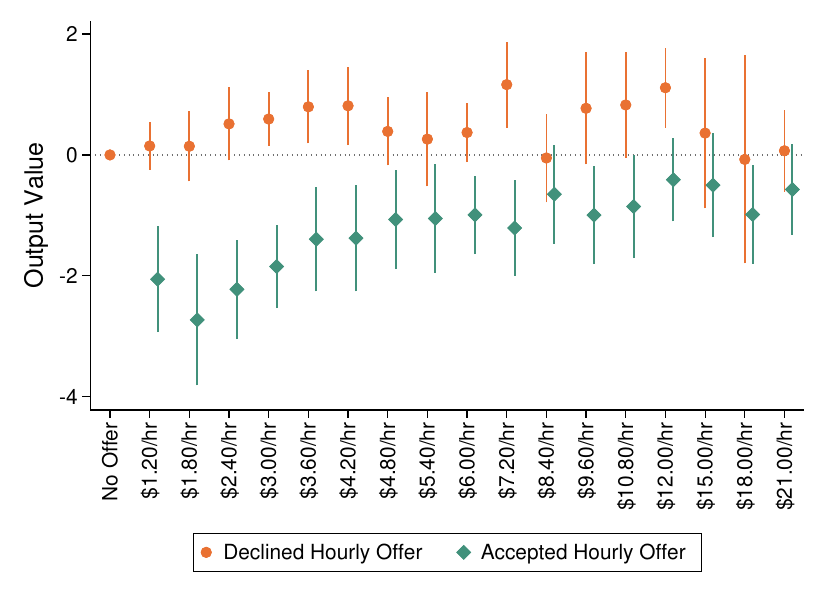}}
    \begin{figurenotes}This figure plots coefficients from an OLS regression of output value against the full set of dummy variables for each experimental wage offer, controlling for log effective wages among hourly workers (inclusive of top-ups) as well as task timing and employment controls. Orange dots represent coefficients on hourly wage offers interacted with an indicator for remaining on the piece rate. Green diamonds represent coefficients on hourly wage offers interacted with an indicator for accepting the offer. Lines represent 90\% confidence intervals.
    \end{figurenotes}
  \end{figure}

Figure~\ref{fig:coef} plots OLS estimates from a modified version of Equation~\eqref{eq:ols} that replaces the linear log-wage-offer term, $W_i$, with a full set of dummy variables for each experimental wage offer. Covariates include log effective wages among hourly workers, task timing, and employment controls. Orange dots represent coefficients on hourly wage offers interacted with an indicator for remaining on the piece rate. Green diamonds represent coefficients on hourly wage offers interacted with an indicator for accepting the offer. The upward slope in both series indicates adverse selection into hourly wages---as wage offers decrease, the most productive workers opt out of hourly work and into the piece rate, resulting in lower average productivity among both hourly and piece-rate workers.

\paragraph{Instrument Validity} In the description above, I interpret my experimental estimates as treatment and selection effects of fixed-wage contracts. In theory, this interpretation could be threatened by behavioral responses to the experimental intervention.
For example, workers' effort might exhibit reference dependence after being ``primed'' with a wage offer before the task begins.\footnote{A large literature shows how ``loss aversion''---a form of reference dependence---can explain the sensitivity of short-term labor supply decisions to benchmark earnings levels \citep{crawford2011nyctaxi,thakral2021adaptive}. While experimental offers are deliberately not framed as potential losses, workers may still feel regret if they earn less than their initial wage offer.} If workers place a premium on earning at least as much as they were offered, their potential piece-rate output, $Y_0$, might increase with the generosity of their experimental wage offer, $W_i$, violating the exclusion restriction. In this case, however, one would expect the density of piece-rate workers' earnings to decline discontinuously just above the wage-offer reference point \citep{mccrary2008manipulation}.\footnote{Recall that the ``counter'' function in my experiment allows workers to observe piece-rate earnings in real time, making output highly manipulable.} Appendix Figure~\ref{fig:lossbunching} reports results from a partial test of this hypothesis and finds no evidence of such discontinuity. Using methods from \citet{cattaneo2020simple}, I estimate piece-rate earnings densities on either side of the wage-offer threshold and find no significant difference (p-value={\plossbunching}), suggesting minimal influence from reference points.\par
A related concern is the potential presence of menu or framing effects \citep{simonson1992choice}. If workers' productivity were influenced not only by their compensation scheme, but also by their ability to \emph{choose} that compensation scheme, then comparisons between treatment and control would be biased. Unless framing effects interact with the amount offered, any such bias would have to be small or negative to produce the statistically insignificant estimated differences between ``No Offer'' control group workers and the piece-rate workers in the lowest offer treatment groups (see Figure~\ref{fig:coef}).

One might also be concerned about framing effects biasing estimates of wage effects. Because variation between offered and effective wages is generated through unexpected top-ups, estimated wage effects could also include behavioral responses to receiving a surprise pay raise.
Note, however, that most behavioral mechanisms like reciprocity through gift exchange \citep{falk2007gift,akerlof1982labor} operate in the same direction as wage effects. The precise null estimates on ``Accepted $\times$ Log Effective Hourly Wage'' in Table~\ref{tab:ASc} likely rule out these mechanisms.\footnote{I can also distinguish pure ``surprise'' effects from true wage effects by testing the significance of the intercept term---a response to surprises should persist for very small wage top-ups, whereas wage effects should reduce to zero as the effective wage approaches the offered wage. Adding an indicator variable for receiving \emph{any} hourly raise (i.e., $R_i\equiv\boldsymbol{1}\left\{\text{Effective Hourly Wage}>\text{Offered Hourly Wage}\right\}$) in Equation~\eqref{eq:ols} yields a null result. That said, I view this test as largely redundant in light of the null wage-effects estimates in Table~\ref{tab:ASc}.}

\paragraph{Summary of Results} To summarize, experimental results provide evidence of both adverse selection and moral hazard in hourly wage contracts---workers who decline more generous wage offers have higher piece-rate productivity, workers who accept less generous wage offers have lower fixed-wage productivity, and the pooled average output of accepters and decliners falls with generosity of the wage they were offered.\footnote{In Appendix Table~\ref{tab:mht}, I report multiplicity-adjusted $p$-values for tests of these three hypotheses using the multiple hypothesis testing procedure from \citet{list2019multiple}.} In the following sections, I develop a model to investigate how these forces affect labor-market equilibrium and worker welfare. I then show how the components of this model correspond to the potential outcome marginal treatment response (MTR) functions under fixed-wage and piece-rate counterfactuals---objects that I estimate in the subsequent MTE analysis.

\section{Model of Asymmetric Information in Wage Contracts}\label{sec:model}
In this section, I present a model of short-term labor markets under asymmetric information. The model borrows from \citet{efc} and \citet{herbst2024opportunity}, who develop models of asymmetric information in health insurance markets and college financing markets, respectively.\footnote{In Appendix~\ref{sec:extensions}, I discuss how the model can be extended to incorporate monitoring costs, partially fixed compensation, multiple contract structures, wage effects, and dynamic contracts.}

Consider a perfectly competitive labor market in which risk-neutral firms face a fixed population of workers.
Assume this population has already been pre-screened, so that workers are observably equivalent from the perspective of firms.\footnote{\label{fn:screen}In my empirical analysis, I allow employers to set wages using public information, $X_i$, like individual work history and approval rating, which can act as signals of future productivity \citep{spence1973job,FarberGibbons1996,gibbons1991layoffs,pallais2016referential}. For simplicity, I omit these ``$X_i$'' terms from the model, which effectively considers a subpopulation of workers with observables matching a particular value, $X_i=x$.} Each worker, $i$, can produce some level of hourly output, $q_i=f\left(\zeta_i,e_i,\nu_i\right)$, which is a function of unobserved worker characteristics ($\zeta_i$), individual effort ($e_i$), and random noise ($\nu_i$). Assume firms know the data-generating process, so they form unbiased beliefs about the distribution of $q_i$ but cannot observe the ex-ante productivity of any individual worker.

Facing this population of observably identical workers with unknown productivity, firms have two options to purchase workers' labor product. One option is to buy worker output at a constant market price of $p$ per unit, either through freelance hiring or more formal piece-rate employment.\footnote{\label{fn:pfix}Note that competitive employers cannot profitably offer piece rates that deviate from this market price, $p$, which is exogenously fixed.
  This model is isomorphic to one in which firms sell earnings insurance at some fixed premium to self-employed workers whose labor product is competitively priced at \$$p$ per unit.
} Alternatively, they can offer a flat, up-front wage, $w$, in exchange for a claim on a worker's hourly output, $q_i$.\footnote{\label{fn:twocontract}I limit the menu to a pure fixed wage and a pure piece rate to reflect the principal compensation structures observed in short-term online labor markets, where work is commonly organized around time- or task-based pay \citep{hornuf2022hourly,upwork2026contracttypes,prolific2026paymentmodel,clickworker2026crowd,hara2018earnings}. As in \citet{efc}, the resulting welfare analysis is conditional on a fixed contract space and is therefore best suited to settings in which that space is limited. Appendix~\ref{sec:extensions} considers mixed and multiple contract structures.} Firms have constant returns to scale, may freely enter, and post fixed wages, $w$, via Bertrand-style competition.

For an individual worker, $i$, I define the reservation wage, $\overline{w}_i$, as the minimum $w$ at which they would accept an hourly contract over selling their labor product at the piece rate. The relative supply of hourly workers is given by
\begin{equation}
  \label{eq:supply}
S(w) \equiv \Pr\left(\overline{w}_i \leq w \right).
\end{equation}
Assuming $S(w)$ is continuous and strictly monotonic ($S(w) > S(w')$ for all $w>w'$), I index workers by a type parameter, $\theta_i\in\left[0,1\right]$, equal to the share of the worker population willing to accept a lower wage than worker $i$'s reservation wage, $\theta_i \equiv S(\overline{w}_i)$. I can then rewrite a worker's reservation wage as a function of their type, $\overline{w}_i=\overline{w}(\theta_i)$, where
\begin{equation}
  \label{eq:wbar}
  \overline{w}(\theta)\equiv S^{-1}(\theta).
\end{equation}

To determine the profit-maximizing wage, firms consider the value of hourly output produced across worker types. I define the \emph{marginal value} of type $\theta$ as
\begin{equation}
  \label{eq:MV}
MV\left(\theta\right)\equiv E\left[Y_i|\theta_i=\theta\right],
\end{equation}
where $Y_i =  pq_i$, the incremental value of output $q_i$ produced by worker $i$ under an hourly contract.
 $MV(\theta)$ equals the expected amount type $\theta$'s hourly output would have earned them under the market piece rate, $p$. However, type $\theta$'s reservation wage may fall below this ``actuarially fair'' wage (i.e., $\overline{w}\left(\theta\right)<MV\left(\theta\right)$), especially if they are risk averse.

If $\overline{w}\left(\theta\right)<MV\left(\theta\right)$, a fully informed employer could profit from offering an hourly wage of $w= \overline{w}\left(\theta\right)$ exclusively to workers of type $\theta$. However, if employers cannot observe types, they cannot prevent some worker $i$ with $\theta_i\neq\theta$ from opting into a contract offered at wage $w= \overline{w}\left(\theta\right)$. In this case, the hourly position would be accepted by all workers $\theta_i\leq \theta$---those with reservation wages no higher than the posted wage. So instead of obtaining type $\theta$'s marginal value, $MV\left(\theta\right)$, the employer would obtain their \emph{average value}, defined as
\begin{equation}
AV\left(\theta\right)\equiv E\left[Y_i|\theta_i \leq \theta\right].\label{eq:AV}
\end{equation}
The average value, $AV\left(\theta\right)$, of type $\theta$ is given by the average value of output produced among all types $\theta_i \leq \theta$. When I account for this selection into contracts, the employer's expected profits from hiring a worker at some wage $w$ are given by
\begin{equation}\label{eq:profit}
\Pi\left(w\right)=S(w) \left(AV\left(\theta^w\right) - w\right),
\end{equation}
where $\theta^w\equiv S(w)$, the worker type with reservation wage equal to $w$.

I assume that the lowest-reservation-wage worker, $\theta=0$, produces a marginal value at least as high as their reservation wage, $MV(0)\geq \overline{w}(0)$.
I further assume that $MV(\theta)$ crosses the supply curve at most once (if $MV(\overline{\theta})<\overline{w}(\overline{\theta})$ for some $\overline{\theta}$, then $MV({\theta})<\overline{w}({\theta})$ for all $\theta>\overline{\theta}$). With these simplifying assumptions in hand, the zero-profit condition implies that the equilibrium share of workers under hourly contracts, $\theta^{EQ}$, is given by
\begin{equation}
\theta^{EQ} = \max\left\{\theta: \overline{w}(\theta)\leq AV(\theta)\right\}.\label{eq:equilibrium}
\end{equation}
Under the regularity assumptions above, this equilibrium exists and is unique. The strict inequality, $\overline{w}(\theta^{EQ})< AV(\theta^{EQ})$, holds only in the full pooling equilibrium,  $\theta^{EQ}=1$, in which all workers receive the fixed wage. Otherwise, firms offer wage contracts up to the point where the marginal worker's reservation wage is exactly equal to the average value of hourly employees' output, so $\overline{w}(\theta^{EQ})=AV(\theta^{EQ})$.
The efficient allocation of hourly contracts, on the other hand, is given by
\begin{equation}
\label{eq:efficient}
\theta^{EF}=\max\left\{\theta: \overline{w}(\theta)\leq MV(\theta)\right\}.
\end{equation}
Under this allocation, $\theta^{EF}$, all workers with reservation wages below their marginal values are hired for fixed-wage positions. This allocation also corresponds to the share of fixed-wage workers under a full-information equilibrium, in which firms can observe workers' reservation wages and offer them type-specific contracts.

\paragraph{Graphical Illustration}
Figure~\ref{fig:model}, Panel~A illustrates the welfare impacts of adverse selection for an example population. An efficient allocation of contracts would lead to hourly employment for all types $\theta\leq \theta^{EF}$, as these workers would accept wages at or below their marginal values ($\overline{w}(\theta)\leq MV(\theta)$). However, while type $\theta^{EF}$'s reservation wage (red line) is equal to their marginal value (blue line), an employer offering an hourly wage of $w=\overline{w}(\theta^{EF})$ would only recoup the average value (green line) among everyone accepting the offer (i.e., all $\theta\leq \theta^{EF}$). The employer could lower their wage offer, but that would drive those with the highest productivity out of the market, further reducing the contract's average value. This process continues across all types for whom $\overline{w}(\theta)>AV(\theta)$, so that the equilibrium share of workers under hourly contracts is $\theta^{EQ}$, where $\overline{w}(\theta^{EQ})=AV(\theta^{EQ})$. In this stylized example, roughly one-third of the population---$\theta\in \left(\theta^{EQ},\theta^{EF}\right)$---cannot obtain hourly employment despite a willingness to work for less than the expected value of their labor product. The result is a welfare loss corresponding to the area of the region shaded in yellow, which is equal to
\begin{equation}\label{eq:dwl}
DWL=\int_{\theta^{EQ}}^{\theta^{EF}}\left(MV(\theta) - \overline{w}(\theta)\right)d\theta.
\end{equation}

In summary, my model shows how worker selection on unobserved productivity can create a gap between the marginal and average values of labor, preventing Pareto-improving exchanges of fixed-wage contracts---workers are paid by the hour if and only if their reservation wage is no higher than the average value of all hourly workers. This information asymmetry reduces total welfare below what it would be under a full-information benchmark.

\subsection{Incorporating Moral Hazard}\label{sec:mh}

Note that the model above allows for moral hazard effects, even if those effects are not explicitly discussed. To see how, consider worker $i$'s potential output values under counterfactual contracts. Specifically, let worker $i$'s output value under the hourly wage, currently represented as $Y_i$, instead be denoted by $Y_{1i}$. Now let $Y_{0i}$ denote the counterfactual output value that worker $i$ would produce if they worked under the piece rate.\footnote{While a worker's output can differ under hourly versus piece-rate contracts, I assume it does not vary with the level of the hourly wage, $w$ (i.e., no wage effects). While my empirical results support this assumption, I nonetheless include a model with wage effects in Appendix~\ref{sec:extensions} for completeness.} The moral hazard effect for worker $i$ is given by the individual treatment effect of the hourly wage, $MH_i\equiv Y_{1i}-Y_{0i}$.\footnote{Strictly speaking, $Y_{1i}-Y_{0i}$ captures worker $i$'s overall output response to the hourly wage contract. I use the terms ``moral hazard'' and ``shirking'' to describe this response, even though it could partly reflect behavioral phenomena not traditionally classified as moral hazard.} This difference in counterfactual outcomes is not explicitly shown in the model because $AV(\theta)$ and $MV(\theta)$ are defined conditional on accepting the hourly contract, and thus depend only on $Y_{1i}$. The profit condition~\eqref{eq:profit} and welfare calculation~\eqref{eq:dwl} are therefore inclusive of any moral hazard effects because $MV\left(\theta\right)\equiv E\left[Y_{1i}|\theta_i=\theta\right] = E\left[Y_{0i}+MH_i|\theta_i=\theta\right]$.\par

While not strictly necessary to calculate welfare loss, explicitly modeling and estimating moral hazard effects is nonetheless important, especially for policy counterfactuals. As I show in Section~\ref{sec:policy}, moral hazard effects are needed to assess the social value of piece-rate taxes, as the public cost of these policies must include the reduced tax revenue from potentially lower earnings among those induced into hourly wage contracts. Separately identifying moral hazard can also be useful in settings where firms can combine fixed base wages with partial piece rates. Relative to pure fixed wages, these mixed contracts can mitigate shirking but do little to prevent adverse selection.\footnote{While these types of mixed contracts appear to be rare in my study setting, my model and empirical design can nonetheless be recast as a market for supplemental hourly wages on top of a preexisting piece rate. Expanding the contract space to allow for a range of such contracts with different piece-rate shares makes estimation considerably more complicated. See Appendix~\ref{sec:extensions} for details.} By separating these two forces, one can construct a counterfactual equilibrium that retains adverse selection while removing moral hazard from worker output.

To explicitly incorporate moral hazard into the model, I split Equations~\eqref{eq:MV} and~\eqref{eq:AV} into two pairs of curves. First, I define marginal values of a type $\theta$ as the conditional means of potential output value with ($Y_{1i}$) and without ($Y_{0i}$) the hourly wage:
\begin{eqnarray}
  MV_1\left(\theta\right)&\equiv& E\left[Y_{1i}|\theta_i=\theta\right]\label{eq:MV1}\\
  MV_0\left(\theta\right)&\equiv& E\left[Y_{0i}|\theta_i=\theta\right].\label{eq:MV0}
\end{eqnarray}
Note that $MV_1(\theta)$ is simply a relabeling of $MV(\theta)$ from Equation~\eqref{eq:MV}---it captures the expected output value under the hourly wage $w=S^{-1}(\theta)$ for the worker who is indifferent between accepting or declining the offer. $MV_0(\theta)$, on the other hand, captures the expected output value of that same worker if they had instead rejected the wage offer $w$ and remained on the piece rate.\footnote{In a loose sense, these two curves can be thought of as bounds on output value under mixed compensation with both fixed and piece-rate components. If the piece-rate component partially mitigates moral hazard effects, the true marginal value curve would lie somewhere between $MV_0(\theta)$ and $MV_1(\theta)$.
} The difference between these two marginal value curves identifies the moral hazard effect for a given type:
\begin{equation}
  \label{eq:MHtheta}
  MH(\theta)\equiv MV_1(\theta) - MV_0(\theta).
\end{equation}
Similarly, the average value curve can be split into two counterfactuals:
\begin{eqnarray}
  AV_1\left(\theta\right)&\equiv& E\left[Y_{1i}|\theta_i \leq \theta\right]\label{eq:AV1}\\
  AV_0\left(\theta\right)&\equiv& E\left[Y_{0i}|\theta_i \leq \theta\right]\label{eq:AV0}.
\end{eqnarray}
$AV_1\left(\theta\right)$ is equivalent to $AV(\theta)$ from Equation~\eqref{eq:AV}; it equals the average value of output among hourly-pay workers with lower reservation wages than type $\theta$. $AV_0\left(\theta\right)$, on the other hand, equals the average value that would be produced by those same workers if they had instead worked under the piece rate.

In the next section, I use MTE methods to estimate both pairs of value curves: $MV_1(\theta)$, $AV_1(\theta)$, $MV_0(\theta)$, and $AV_0(\theta)$. These estimates allow me to calculate counterfactual equilibria and quantify the type-specific moral hazard effect, $MH(\theta)$, in Equation~\eqref{eq:MHtheta}. As I show in Section~\ref{sec:policy}, this moral hazard effect is an important component of policy evaluation.

\section{Estimating the Model using Marginal  Treatment Effects}\label{sec:estmodel}

The model above shows how the welfare effects of asymmetric information depend on counterfactual distributions of workers' marginal values across a range of reservation wages. In this section, I show how I can semi-parametrically identify these marginal values using a marginal treatment effects (MTE) framework.

\begin{figure}
  \centering \caption{Model of Asymmetric Information in Wage Contracts}
  \label{fig:model}
  \begin{subfigure}{\textwidth}
    \vspace{-.1cm}
    \includegraphics[scale=0.43,trim=1.1cm 0.8cm 2cm 1.8cm,clip]{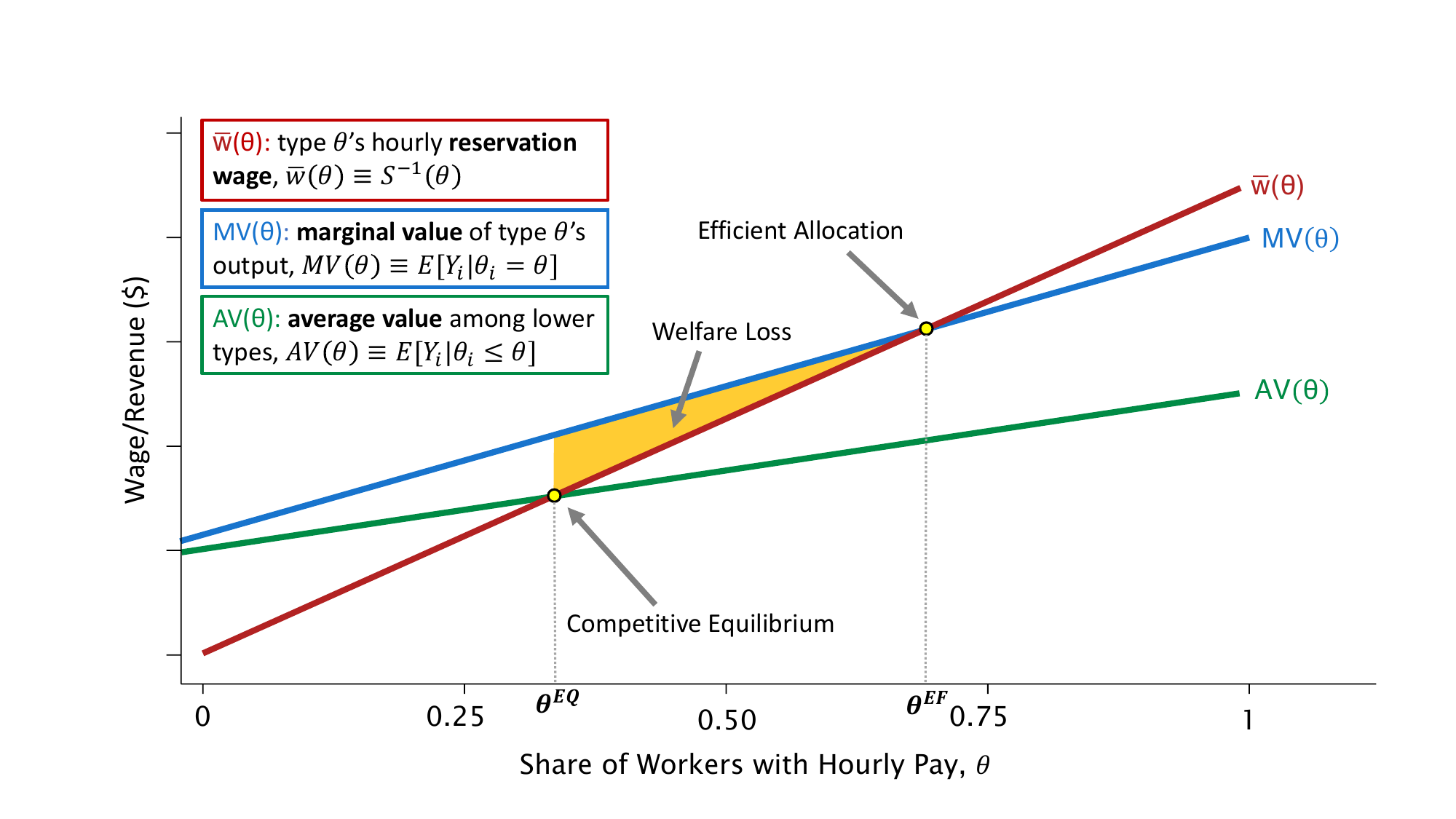}
    \caption{Welfare Loss from Adverse Selection}
  \end{subfigure}
  \\\vspace{.2cm}
  \begin{subfigure}{\textwidth}
    \vspace{-.1cm}
    \includegraphics[scale=0.43,trim=1.1cm 0.8cm 0.1cm 1.8cm,clip]{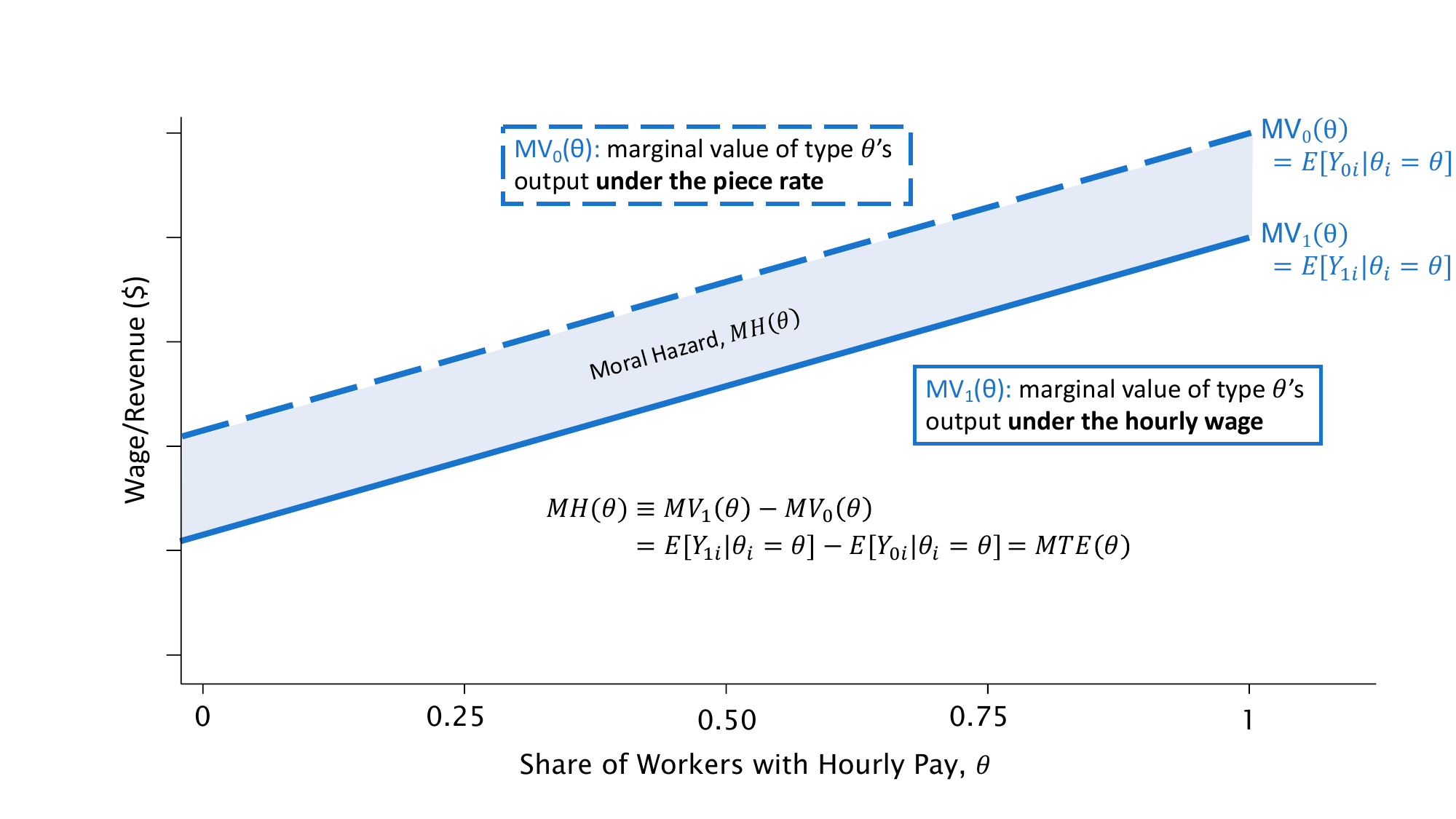}
\caption{Incorporating Moral Hazard}
  \end{subfigure}
  \begin{figurenotes}
This figure provides a graphical example of my model. Panel~A illustrates welfare loss from adverse selection. On the horizontal axis, types, $\theta$, are enumerated in ascending order based on their hourly reservation wage, $\overline{w}_i$. The blue line plots the marginal value curve, $MV\left(\theta\right)\equiv E[Y_i|\theta_i=\theta]$, where $Y_i$ denotes output value under the hourly wage. The red line plots hourly reservation wage, $\overline{w}\left(\theta\right)\equiv S^{-1}(\theta)$. The green line plots the average value curve, $AV\left(\theta\right)\equiv E[Y_i|\theta_i\leq\theta]$. The yellow region corresponds to the welfare loss from adverse selection into hourly wages. Panel~B incorporates moral hazard into the model. The solid and dashed blue lines denote marginal values of potential output under hourly wages, $MV_1\left(\theta\right)\equiv E[Y_{1i}|\theta_i=\theta]$, and under the piece rate, $MV_0\left(\theta\right)\equiv E[Y_{0i}|\theta_i=\theta]$, respectively. The difference between the two marginal value curves identifies the moral hazard effect for a given type, $MH(\theta)\equiv MV_1(\theta) - MV_0(\theta)$, which is equivalent to the marginal treatment effect of the hourly contract among those whose resistance to treatment (quantile reservation wage, $\theta_i\equiv S(\overline{w}_i)$) is equal to the propensity score (share of hourly workers, $\theta=S(w)$) for their assigned instrument (wage offer, $W_i$).
  \end{figurenotes}
  \end{figure}

As in Section~\ref{sec:experiment}, consider a potential outcomes framework in which treatment corresponds to working under an hourly contract. Let experimental wage offers, $w$, serve as a continuous instrument for taking up that treatment condition. Adopting the parlance of the causal inference literature, a worker's  quantile reservation wage, $\theta_i\equiv S(\overline{w}_i)$, is their ``resistance to treatment'' \citep{bjorklund1987estimation,heckman1999local,heckman2005structural,heckman2007econometric}. Under this framing, the marginal values defined in Equations~\eqref{eq:MV1} and~\eqref{eq:MV0} are exactly the treated and untreated marginal treatment response (MTR) functions, $MTR_1(\theta)\equiv E[Y_{1i}|\theta_i=\theta]=MV_1(\theta)$ and $MTR_0(\theta)\equiv E[Y_{0i}|\theta_i=\theta]=MV_0(\theta)$.\footnote{\label{fn:condition}To isolate selection on factors that cannot be observed by firms, my estimation procedure conditions on a vector of screenable characteristics. As in Section~\ref{sec:model}, I suppress these ``$X_i$'' terms for expositional simplicity;  Appendix~\ref{sec:curvedetails} describes the additive-separable specification used to construct value and reservation-wage curves for each observed $X_i$.} The moral hazard effect in Equation~\eqref{eq:MHtheta} is therefore equivalent to the marginal treatment effect of the hourly contract,
\begin{equation}
  \label{eq:mh}
MTE(\theta)\equiv E[Y_{1i}-Y_{0i}|\theta_i=\theta]= MH(\theta).
\end{equation}
As Figure~\ref{fig:model}, Panel~B illustrates, this marginal treatment effect measures the average effect of treatment (hourly contract) among those with a given resistance to treatment (quantile reservation wage, $\theta_i$).\footnote{\citet{kowalski2023reconciling} uses similar insights from the MTE literature to show how adverse selection on potential outcomes can explain differences in the estimated treatment effects of health insurance in Oregon and Massachusetts.}

The correspondence above means I can apply insights from the MTE literature to identify the model with minimal parametric assumptions. First, note that the propensity score of the wage-offer instrument is equivalent to the supply curve, $S(w)$, in Equation~\eqref{eq:supply}. It can be straightforwardly identified as the share of accepters across wage offers:
\begin{equation}
  \label{eq:estsupply}
S(w) \equiv \Pr\left(\overline{w}_i \leq w \right)=\Pr(D_i=1|W_i=w).
\end{equation}
Next, at each propensity score, $\theta = S(w)$, average value under both hourly and piece-rate contracts can be identified from the conditional means of output value among respective accepters and decliners of the corresponding wage offer, $w$. So the average value curve under the hourly wage equals the average output value among those who accept the hourly wage offer that induces a share $\theta$ of workers into the hourly contract:
\begin{equation}
  AV_1\left(\theta\right) \equiv E\left[Y_{1i}|\theta_i \leq \theta\right] = E\left[Y_{i}|S(W_i)=\theta,D_i=1\right].\label{eq:estAV1}
\end{equation}
Likewise, the average value curve under the piece rate is derived from the average output value among those who \emph{decline} the hourly wage offer that is accepted by a share $\theta$ of workers:
\begin{equation}
  AV_0\left(\theta\right) \equiv E\left[Y_{0i}|\theta_i \leq \theta\right] = \frac{E\left[Y_{i}|S(W_i)=0\right] - E[Y_{i}|S(W_i)=\theta,D_i=0](1-\theta)}{\theta},\label{eq:estAV0}
\end{equation}
where $E\left[Y_{i}|S(W_i)=0\right]$ is the average output value of workers in the control group, who all work under the piece rate.\footnote{By offering no alternative to the piece rate, the control condition effectively allows for ``identification at infinity''---the average piece-rate output among workers of all reservation wages \citep{heckman1990varieties,chamberlain1986189}.} Finally, marginal values can be identified by separately differentiating take-up weighted conditional means for decliners and accepters of each offer:
\begin{eqnarray}
  MV_1(\theta)&=&\frac{\partial \left(E\left[Y_{1i}|\theta_i\leq\theta\right]\theta\right)}{\partial \theta} = \frac{\partial \left(E\left[Y_{1i}|S(W_i)=\theta, D_i=1\right]\theta\right)}{\partial \theta}\label{eq:dmu1}\\
  MV_0(\theta)&=&-\frac{\partial\left( E\left[Y_{0i}|\theta_i>\theta\right](1-\theta)\right)}{\partial \theta}\nonumber\\
  &=&-\frac{\partial\left( E\left[Y_{i}|S(W_i)=\theta,D_i=0\right](1-\theta)\right)}{\partial \theta}.\label{eq:dmu0}
\end{eqnarray}
Intuitively, Equations~\eqref{eq:dmu1} and~\eqref{eq:dmu0} identify marginal values by differentiating total value, $TV(\theta)\equiv AV(\theta)\theta$, with respect to $\theta$ under hourly and piece-rate counterfactuals.
To estimate these objects, I apply the local polynomial regression approach from \citet{carneiro2011estimating}. Estimation details are provided in Appendix~\ref{sec:curvedetails}.

\subsection{Results}\label{sec:results_marginal}
Figure~\ref{fig:curves} plots semi-parametric estimates of supply and value curves under hourly wages. On the horizontal axis, the type parameter, $\theta$, corresponds to  quantiles of workers' hourly reservation wages.\footnote{As discussed in Footnote~\ref{fn:condition}, reservation-wage and value curves are estimated conditional on a vector of employer-screenable characteristics, $X_i$. Lines plotted in Figure~\ref{fig:curves} report these conditional functions evaluated at the sample mean of $X_i$. Meanwhile, point estimates of welfare loss and equilibrium shares report conditional values from Equations~\eqref{eq:equilibrium}--\eqref{eq:dwl}, calculated at each observed $X_i$ and averaged over the empirical distribution of $X_i$.  In other words, Figure~\ref{fig:curves} represents a reference market for individuals with the average covariate profile, whereas reported point estimates aim to capture population-level quantities by averaging over all $X_i$-specific markets. Details on the conditioning procedure are provided in Appendix~\ref{sec:curvedetails}.} The red line plots hourly reservation wage at each quantile, $\overline{w}\left(\theta\right)$, which equals the inverse of the labor supply curve implied by first-stage logit estimates from Table~\ref{tab:Sc}, $\overline{w}\left(\theta\right)\equiv S^{-1}(\theta)$.\footnote{Appendix Figure~\ref{fig:curvesfullsemi} reports analogous estimates from an alternative specification that replaces the linear log-wage term in the first-stage logit with a monotone increasing spline. Appendix~\ref{sec:curvedetails} provides estimation details.} The green and blue lines plot the average and marginal value curves under hourly wages, $AV_1\left(\theta\right)\equiv E[Y_{1i}|\theta_i\leq\theta]$ and $MV_1\left(\theta\right)\equiv E[Y_{1i}|\theta_i=\theta]$, respectively.
Both value curves slope upward over most of the observed support, consistent with the negative selection into fixed wages observed in Section~\ref{sec:results}.
Appendix Figure~\ref{fig:curvesnomh} reports corresponding estimates using value curves under piece-rate counterfactuals, $AV_0\left(\theta\right)$ and $MV_0\left(\theta\right)$.

\paragraph{Full Information Benchmark} Figure~\ref{fig:curves} shows that \fp[0]{100*\popthetaeffHcorvalueXXsemi} percent (SE=\fp[2]{100*{\popthetaeffHsecorvalueXXsemi}}) of workers produce labor at marginal values, $MV(\theta)$, that exceed their hourly reservation wages, $\overline{w}(\theta)$. This share represents the equilibrium that would exist if employers were fully informed of workers' reservation wages and set fixed wages to incorporate their moral hazard response.
Because a worker's marginal value is equivalent to their expected output times the piece rate, these workers place an implicit premium on fixed wages over output-based pay.
While this behavior is consistent with risk-aversion, it is also consistent with a variety of behavioral mechanisms.\footnote{The willingness of workers to accept
fixed wages below their actuarially fair values is consistent with existing
evidence of ostensibly risk-averse behavior in low-stakes settings
\citep{falk2018global,AndersonRisk,harrison2007estimating}. Under standard utility functions, such behavior implies implausibly high levels of risk aversion in higher-stakes environments \citep{rabin2000risk}, leaving researchers to rely on behavioral explanations like probability weighting \citep{kahneman1979,prelec1998}, mental accounting \citep{rabin2009,barberis2006}, and reference dependence \citep{kahneman1991}. Systematically biased beliefs provide another potential explanation for seemingly risk-averse decisions---a risk-neutral worker might prefer $w<MV(\theta)$ if they undervalue their productive potential. While the lack of bunching in Appendix Figure~\ref{fig:lossbunching} provides partial evidence against reference dependence, my results cannot rule out behavioral mechanisms more broadly.} Importantly, my design allows me to remain agnostic about which mechanisms drive workers' preferences for fixed wages. Because semi-parametric estimates of marginal and average value curves rely entirely on revealed preferences over contracts, I can determine both competitive equilibrium and its full-information counterfactual without imposing explicit structure on worker utility or the distribution of parameters governing it. Workers' choices could be driven by any combination of risk aversion, myopia, or other behavioral phenomena, and the market outcomes of those choices would remain the same.\footnote{Even if they predict the same equilibria, behavioral explanations may have different welfare implications depending upon which choices reflect ``true'' preferences and which reflect ``mistakes'' \citep{goldin2022defaults}. Rather than weigh in on this question, this paper takes a classical approach to welfare and policy analysis, assuming that a worker's willingness to accept wage offers below their breakeven value corresponds to a revealed preference for fixed wages rather than systematic errors.}

\paragraph{Partial Unraveling from Adverse Selection} Despite most workers' willingness to sacrifice expected earnings in exchange for a fixed wage contract, the most productive of these workers remain on piece rates due to adverse selection. This inefficiency can be seen in the divergence between the marginal and average value curves in Figure~\ref{fig:curves}. In an equilibrium with both adverse selection and moral hazard, only \fp[0]{100*\popthetaeqHcorvalueXXsemi} percent (SE=\fp[2]{100*\popthetaeqHsecorvalueXXsemi}) of workers can find hourly positions, compared to the full-information benchmark of \fp[0]{100*\popthetaeffHcorvalueXXsemi} percent (SE=\fp[2]{100*{\popthetaeffHsecorvalueXXsemi}}). The resulting welfare loss is an estimated \$\fp{\popDWLHcorvalueXXsemi} (SE=\$\fp[3]{\popDWLHsecorvalueXXsemi}) per hour of labor, represented by the shaded region in Figure~\ref{fig:curves}. Using a back-of-the-envelope estimate of roughly 35 billion annual paid worker-hours in online clerical and data-entry gig work, this per-hour loss implies aggregate annual welfare costs of about \$\fp[1]{35*\popDWLHcorvalueXXsemi} billion globally.\footnote{\label{fn:welfarescale}I obtain the 35 billion annual-hours estimate by combining the lower end of the World Bank's estimate of 154 to 435 million global online gig workers with the World Bank's estimate that clerical and data-entry tasks account for 23 percent of online labor demand, and the ILO's estimate that online platform workers spend an average of 27 hours per week on platform work, of which roughly 8 hours are unpaid \citep{worldbank2024workingwithoutborders,ilo2021digitalplatforms}. This yields $154\times 0.23\times (27-8) \times 52\text{ million} \approx 35$ billion annual paid worker-hours. For comparison, the broader U.S. category of office and administrative support occupations accounted for 18.5 million jobs, or 12.2 percent of total employment, in 2023; multiplying by 1{,}900 to 2{,}080 hours per job implies roughly 35 to 38 billion annual paid hours in the United States alone \citep{bls2024officeadmin}. A mechanical extrapolation of that employment share to the 3.51 billion employed persons reported worldwide in 2024 implies on the order of 0.8 trillion annual paid hours globally \citep{ilo2024weso}.}

Appendix Figure~\ref{fig:curvesnomh} reports corresponding equilibrium and welfare estimates under a counterfactual that eliminates moral hazard. Estimated welfare loss under this no-shirking counterfactual is \$\fp{\popDWLPcorvalueXXsemi} (SE=\$\fp[3]{\popDWLPsecorvalueXXsemi})---slightly smaller and less precise than estimates inclusive of moral hazard responses.\footnote{This exercise holds fixed the hourly reservation-wage curve, $\overline{w}(\theta)$, implicitly assuming that reservation wages remain unchanged when workers' scope to shirk is removed. Under the weaker assumption that the reservation wages differ in levels but preserve workers' rank ordering, $MV_0(\theta)$ and $AV_0(\theta)$ remain correctly indexed, so their gap continues to characterize selection absent moral hazard.
}

\begin{figure}[p]
  \centering \caption{Estimates of Marginal Treatment Effects (Moral Hazard)}
  \label{fig:curvesMH}
    \includegraphics[width=\columnwidth,trim=.4cm .4cm 0cm 0cm,clip]{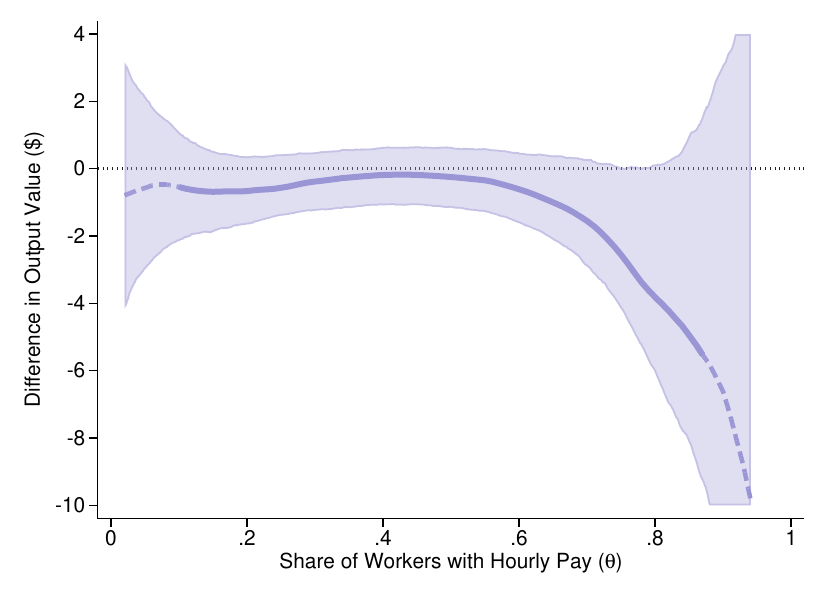}

  \begin{figurenotes}This figure plots estimated marginal treatment effects of hourly wages on worker output value. Estimates are obtained using local polynomial regressions of worker output value against propensity score (i.e., hourly supply share), as described in Section~\ref{sec:estmodel}. The solid line denotes $MH(\theta)\equiv MV_1(\theta)-MV_0(\theta)$---the difference in the marginal worker's potential output value under an hourly wage versus the piece rate. Since the underlying value curves are estimated conditional on employer-screenable characteristics, $X_i$, including task timing and employment history, the plotted marginal treatment effect is evaluated at $X=\overline X$, the sample means of these covariates.
    Value curves are estimated using a second-degree local polynomial regression (bandwidth=0.2) of residualized hourly output value against predicted hourly supply. Dashed portions of the line represent regions outside the support of observed propensity scores over which local polynomials were extrapolated. The shaded region represents a 90\% confidence interval, constructed via bootstrap resampling.
  \end{figurenotes}
\end{figure}

\paragraph{Selection on Moral Hazard} Figure~\ref{fig:curvesMH} plots the difference between hourly-wage marginal values, $MV_1(\theta)$, and piece-rate marginal values, $MV_0(\theta)$, across the distribution of worker types. This difference, $MV_1(\theta)- MV_0(\theta)$, is equivalent to the marginal treatment effect of hourly wages. In the context of the model, this curve represents how the (marginal) moral hazard effect of an hourly wage contract changes with workers' quantile reservation wage. Its shape suggests that selection on moral hazard is non-linear: Among workers with below-median reservation wages, those with marginally stronger preferences for fixed wages (lower $\theta$) have a slightly higher propensity to shirk (more negative treatment effect), consistent with patterns of selection in insurance markets \citep{einav2013selection}. Among those with above-median reservation wages, however, shirkers tend to prefer piece rates---those most resistant to hourly wages are most likely to shirk once they are paid hourly. This pattern could be explained by heterogeneous risk preferences: If tolerance for earnings risk correlates with tolerance for reputational risk from excessively low output, those who prefer the variability of piece-rate pay may also be more likely to shirk under fixed wages because they are less fearful of poor performance reviews.

Regardless of the underlying mechanism, moral hazard estimates play an important role in policy evaluation. Piece-rate taxes might mitigate adverse selection into fixed wages, but such policies must balance the welfare benefits of inducing the marginal worker into fixed wages with the distortionary costs of encouraging that worker to shirk. I explore this tradeoff in the following section.

\section{Policy Implications: MCPF of Piece-Rate Taxes}\label{sec:policy}

In this section, I evaluate policies aimed at mitigating the welfare loss estimated in Section~\ref{sec:estmodel}. Specifically, I estimate the marginal value of public funds (MVPF) for piece-rate taxes in data-entry work.\footnote{In Appendix~\ref{sec:mvpfdetails}, I calculate MVPFs for hourly wage subsidies as an alternative to piece-rate taxes. Results are reported in Appendix Figure~\ref{fig:mvpfsub}.}
The MVPF is defined as the change in social welfare per dollar of government expenditure \citep{hendren2020unified}. Under this measure, corrective tax policies are evaluated not by the size of the welfare loss they combat but by how efficiently that loss can be recovered.
Note that, for a revenue-generating policy like piece-rate taxes, the MVPF is equivalent to the marginal cost in social welfare per dollar of increased revenue raised under the policy.\footnote{For revenue-raising policies, a lower-valued MVPF implies that piece-rate taxes are a more efficient source of public funds, with values below one implying they are more socially efficient than a non-distortionary tax \citep{hendren2024audit}.} For this reason, I refer to this quantity as the marginal \emph{cost} of public funds (MCPF) associated with the tax.

\subsection{MCPF Derivation}
Consider an ad-valorem piece-rate tax, $\rho$, assessed on the value of labor product produced under the piece rate. The MCPF of this tax is given by
\begin{equation}\label{eq:mcpfagg}
MCPF_{\text{Tax}}(\rho) \equiv \frac{WTP(\rho)}{NR(\rho)},
\end{equation}
where $WTP(\rho)$ is private willingness-to-pay to \emph{avoid} the tax, and $NR(\rho)$ is net government revenue raised by the tax.\footnote{\label{fn:mcpfcondition}Similar to Sections~\ref{sec:model} and~\ref{sec:estmodel}, this section simplifies exposition by suppressing conditioning on employer-screenable characteristics, $X_i$. In the estimates, one common tax rate applies to all observed covariate profiles; the conditional willingness-to-pay and fiscal components are averaged over the empirical distribution of $X_i$ before the MCPF ratio is formed. See Appendix~\ref{sec:mvpfdetails}, especially Equation~\eqref{eq:mcpfaggregation}, for details.} Importantly, $NR(\rho)$ includes both the tax's direct transfer to the government and any fiscal externalities it imposes on long-term revenue.

To calculate welfare costs, $WTP(\rho)$, and net revenue, $NR(\rho)$, I parameterize the policy as a wedge on piece-rate transactions: For each dollar of labor product compensated under the piece rate, the government collects an additional $\rho$ dollars in tax revenue. Under this wedge, procuring labor product through the piece rate becomes $1+\rho$ times as expensive as in the baseline, so the equilibrium condition under tax $\rho$ is given by $\overline{w}(\theta^{\rho})=(1+\rho)AV_1(\theta^{\rho})$, where $\theta^{\rho}$ is the equilibrium share of hourly workers under the tax.\footnote{Recall that the marginal and average values of hourly workers' labor product are defined as the amount the firm saves by not procuring that labor product directly through the piece rate. Under this tax wedge, each dollar of piece-rate labor product generates an additional $\rho$ dollars of tax liability, so the corresponding hourly value curves scale by a factor of $1+\rho$. Note that, for simplicity, this equilibrium condition excludes the limiting case with full pooling ($\theta^{\rho}=1$) and assumes that firms bear the entire cost of the piece-rate tax, keeping the price of labor product fixed at $p$. See Appendix~\ref{sec:mvpfdetails} for details.}

Let $\tau$ represent the baseline income tax rate,\footnote{I assume worker and firm income face a common effective tax rate, which is calibrated to $\tau=0.2$. See Appendix~\ref{sec:mvpfdetails} for details.} and assume piece-rate-tax payments are deductible from the preexisting tax base. The direct fiscal consequence of the piece-rate tax would be to increase government revenue by $(1-\tau)\rho MV_0(\theta)$ for all types $\theta \in \left[ \theta^\rho,1\right]$, reflecting total tax receipts from piece-rate production in the tax equilibrium. However, this increased revenue is partially offset by the indirect costs of inducing more workers into hourly pay---a tax on piece rates could lead to lower earnings and decreased tax revenue. Net government revenue, $NR(\rho)$, from the piece-rate tax is therefore given by
\begin{equation}
  \label{eq:revenue}
  NR(\rho) = \underbrace{(1-\tau)\int^1_{\theta^\rho}\rho MV_0(\theta)d\theta}_{\text{Direct Revenue from Transfer}}  + \underbrace{\int_{\theta^{EQ}}^{\theta^\rho}\tau MH(\theta)d\theta}_{\text{Fiscal Externality from Moral Hazard}}.
\end{equation}

The social cost of raising $NR(\rho)$ in net revenue is given by private willingness-to-pay to avoid the tax. This willingness-to-pay, $WTP(\rho)$, has two components: First, piece-rate production with $\theta \in \left[\theta^\rho,1\right]$ would avoid direct tax payments of $(1-\tau)\rho MV_0(\theta)$ in the absence of the tax. Second, workers with $\theta \in \left[\theta^{EQ},\theta^\rho\right]$ would lose $MV_1(\theta) - \overline{w}(\theta)$ of the implied value they place on hourly positions, scaled by $1-\tau$ to convert that value to after-tax dollars. The total welfare cost of the piece-rate tax, $\rho$, is given by
\begin{equation}
  \label{eq:wtprho}
WTP(\rho) = \underbrace{(1-\tau)\int^1_{\theta^\rho}\rho MV_0(\theta)d\theta}_{\text{Direct Tax Savings}}  - \underbrace{(1-\tau)\int_{\theta^{EQ}}^{\theta^\rho}\left(MV_1(\theta) - \overline{w}(\theta)\right)d\theta}_{\text{Lost Value of Fixed Wages}}.
\end{equation}

With Equations~\eqref{eq:revenue} and~\eqref{eq:wtprho} in hand, I can write $MCPF_{\text{Tax}}(\rho)$---the MCPF of piece-rate tax $\rho$---as
\begin{equation}
  \label{eq:mcpf}
  MCPF_{\text{Tax}}(\rho) \equiv \frac{WTP(\rho)}{NR(\rho)} =  \frac{\int^1_{\theta^\rho}\rho MV_0(\theta)d\theta  - \int_{\theta^{EQ}}^{\theta^\rho}\left(MV_1(\theta) - \overline{w}(\theta)\right)d\theta}{\int^1_{\theta^\rho}\rho MV_0(\theta)d\theta  + \int_{\theta^{EQ}}^{\theta^\rho}\frac{\tau}{1-\tau} MH(\theta)d\theta}.
\end{equation}
Equation~\eqref{eq:mcpf} reveals the trade-off faced by policymakers promoting hourly wage contracts. The marginal social benefit of an additional hourly contract depends on the relative magnitudes of its insurance value to the marginal worker and that worker's propensity to shirk. More generally, this trade-off highlights the importance of separating adverse selection from moral hazard in markets with asymmetric information---mistaking one for the other can lead to suboptimal policy decisions.

The green line in Figure~\ref{fig:mvpf} plots estimates of $MCPF_{\text{Tax}}(\rho)$. Estimated MCPFs increase with the size of the tax because the first workers induced into hourly contracts are willing to pay the highest implicit premia for fixed wages among non-hourly workers. Positive values of $\rho$ up to 5 percent generate revenue at net social costs no higher than \$0.90 per dollar.

\subsection{Calculating Optimal Piece-Rate Tax}

While the above analysis helps identify the range of potential MCPFs associated with piece-rate taxes, it does not solve for the welfare-maximizing tax rate, $\rho^*$. In general, comparisons of MVPFs between mutually exclusive policies that endogenously differ in scale can lead to suboptimal policy choices. In this instance, the lowest-MCPF tax would generate a vanishingly small amount of total revenue because the first person induced into the hourly wage has the highest possible welfare gain from leaving the piece rate.

To determine the optimal tax, I minimize net welfare costs:
\begin{equation}
  \label{eq:optimal}
  \rho^*\equiv \argmin_\rho \left\{ WTP(\rho) - \eta NR(\rho)\right\},
\end{equation}
where $\eta$ represents the marginal value of government revenue---the highest MVPF among policies to which tax revenue might be allocated. At an interior, differentiable optimum, the first-order condition for \eqref{eq:optimal} implies
\begin{equation}\label{eq:rhostar}
  MCPF_{d\text{Tax}}({\rho^*})\equiv\frac{WTP'(\rho^*)}{NR'(\rho^*)}=\eta.
\end{equation}
$MCPF_{d\text{Tax}}({\rho})$ is the MCPF for a \emph{marginal increase} in the piece-rate tax. Equation~\eqref{eq:rhostar} provides a prescription for achieving the optimal tax rate---the one that minimizes net welfare costs. Policymakers should increase the tax until the MCPF of a marginally higher tax equals the marginal value of a one-dollar increase in government revenue---the MVPF of the policy or tax cut to which the revenue might be directed.

\begin{figure}[p]
  \centering \caption{Estimated MCPFs for Piece-Rate Taxes}
  \label{fig:mvpf}
  \centering
    \includegraphics[scale=1.1,trim=.4cm .4cm 0cm 0cm,clip]{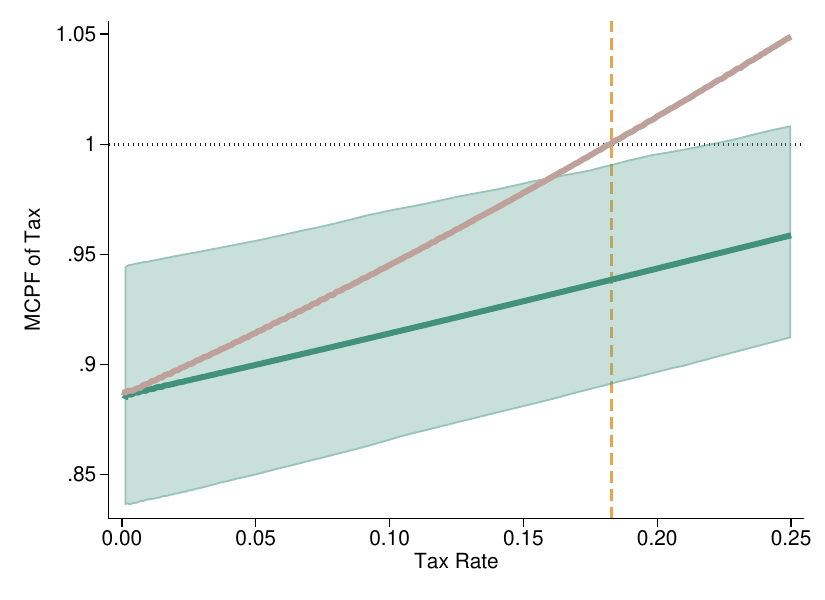}
  \vspace{.12cm}
  \begin{figurenotes}This figure plots estimated marginal costs of public funds (MCPFs) for piece-rate taxes. The green line plots estimated MCPFs associated with hypothetical piece-rate tax rates denoted on the horizontal axis, and the pink line plots estimated MCPFs associated with a marginal increase in the tax rate. At each tax rate, willingness to pay and net revenue are averaged over the empirical distribution of employer-screenable characteristics before their ratio is formed. The vertical dashed line denotes the optimal tax rate assuming the marginal value of tax receipts equals one. The shaded region represents the pointwise 90\% confidence interval for the level MCPF, constructed via bootstrap resampling.
  \end{figurenotes}
  \end{figure}

Figure~\ref{fig:mvpf} plots the marginal MCPF, $MCPF_{d\text{Tax}}(\rho)$, in pink. The MCPF of marginally higher taxes increases with the tax rate, reaching one at a piece-rate tax of \fp[0]{100*\poptaxstarcorvalueXXsemi} percent (SE=\fp[2]{100*\poptaxstarsecorvalueXXsemi}) and rising thereafter. In other words, the socially optimal piece-rate tax is \fp[0]{100*\poptaxstarcorvalueXXsemi} percent, even when the revenue it raises generates net-zero social return ($\eta=1$). This occurs because the piece-rate tax's distortionary effects are outweighed by the welfare gains it creates from promoting fixed wages. If I allow for a marginal value of government funds greater than one ($\eta>1$), the optimal tax would increase from \fp[0]{100*\poptaxstarcorvalueXXsemi} percent to the value of $\rho$ at which $MCPF_{d\text{Tax}}(\rho)=\eta$. I provide detailed derivations of both $MCPF_{\text{Tax}}(\rho)$ and $MCPF_{d\text{Tax}}({\rho})$ in Appendix~\ref{sec:mvpfdetails}.

In conclusion, MCPFs for piece-rate taxes compare favorably with many alternative revenue sources. By mitigating adverse selection costs, taxing output-based pay at a rate of \fp[0]{100*\poptaxstarcorvalueXXsemi} percent or less can raise government revenue at least as efficiently as a distortionless tax. At these levels, each dollar of piece-rate tax revenue carries a net social cost as low as \$0.88 and no higher than \$1.00. While these point estimates are specific to the data-entry workers in my setting, they may nonetheless inform priors on the welfare impact of taxing tips, commissions, or bonuses in other labor markets. For example, my findings suggest that the existing tax-favored treatment of tipped compensation in the tax code is likely to amplify the welfare costs of adverse selection into fixed wages.\footnote{IRS Code Section~45B provides food-and-beverage businesses with an income-tax credit for employer FICA taxes paid on certain tipped compensation \citep{rane2024consequences}, while the 2025 One Big Beautiful Bill Act created separate deductions for qualified tips and qualified overtime compensation \citep{irsOBBBA2025}.}
In the following section, I discuss these broader implications and explain how this paper's empirical framework can be used in other labor markets.

\section{Implications for Other Labor Markets}\label{sec:discuss}
Any job with an uncertain, effort-intensive labor product is susceptible to the forces of moral hazard and adverse selection.
Perhaps as a consequence, labor markets most vulnerable to these forces are often dominated by self-employment, freelance work, or piece-rate compensation.\footnote{Failed attempts at ``no-tipping restaurants'' provide a potential case study: After replacing tipped earnings with fixed wages, many restaurants struggled to maintain profitability and ultimately reverted course, often citing an exodus of qualified servers as the cause \citep{kadvany2022zuni,nytimes2020meyer,dunn2018limitations}. One San Francisco restaurant noted that ``[Server positions] have been harder to fill. Many veteran servers weren't interested, saying they could make double elsewhere with tips...As a result, many of the people who work in the dining room started with little to no restaurant experience'' \citep{kadvany2022good}.} While this paper sheds light on the potential distortionary effects of asymmetric information in these settings, my estimates are not intended to measure welfare losses beyond data-entry or clerical gig work. Such limits to generalizability are ubiquitous in applied research on worker incentives---whether they come from agricultural workers \citep{brune2022peers,bandiera2010social}, call centers \citep{mas2017valuing,nagin2002}, cashiers \citep{mas2009peers}, or automotive glass repairers \citep{lazear2000}, specific estimates of parameters concerning worker productivity are usually difficult to generalize beyond narrowly defined labor markets.\footnote{For example, \citet{herbst2015peer} find that estimates of peer effects on worker output vary dramatically between individual studies.} While my experiment is not immune to these limitations, it has several features that help maximize external validity. As discussed in Section~\ref{sec:setting}, the experimental job posting was designed to mimic a private firm seeking ``AI Taskers,'' and the typing task resembles those frequently requested on online labor platforms. Moreover, participants were not informed that they were part of an experiment until after they performed the task.

In Appendix~\ref{sec:external}, I further assess the external validity of my design using the SANS conditions from \citet{list2020trial}. This assessment establishes how, at a minimum, my results speak to information asymmetries in the large and growing online labor markets for data-labeling services. I also discuss how my estimates might compare to those one would find in other labor markets, especially markets for short-term gig work. My results can be informative of potential distortions in these markets, even if my point estimates only apply to a specific task and population of workers. Nonetheless, policies aimed at addressing information asymmetries outside of data-entry work warrant further evidence targeted to those settings. This paper provides a flexible framework for gathering this evidence. The following calibration exercise provides an example application.

\subsection{Application to  Rideshare Drivers: Calibration from  \texorpdfstring{\citet{angrist2021uber}}{Angrist et al. (2021)}}\label{sec:uber}

To demonstrate how my framework might be applied to other settings, I calibrate my model to approximate welfare loss from adverse selection in the market for rideshare drivers. My calibration exercise uses estimates from \citet{angrist2021uber}, who conduct an experiment that offers Uber drivers the opportunity to increase their commission on ride fares in exchange for making fixed weekly payments that do not vary with earnings. Offering this ``taxi-style'' compensation structure effectively gives the driver an opportunity to buy back a share of the fare revenue they generate. Viewed through this lens, opting into a taxi-style lease is the same as opting \emph{out} of a (partial) fixed wage contract---a driver who accepts a lease offer gives up some fixed weekly income in favor of a higher share of their labor product. This equivalence means I can calibrate my model using estimates of selection \emph{out of} taxi leases from \citet{angrist2021uber} to approximate welfare losses from inefficient compensation in the market for rideshare drivers.\footnote{Calibration details are provided in Appendix~\ref{sec:uberapp}.}

Appendix Figure~\ref{fig:uber} plots results from this exercise. The horizontal axis denotes the share of drivers declining weekly lease offers. The red curve plots the inverse supply of lease-free drivers, $\overline{w}_B\left(\theta\right)\equiv S^{-1}(\theta)$. The blue line plots the marginal value curve, $MV\left(\theta\right)\equiv E[Y_i|\theta_i=\theta]$, where $Y_i$ denotes driver $i$'s weekly revenue from ride fares. The green line plots the average value curve, $AV\left(\theta\right)\equiv E[Y_i|\theta_i\leq\theta]$. I calibrate the value curve from average fare revenues among lease accepters and decliners, dividing reported dollar gains by each offer's fee reduction and averaging across offers using driver counts. The equilibrium share of lease-free contracts is $\theta^{EQ}=\fp[1]{100*\ubereq}$ percent, while the efficient share of lease-free contracts one would expect in a full-information equilibrium is $\theta^{EF}=\fp[1]{100*\ubereff}$ percent. The difference of $\theta^{EF} - \theta^{EQ} = \fp[1]{100*(\ubereff-\ubereq)}$ percentage points represents potential drivers who would generate profitable levels of fare revenue if they were charged a lower proportional fee or offered some additional fixed weekly compensation that placed a higher value on their labor product. However, the cost of adverse selection among less productive drivers would make offering this compensation unprofitable. The corresponding area, shaded in yellow and expressed in fare-revenue units, is $\$\fp[2]{\uberdwl}$ per week, or approximately $\$\fp[2]{\uberdwl/\hoursperweek}$ per hour worked.
Scaling by the average fee reduction of $\fp[1]{100*\feereduction}$ percentage points yields a welfare loss of approximately $\$\fp[2]{\feereduction*\uberdwl/\hoursperweek}$ per hour worked, comparable to my estimates of welfare loss in the market for online data-entry.

There are several limitations to this calibration exercise. Because \citet{angrist2021uber} do not report results separately by experimental strata, treatment condition, and take-up status, my calibrated curves rely on transformations across a small number of aggregated point estimates. Linear extrapolation of market-wide selection patterns from these points comes with strong functional-form assumptions, and the absence of standard errors prevents any meaningful discussion of statistical significance. Moreover, as I discuss in Section~\ref{sec:intro}, a precise accounting for welfare costs of selection must consider the value of labor product ``off the equilibrium path,'' but the experimental sample in \citet{angrist2021uber} consists of \emph{existing} Uber drivers. In this sample, observed selection into taxi-style lease offers likely excludes drivers who desire it most---the ones who are already working as taxi drivers! Entry into platform work also often follows adverse changes in workers' outside income, suggesting that samples of incumbent gig workers may be selected on the outside options that govern contract take-up \citep{koustas2019bigdata,garin2020new}. More generally, because the welfare consequences of asymmetric information require market-wide estimates of treatment and selection, studies of workers under existing employment contracts are likely to understate the effects of information asymmetries.
Nonetheless, this calibration exercise provides an example of how my framework can be applied to a variety of labor markets where worker selection on potential productivity can prevent efficient contracting.

\section{Conclusion}\label{sec:conclusion}

This paper uses an experimental approach to estimate the equilibrium and welfare effects of moral hazard and adverse selection in fixed-wage contracts. My experiment offers workers a choice between a piece rate and a randomized hourly wage, allowing me to separately identify selection and treatment effects of wage contracts. Using experimental wage offers as an instrument for hourly wage take-up, I find evidence of both moral hazard and adverse selection: Hourly wage contracts reduce worker productivity by an estimated \fp{-100*\bTEccorvalue/\meancorvalue} percent relative to the mean. Meanwhile, a 10 percent increase in the hourly wage offer attracts a marginal worker whose productivity is higher by \fp{20*\bAScHcorvalue/\meancorvalue} percent of mean worker output.

I place these experimental estimates into a theoretical framework that shows how the provision of hourly employment contracts is determined by two factors: a worker's reservation wage for fixed compensation and the average output of workers with comparatively lower reservation wages. I semi-parametrically identify these objects using a marginal treatment effects (MTE) framework. My estimates imply that information asymmetries lead to an underprovision of fixed-wage contracts and a welfare loss of \$\fp{\popDWLHcorvalueXXsemi} per hour worked. To investigate the policy implications of this welfare loss, I calculate the marginal costs of public funds (MCPFs) for taxes on performance-based pay. My estimates suggest that an  \fp[0]{100*\poptaxstarcorvalueXXsemi}-percent tax on performance-based pay can efficiently raise government revenue by correcting the market inefficiencies associated with adverse selection.

My findings point to several directions for future research. As I discuss in Section~\ref{sec:discuss}, my framework can be applied to a variety of other labor markets, especially those characterized by self-employment, freelance work, or piece-rate compensation. It could also be applied to alternative contracts in the broader labor market. For example, \citet{ganong2025earnings} document large month-to-month volatility in earnings among stably employed hourly workers, apparently driven by fluctuations in firms' demand for hours. My methods could help determine why employers neglect to insure these workers through monthly or annual contracts. Future work could also extend my tax analysis to other labor-market policies promoting fixed wages. A binding minimum wage, for instance, may act as an ``insurance mandate'' that pools workers with different latent productivities and mitigates adverse selection.
Designing policies to better address information asymmetry problems in wage contracts could meaningfully improve the lives of millions of workers.

\clearpage
\section*{Figures and Tables}
\renewcommand{\footnotesize}{\fontsize{9pt}{11pt}\selectfont}

   \clearpage

  \clearpage

  \clearpage

  \clearpage
  \begin{figure}[p]
  \centering \caption{Estimates of Marginal and Average Value Curves}
  \label{fig:curves}
  \centering
  \begin{overpic}[width=\columnwidth,trim=.4cm .2cm .2cm .1cm,clip]{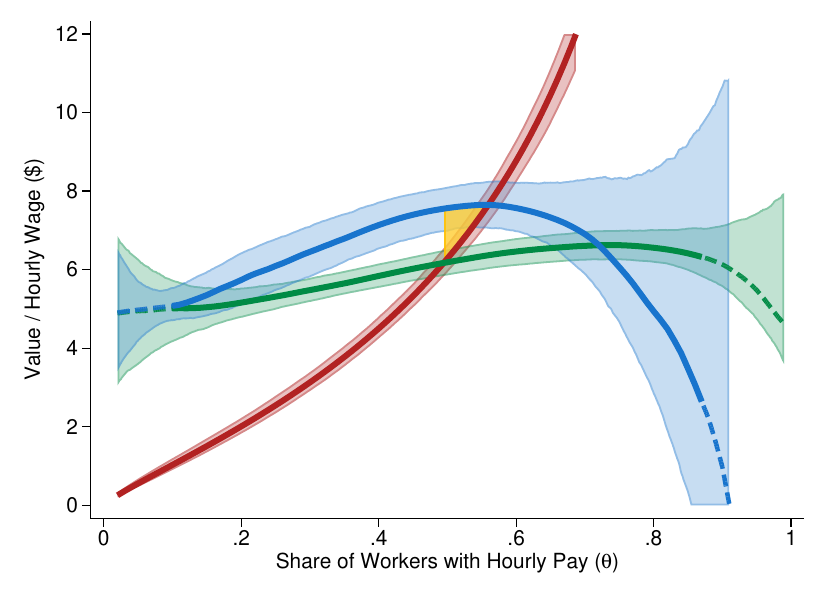}
  \put(74, 68){\wbar}
  \put(48, 52.5){\mv{1}}
  \put(32, 32.5){\av{1}}

\end{overpic}
  \begin{figurenotes}This figure plots estimated supply and value curves under hourly wages, where output values reflect the number of correctly typed sentences multiplied by the piece rate. The blue and green lines plot semi-parametric estimates of the marginal value, $MV_1\left(\theta\right)$, and average value, $AV_1\left(\theta\right)$, under hourly wages, as defined in Figure~\ref{fig:model}. The red line plots the estimated hourly reservation-wage curve implied by a logit regression of hourly take-up against log experimental wage offers. The area in yellow denotes estimated welfare loss from adverse selection into hourly wages. Since curves are estimated conditional on task timing and employment history, plotted lines are evaluated at the sample means of these covariates. Value curves are estimated using a second-degree local polynomial regression (bandwidth=0.2) of residualized hourly output value against predicted hourly supply. Dashed portions of the value curves represent regions outside the support of observed propensity scores over which local polynomials were extrapolated. Shaded regions represent 90\% confidence intervals, constructed via bootstrap resampling. Note that standard errors on welfare estimates account for mechanically correlated sampling error between $MV(\theta)$ and $AV(\theta)$ curves.\end{figurenotes}
\end{figure}

  \clearpage

\clearpage

  \clearpage
  \begin{table}[p]
  \centering
  \caption{Experimental Group Assignment}
  \label{tab:groups}
  \resizebox{.8\columnwidth}{!}{\def \No {No } \begin{tabular}{lcc}\toprule Hourly Wage Offer & Piece-Rate Offer & Number of Participants\\ \cmidrule{1-3}
\No Hourly Offer&\$0.03 per sentence&302\\
\$1.20/hr&\$0.03 per sentence&300\\
\$1.80/hr&\$0.03 per sentence&101\\
\$2.40/hr&\$0.03 per sentence&103\\
\$3.00/hr&\$0.03 per sentence&304\\
\$3.60/hr&\$0.03 per sentence&100\\
\$4.20/hr&\$0.03 per sentence&99\\
\$4.80/hr&\$0.03 per sentence&101\\
\$5.40/hr&\$0.03 per sentence&101\\
\$6.00/hr&\$0.03 per sentence&305\\
\$7.20/hr&\$0.03 per sentence&100\\
\$8.40/hr&\$0.03 per sentence&102\\
\$9.60/hr&\$0.03 per sentence&101\\
\$10.80/hr&\$0.03 per sentence&100\\
\$12.00/hr&\$0.03 per sentence&305\\
\$15.00/hr&\$0.03 per sentence&100\\
\$18.00/hr&\$0.03 per sentence&102\\
\$21.00/hr&\$0.03 per sentence&304\\
\midrule & &\emph{Total:} 3030\phantom{\emph{Total:} }\\ \bottomrule \end{tabular}
}
\begin{tablenotes}
  This table summarizes the treatment conditions and sample sizes for each experimental group in the experiment. \emph{Piece-rate offer} denotes the performance-based bonus offer, which is awarded on a per-sentence basis and common across all experimental groups. \emph{Hourly wage offer} denotes the fixed-rate compensation offered to workers for the five-minute task, prorated to one hour. Note that both piece rate and hourly wage options are offered as supplements to a base wage of \$12.00 per hour.
\end{tablenotes}
\vspace{.1cm}
\end{table}

\clearpage
\begin{table}[p]
  \centering \caption{Summary Statistics}
  \label{tab:summ}
\resizebox{\columnwidth}{!}{
  \begin{tabular}{c}
    \toprule
    \begin{tabular}{M{.26\columnwidth}}
      Category
    \end{tabular}
    \begin{tabular}{M{.35\columnwidth}*2{M{.1\columnwidth}}}
      Variable & Mean & SD
    \end{tabular}
    \\\cmidrule{1-1}
    \begin{tabular}{M{.26\columnwidth}}
      \emph{Panel A:} \\
      \emph{Task Performance}\\
      \ \
    \end{tabular}
    \begin{tabular}{p{.35\columnwidth}*2{M{.1\columnwidth}}}
      Accepted Hourly Offer&       0.486&       0.500\\
Completed Sentences &       21.98&       8.143\\
Correct Sentences   &       19.45&       8.480\\
Output Value        &       7.003&       3.053\\
Finished            &       0.986&       0.117\\

    \end{tabular}\vspace{-.5cm}
    \\\cmidrule{1-1}
    \begin{tabular}{M{.26\columnwidth}}
\emph{Panel B:}\\
      \emph{Demographics \& Employment}\\
      \ \
    \end{tabular}
    \begin{tabular}{p{.35\columnwidth}*2{M{.1\columnwidth}}}
      Age                 &       37.23&       12.18\\
Female              &       0.643&       0.479\\
Minority            &       0.357&       0.479\\
Employed            &       0.685&       0.465\\
Student             &       0.187&       0.390\\
Number of Previous Tasks&      1281.6&      1746.4\\

    \end{tabular}\vspace{-.5cm}
    \\\bottomrule
  \end{tabular}
}
 \begin{tablenotes} This table reports summary statistics for the experimental sample. Panel~A reports statistics on variables related to experimental task performance and experience. Panel~B reports demographic and employment information. The total number of participating workers is \fp[0]{\N}. ``Accepted Hourly Offer'' is calculated among the \fp[0]{\N-\Ncontrol} workers receiving an offer. \end{tablenotes}
\end{table}

\clearpage
\begin{table}[p]
  \centering
  \caption{Logit Estimates of Hourly Supply}
  \label{tab:Sc}
\resizebox{\columnwidth}{!}{{
\def\sym#1{\ifmmode^{#1}\else\(^{#1}\)\fi}
\begin{tabular}{l*{4}{d}}
\toprule
                    &\multicolumn{1}{c}{(1)}&\multicolumn{1}{c}{(2)}&\multicolumn{1}{c}{(3)}&\multicolumn{1}{c}{(4)}\\
                    &\multicolumn{1}{c}{Accepted Offer}&\multicolumn{1}{c}{Accepted Offer}&\multicolumn{1}{c}{Accepted Offer}&\multicolumn{1}{c}{Accepted Offer}\\
\midrule
Log Hourly Wage Offer&       1.202\sym{***}&       1.205\sym{***}&       1.216\sym{***}&       1.248\sym{***}\\
                    &    (0.0555)         &    (0.0555)         &    (0.0561)         &    (0.0579)         \\
\addlinespace
Number of Previous Tasks/1000&                     &                     &      0.0145         &    -0.00270         \\
                    &                     &                     &    (0.0252)         &    (0.0261)         \\
\addlinespace
Age                 &                     &                     &                     &      0.0292\sym{***}\\
                    &                     &                     &                     &   (0.00428)         \\
\addlinespace
Female              &                     &                     &                     &       0.281\sym{***}\\
                    &                     &                     &                     &    (0.0957)         \\
\midrule
Task Controls       &\multicolumn{1}{c}{No}         &\multicolumn{1}{c}{Yes}         &\multicolumn{1}{c}{Yes}         &\multicolumn{1}{c}{Yes}         \\
Employment Controls &\multicolumn{1}{c}{No}         &\multicolumn{1}{c}{No}         &\multicolumn{1}{c}{Yes}         &\multicolumn{1}{c}{Yes}         \\
Demographic Controls&\multicolumn{1}{c}{No}         &\multicolumn{1}{c}{No}         &\multicolumn{1}{c}{No}         &\multicolumn{1}{c}{Yes}         \\
\emph{N}            &\multicolumn{1}{c}{2728}         &\multicolumn{1}{c}{2728}         &\multicolumn{1}{c}{2728}         &\multicolumn{1}{c}{2728}         \\
\bottomrule
\end{tabular}
}
}
 \begin{tablenotes}
This table reports estimated coefficients from logistic regressions of hourly contract acceptance against log wage offers, excluding control-group workers who were only offered a piece rate. Columns (2)--(4) add control variables for the categories of observable characteristics listed in the bottom panel. Task controls include experimental-wave indicators and start time. Employment controls include employment-status indicators, student enrollment status, and number of previous tasks completed on Prolific. Demographic controls include race, gender, and age. \starnote
\end{tablenotes}
\end{table}

\clearpage
\begin{table}[p]
  \centering
  \caption{2SLS Estimates of Treatment Effects of Hourly Wages on Output Value}
  \label{tab:TEc}
  \resizebox{\columnwidth}{!}{{
\def\sym#1{\ifmmode^{#1}\else\(^{#1}\)\fi}
\begin{tabular}{l*{4}{d}}
\toprule
                    &\multicolumn{1}{c}{(1)}&\multicolumn{1}{c}{(2)}&\multicolumn{1}{c}{(3)}&\multicolumn{1}{c}{(4)}\\
                    &\multicolumn{1}{c}{Output Value}&\multicolumn{1}{c}{Output Value}&\multicolumn{1}{c}{Output Value}&\multicolumn{1}{c}{Output Value}\\
\midrule
Accepted Hourly Offer&      -0.612\sym{***}&      -0.606\sym{***}&      -0.591\sym{***}&      -0.475\sym{**} \\
                    &     (0.215)         &     (0.208)         &     (0.207)         &     (0.193)         \\
\addlinespace
Number of Previous Tasks/1000&                     &                     &       0.178\sym{***}&       0.178\sym{***}\\
                    &                     &                     &    (0.0338)         &    (0.0325)         \\
\addlinespace
Age                 &                     &                     &                     &     -0.0429\sym{***}\\
                    &                     &                     &                     &   (0.00446)         \\
\addlinespace
Female              &                     &                     &                     &       0.557\sym{***}\\
                    &                     &                     &                     &     (0.114)         \\
\midrule
Task Controls       &\multicolumn{1}{c}{No}         &\multicolumn{1}{c}{Yes}         &\multicolumn{1}{c}{Yes}         &\multicolumn{1}{c}{Yes}         \\
Employment Controls &\multicolumn{1}{c}{No}         &\multicolumn{1}{c}{No}         &\multicolumn{1}{c}{Yes}         &\multicolumn{1}{c}{Yes}         \\
Demographic Controls&\multicolumn{1}{c}{No}         &\multicolumn{1}{c}{No}         &\multicolumn{1}{c}{No}         &\multicolumn{1}{c}{Yes}         \\
Dep. Mean           &\multicolumn{1}{c}{7.00}         &\multicolumn{1}{c}{7.00}         &\multicolumn{1}{c}{7.00}         &\multicolumn{1}{c}{7.00}         \\
R-squared           &\multicolumn{1}{c}{0.040}         &\multicolumn{1}{c}{0.088}         &\multicolumn{1}{c}{0.104}         &\multicolumn{1}{c}{0.221}         \\
\emph{N}            &\multicolumn{1}{c}{3030}         &\multicolumn{1}{c}{3030}         &\multicolumn{1}{c}{3030}         &\multicolumn{1}{c}{3030}         \\
\bottomrule
\end{tabular}
}
}
\begin{tablenotes}
  This table reports estimated coefficients from two-stage least-squares regressions of residualized output value, where output value equals correctly typed sentences per hour times \$0.03, against an indicator for accepting an hourly wage offer. I partial out wage effects by regressing output value against treatment offers and log effective hourly wages among hourly workers, then subtracting the demeaned wage effect implied by the coefficient on log effective hourly wages. I then instrument for hourly wage take-up with log wage offer and an indicator variable for being in the no-offer control group. Columns (2)--(4) add control variables for the categories of observable characteristics listed in the bottom panel. Task controls include experimental-wave indicators and start time. Employment controls include employment-status indicators, student enrollment status, and number of previous tasks completed on Prolific. Demographic controls include race, gender, and age. \starnote
\end{tablenotes}
\end{table}

\clearpage
\begin{table}[p]
  \centering
  \caption{OLS Estimates of Selection on Output Value by Wage Offer}
  \label{tab:ASc}
  \resizebox{\columnwidth}{!}{{
\def\sym#1{\ifmmode^{#1}\else\(^{#1}\)\fi}
\begin{tabular}{l*{4}{d}}
\toprule
                    &\multicolumn{1}{c}{(1)}&\multicolumn{1}{c}{(2)}&\multicolumn{1}{c}{(3)}&\multicolumn{1}{c}{(4)}\\
                    &\multicolumn{1}{c}{Output Value}&\multicolumn{1}{c}{Output Value}&\multicolumn{1}{c}{Output Value}&\multicolumn{1}{c}{Output Value}\\
\midrule
Accepted Hourly Offer&      -2.664\sym{***}&      -2.552\sym{***}&      -2.513\sym{***}&      -2.387\sym{***}\\
                    &     (0.340)         &     (0.333)         &     (0.334)         &     (0.312)         \\
\addlinespace
Declined $\times$ Log Hourly Wage Offer&      0.0927         &       0.119         &       0.141         &       0.163\sym{*}  \\
                    &     (0.100)         &    (0.0970)         &    (0.0965)         &    (0.0906)         \\
\addlinespace
Accepted $\times$ Log Hourly Wage Offer&       0.666\sym{***}&       0.610\sym{***}&       0.605\sym{***}&       0.555\sym{***}\\
                    &     (0.121)         &     (0.117)         &     (0.117)         &     (0.110)         \\
\addlinespace
Accepted $\times$ Log Effective Hourly Wage&      -0.109         &     -0.0860         &     -0.0815         &     -0.0410         \\
                    &     (0.126)         &     (0.123)         &     (0.123)         &     (0.116)         \\
\addlinespace
Control Group (No Hourly Offer)&      -0.361         &      -0.323         &      -0.263         &      -0.263         \\
                    &     (0.234)         &     (0.226)         &     (0.226)         &     (0.209)         \\
\addlinespace
Number of Previous Tasks/1000&                     &                     &       0.180\sym{***}&       0.177\sym{***}\\
                    &                     &                     &    (0.0326)         &    (0.0313)         \\
\addlinespace
Age                 &                     &                     &                     &     -0.0359\sym{***}\\
                    &                     &                     &                     &   (0.00430)         \\
\addlinespace
Female              &                     &                     &                     &       0.624\sym{***}\\
                    &                     &                     &                     &     (0.111)         \\
\midrule
Task Controls       &\multicolumn{1}{c}{No}         &\multicolumn{1}{c}{Yes}         &\multicolumn{1}{c}{Yes}         &\multicolumn{1}{c}{Yes}         \\
Employment Controls &\multicolumn{1}{c}{No}         &\multicolumn{1}{c}{No}         &\multicolumn{1}{c}{Yes}         &\multicolumn{1}{c}{Yes}         \\
Demographic Controls&\multicolumn{1}{c}{No}         &\multicolumn{1}{c}{No}         &\multicolumn{1}{c}{No}         &\multicolumn{1}{c}{Yes}         \\
Dep. Mean           &\multicolumn{1}{c}{7.00}         &\multicolumn{1}{c}{7.00}         &\multicolumn{1}{c}{7.00}         &\multicolumn{1}{c}{7.00}         \\
R-squared           &\multicolumn{1}{c}{0.075}         &\multicolumn{1}{c}{0.121}         &\multicolumn{1}{c}{0.137}         &\multicolumn{1}{c}{0.254}         \\
\emph{N}            &\multicolumn{1}{c}{3030}         &\multicolumn{1}{c}{3030}         &\multicolumn{1}{c}{3030}         &\multicolumn{1}{c}{3030}         \\
\bottomrule
\end{tabular}
}
}
  \begin{tablenotes}
    This table reports estimated coefficients from OLS regressions of output value (number of correctly typed sentences per hour times \$0.03) against log hourly wage offers interacted with acceptance status, adjusting for log effective wages among hourly workers (see Equation~\eqref{eq:ols} in the text). The coefficient on ``Declined $\times$ Log Hourly Wage Offer'' captures the change in output value among piece-rate workers for each unit increase in their log hourly wage offer. The coefficient on ``Accepted $\times$ Log Hourly Wage Offer'' captures the change in output value among hourly workers for each unit increase in their log hourly wage offer. The coefficient on ``Accepted $\times$ Log Effective Hourly Wage'' captures the change in output value among hourly workers for each unit increase in the log hourly wage they are \emph{paid}, conditional on the wage they are \emph{offered}. Columns (2)--(4) add control variables for the categories of observable characteristics listed in the bottom panel. Task controls include experimental-wave indicators and start time. Employment controls include employment-status indicators, student enrollment status, and number of previous tasks completed on Prolific. Demographic controls include race, gender, and age. \starnote
\end{tablenotes}
\end{table}

\clearpage
\bibliographystyle{econ}
\singlespacing

\renewcommand{\bibfont}{\footnotesize}
\renewcommand{\bibsection}{\section*{\small References}}
\setlength{\bibsep}{0pt}
\setlength{\bibhang}{1em}
\bibliography{references}

\global\long\def\thefigure{\Alph{section}\arabic{figure}}\setcounter{secnumdepth}{1} \setcounter{figure}{0}
\global\long\def\thetable{\Alph{section}\arabic{table}}\setcounter{table}{0}

\clearpage\begin{appendices}

  \section{Additional Figures and Tables}\label{sec:addfigtables}

\begin{figure}[h]
  \centering \caption{Counterfactual Equilibrium without Moral Hazard}
  \label{fig:curvesnomh}
  \begin{overpic}[width=\columnwidth,trim=.4cm .2cm .2cm .1cm,clip]{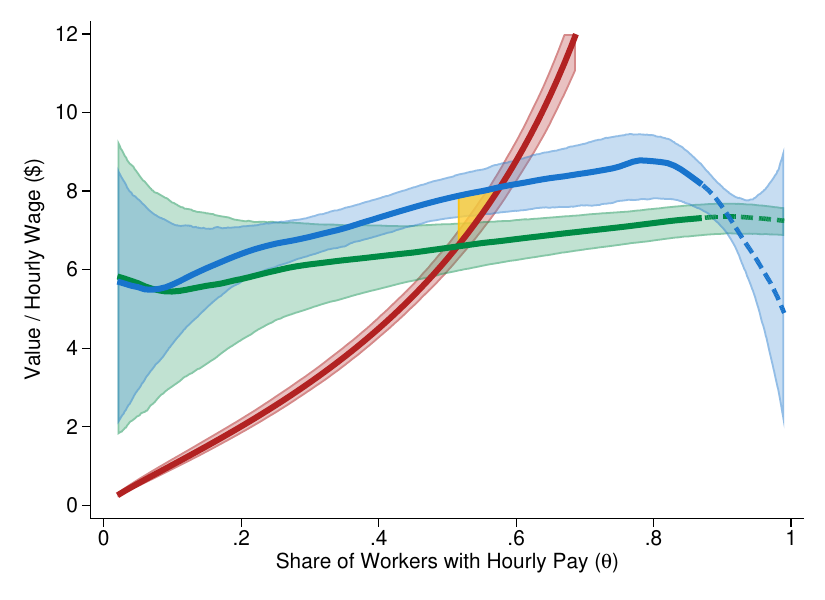}
  \put(74, 68){\wbar}
  \put(50, 54){\mv{0}}
  \put(80, 41){\av{0}}
\end{overpic}
  \begin{figurenotes}This figure reports counterfactual equilibrium and welfare estimates in the absence of moral hazard. The blue and green lines plot semi-parametric estimates of the marginal value, $MV_0\left(\theta\right)$, and average value, $AV_0\left(\theta\right)$, under the piece rate. The red line reproduces the estimated hourly reservation-wage curve from Figure~\ref{fig:curves}, holding the worker supply curve fixed at its estimated baseline. The area in yellow denotes the resulting welfare loss. Since curves are estimated conditional on task timing and employment history, plotted lines are evaluated at the sample means of these covariates. Value curves are estimated using a second-degree local polynomial regression (bandwidth=0.2) of residualized hourly output value against predicted hourly supply. Dashed portions of the value curves represent regions outside the support of observed propensity scores over which local polynomials were extrapolated. Shaded regions represent 90\% confidence intervals, constructed via bootstrap resampling.\end{figurenotes}
\end{figure}

\vspace{2.7cm}
\begin{figure}[h]
  \centering{}\protect\caption{Test for Bunching at Wage-Offer Reference Points}\label{fig:lossbunching} \resizebox{\columnwidth}{!}{\includegraphics{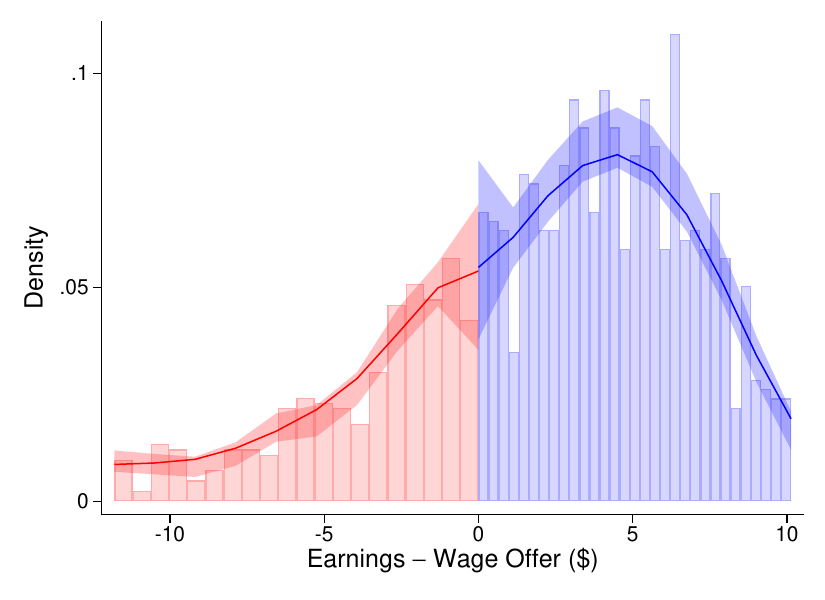}}
  \begin{figurenotes}This figure reports estimated densities of realized hourly earnings minus initial wage offers among workers in all treatment-offer groups who declined their wage offers in favor of the piece rate. The red and blue lines denote separate estimated densities for earnings below and above initial wage offers. Second-order local polynomial density estimates are based on a triangular kernel method from \citet{cattaneo2020simple}, using optimal bandwidths of {\bwllossbunching} (left) and {\bwrlossbunching} (right) of the cutoff. The p-value for the continuity test at the cutoff is \plossbunching. The shaded regions indicate 90\% confidence intervals computed from bias-corrected jackknife standard errors.
    \end{figurenotes}
  \end{figure}

\clearpage
\begin{figure}[p]
  \centering \caption{Asymmetric Information with Input and Output Monitoring Costs}
  \label{fig:model_mon}
  \begin{subfigure}{\textwidth}  \centering
    \includegraphics[width=\linewidth,trim=2cm 0.8cm 2cm 1.8cm,clip]{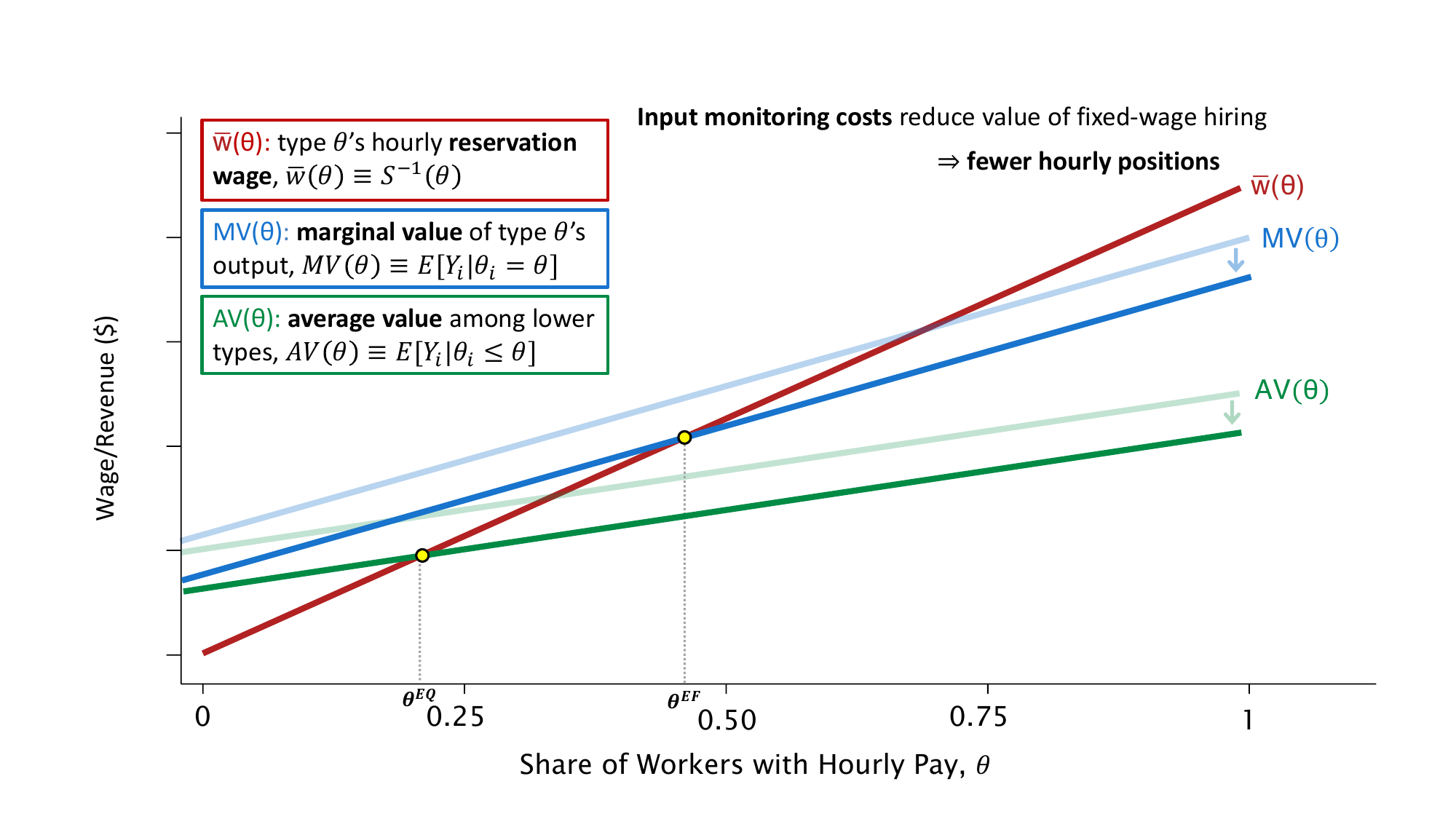}
    \caption{\large{Input Monitoring Costs}}
  \end{subfigure}\\\vspace{.5cm}
  \begin{subfigure}{\textwidth}  \centering
    \includegraphics[width=\linewidth,trim=2cm 0.8cm 0cm 1.8cm,clip]{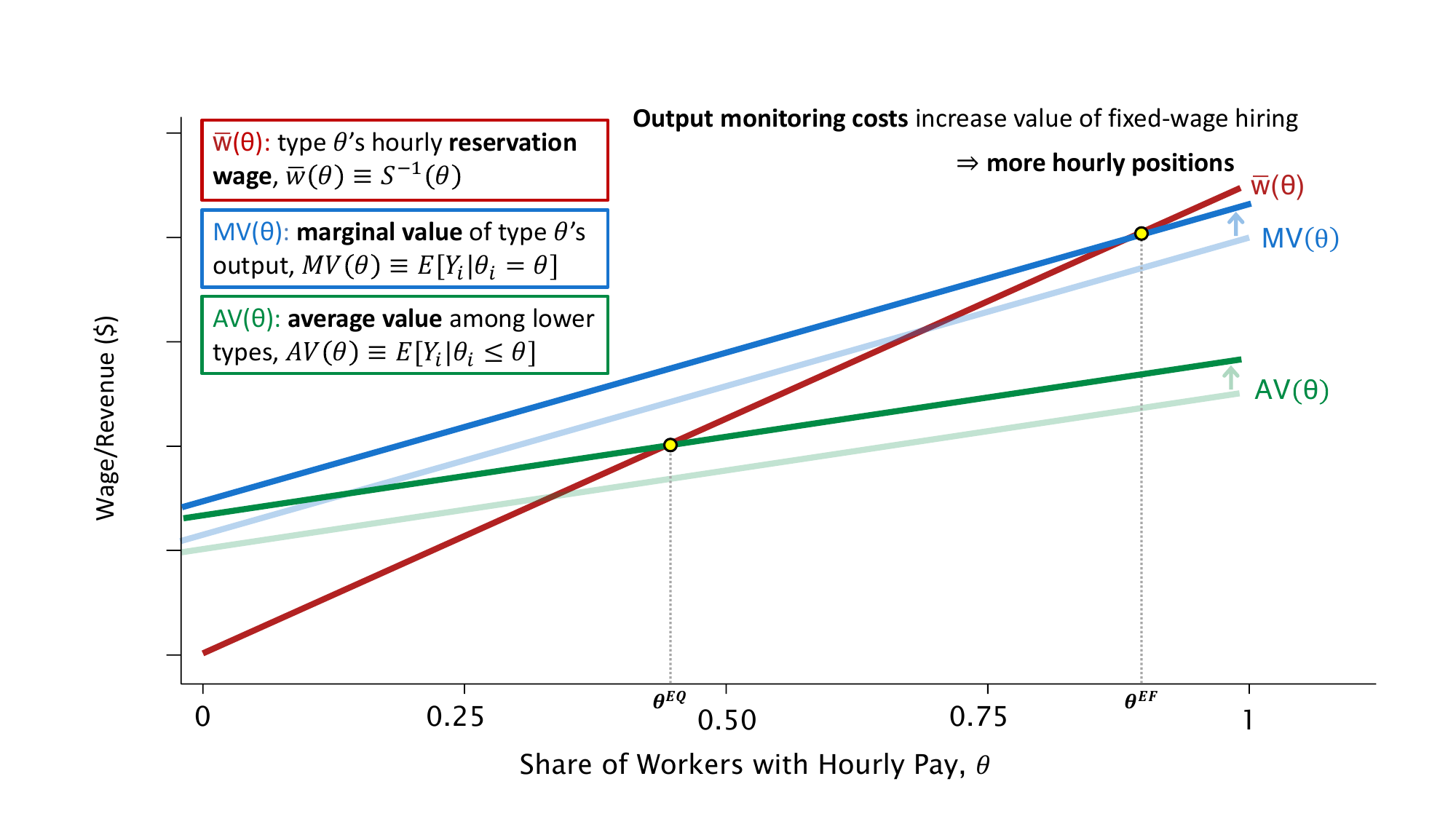}
    \caption{\large{Output Monitoring Costs}}
  \end{subfigure}
  \begin{figurenotes}[.93]This figure provides a graphical representation of a market for hourly wages under asymmetric information with input and output monitoring costs. \end{figurenotes}
\end{figure}

  \clearpage
  \begin{figure}[p]
  \centering \caption{Estimates of Marginal and Average Value Curves with Monotone-Spline First Stage}
  \label{fig:curvesfullsemi}
  \begin{subfigure}{\textwidth}
    \centering
\begin{overpic}[scale=1.05,trim=.4cm 1.7cm .5cm 1.5cm,clip]{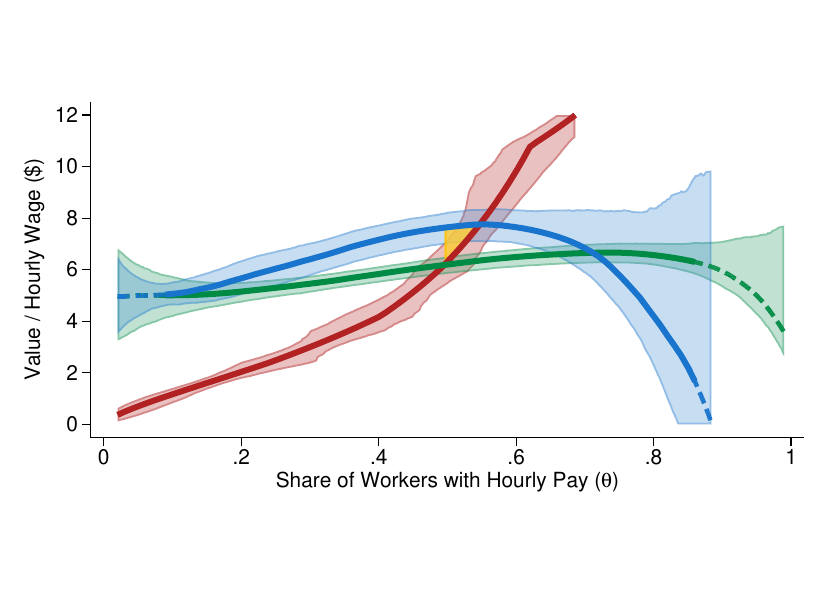}
  \put(76, 49.5){\wbar}
  \put(48, 39){\mv{1}}
  \put(32, 23.5){\av{1}}
\end{overpic}
    \caption{Potential Value Under Hourly Wage}
  \end{subfigure}\\
  \begin{subfigure}{\textwidth}
    \centering
\begin{overpic}[scale=1.05,trim=.4cm 1.7cm .5cm 1.5cm,clip]{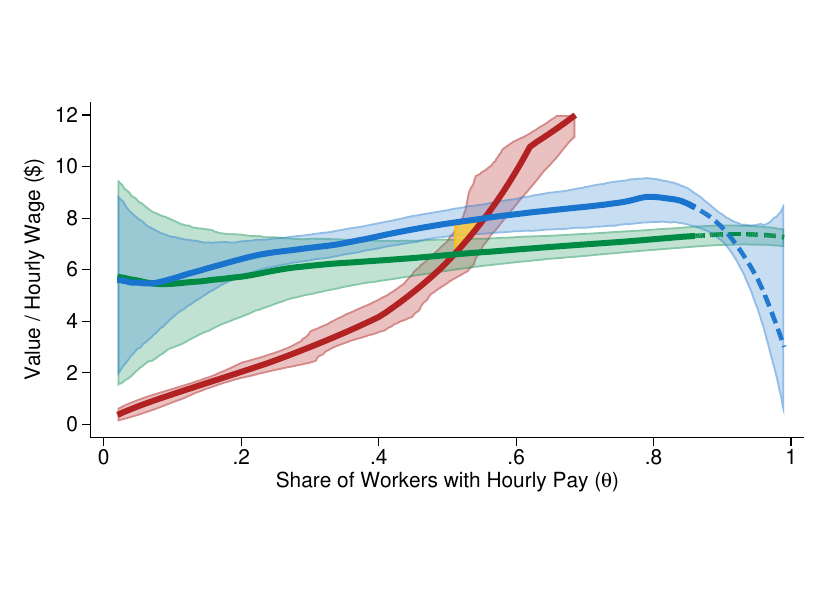}
  \put(76, 49.5){\wbar}
  \put(47.5, 39.5){\mv{0}}
  \put(80, 29){\av{0}}
\end{overpic}
    \caption{Potential Value Under Piece Rate}
  \end{subfigure}
  \begin{figurenotes}This figure reproduces the value-curve estimates in Figure~\ref{fig:curves} and Appendix Figure~\ref{fig:curvesnomh} using an alternative first-stage specification that replaces the linear log-wage term in the first-stage logit with a monotone increasing spline. All plotted curves are evaluated at the sample means of employer-screenable characteristics. The implied hourly reservation-wage curves are plotted in red, while blue and green lines plot semi-parametric estimates of marginal value and average value curves under hourly-wage (top panel) and piece-rate (bottom panel) counterfactuals. Areas in yellow illustrate welfare loss at the mean covariate profile. Averaging profile-specific losses yields \$\fp{\popDWLHcorvalueXXfullsemi} (SE=\$\fp[3]{\popDWLHsecorvalueXXfullsemi}) under hourly-wage potential values and \$\fp{\popDWLPcorvalueXXfullsemi} (SE=\$\fp[3]{\popDWLPsecorvalueXXfullsemi}) under piece-rate potential values. Shaded regions represent 90\% confidence intervals, constructed via bootstrap resampling. Monotone splines are estimated using five interior knots placed at equally spaced quantiles of log wage offers. See Appendix~\ref{sec:curvedetails} for estimation details.
\end{figurenotes}
\end{figure}

  \clearpage
  \begin{figure}[p]
  \centering \caption{Estimated MVPFs for Hourly-Wage Subsidies}
  \label{fig:mvpfsub}
    \includegraphics[scale=1,trim=0cm .4cm 0cm 0cm,clip]{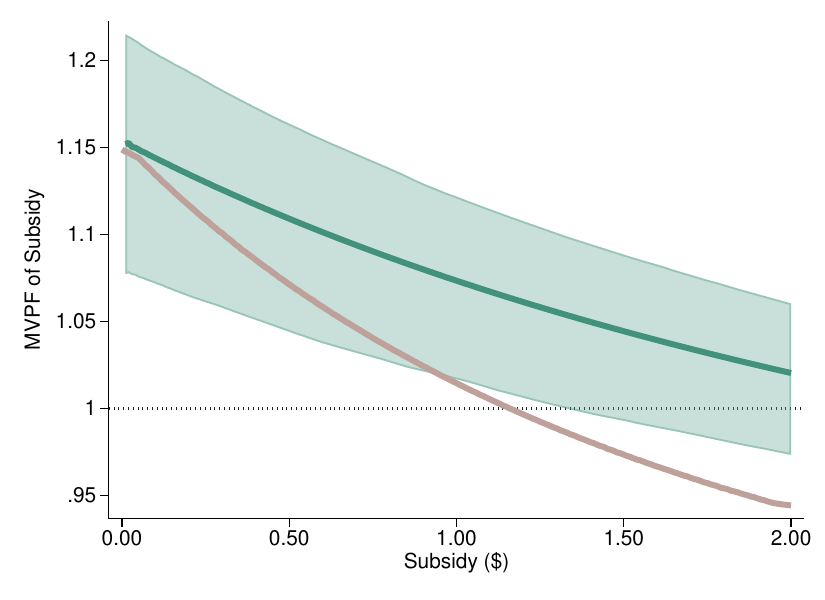}
  \vspace{.12cm}
  \begin{figurenotes}This figure plots estimated marginal values of public funds (MVPFs) for hourly-wage subsidies. The green line plots estimated MVPFs associated with hypothetical hourly-wage subsidies (in dollars per hour worked) denoted on the horizontal axis, and the pink line plots estimated MVPFs associated with a marginal increase in the subsidy. At each subsidy level, willingness to pay and net cost are averaged over the empirical distribution of employer-screenable characteristics before their ratio is formed. MVPFs are constructed as in Appendix~\ref{sec:mvpfdetails}. The shaded region represents the pointwise 90\% confidence interval for the level MVPF, constructed via bootstrap resampling.
  \end{figurenotes}
  \end{figure}

  \clearpage
  \begin{figure}[p]
\caption{Calibrated Value Curves for Lease-Free Rideshare Driving}
  \label{fig:uber}
\resizebox{\columnwidth}{!}{
\begin{overpic}[trim=.42cm .4cm 0cm 0cm,clip]{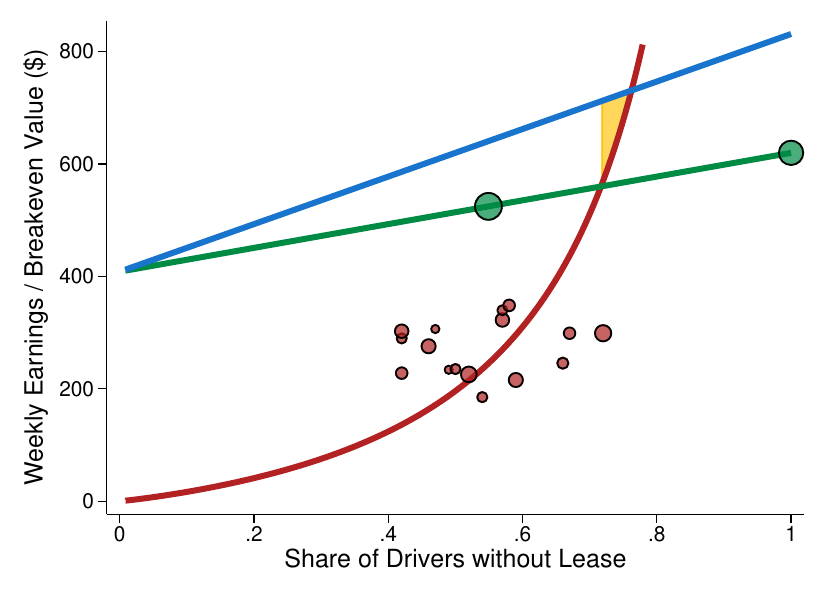}
  \put(76, 69.5){\wbarb}
  \put(101, 70){\mv{}}
  \put(101, 57){\av{}}
\end{overpic}
}
\begin{figurenotes}This figure plots reservation wages and value curves for lease-free rideshare contracts, calibrated using estimates from \citet{angrist2021uber}. The horizontal axis denotes the share of drivers declining weekly lease offers. The red curve plots the inverse supply of lease-free drivers, $\overline{w}_B\left(\theta\right)\equiv S^{-1}(\theta)$---the breakeven value, $w_B$, at which a $1-\theta$ share of drivers will accept taxi leases. The blue line plots the marginal value curve, $MV\left(\theta\right)\equiv E[Y_i|\theta_i=\theta]$, where $Y_i$ denotes driver $i$'s weekly revenue from ride fares. The green line plots the average value curve, $AV\left(\theta\right)\equiv E[Y_i|\theta_i\leq\theta]$. Red circles plot lease opt-out rates against experimental breakeven offers residualized by strata- and week-fixed effects and recentered at their overall mean. Green circles denote estimates of average fare revenues earned by decliners of lease offers and all drivers, respectively, constructed from fee-adjusted gains and driver-count-weighted averages across offers. Circle sizes reflect frequency weights based on reported observation counts. The yellow region corresponds to the welfare loss from adverse selection into lease-free rideshare contracts expressed in fare-revenue units. See Appendix~\ref{sec:uberapp} for details.
  \end{figurenotes}
\end{figure}

\clearpage

\clearpage
  \begin{table}[p]
    \centering
    \caption{Balance Test}
    \label{tab:balance}
  \resizebox{\columnwidth}{!}{{
\def\sym#1{\ifmmode^{#1}\else\(^{#1}\)\fi}
\begin{tabular}{l*{3}{d}}
\toprule
                    &\multicolumn{1}{c}{(1)}&\multicolumn{1}{c}{(2)}&\multicolumn{1}{c}{(3)}\\
                    &\multicolumn{1}{c}{Any Hourly Wage Offer}&\multicolumn{1}{c}{Experimental Wage Offer}&\multicolumn{1}{c}{Output Value}\\
\midrule
Number of Previous Tasks/1000&     0.00460         &     -0.0446         &       0.195\sym{***}\\
                    &   (0.00316)         &    (0.0718)         &    (0.0321)         \\
\addlinespace
Age                 &   0.0000335         &     0.00421         &     -0.0599\sym{***}\\
                    &  (0.000520)         &    (0.0109)         &   (0.00477)         \\
\addlinespace
Female              &    -0.00469         &       0.156         &       0.567\sym{***}\\
                    &    (0.0118)         &     (0.257)         &     (0.113)         \\
\addlinespace
Minority            &     0.00793         &     0.00323         &      -1.040\sym{***}\\
                    &    (0.0117)         &     (0.263)         &     (0.114)         \\
\addlinespace
Employed            &      0.0173         &      -0.234         &       0.144         \\
                    &    (0.0135)         &     (0.287)         &     (0.127)         \\
\addlinespace
Student             &      0.0106         &       0.366         &      -0.538\sym{***}\\
                    &    (0.0162)         &     (0.355)         &     (0.156)         \\
\midrule
F-statistic         &\multicolumn{1}{c}{0.707}         &\multicolumn{1}{c}{0.872}         &\multicolumn{1}{c}{30.456}         \\
$p$-value           &\multicolumn{1}{c}{0.733}         &\multicolumn{1}{c}{0.568}         &\multicolumn{1}{c}{0.000}         \\
\emph{N}            &\multicolumn{1}{c}{3030}         &\multicolumn{1}{c}{2728}         &\multicolumn{1}{c}{3030}         \\
\bottomrule
\end{tabular}
}
}
  \begin{tablenotes} This table reports results from a test of balanced treatment for experimental hourly wage offers. Columns 1 and 2 report estimated coefficients from OLS regressions of experimental group assignment measures against the baseline demographic variables reported in the leftmost column. The outcome variable in Column~1 is an indicator for receiving any experimental wage offer. The outcome variable in Column~2 is the level of wage offer received among workers not in the control group. Column 3 reports estimated coefficients from the same specification, but with output value as the dependent variable. The bottom rows report F-statistics and $p$-values from a test of joint significance for all right-hand side variables. \starnote
    \end{tablenotes}
  \end{table}

\clearpage
\begin{table}[p]
  \centering
  \caption{2SLS Estimates of Treatment Effects of Hourly Wages on Output Value without Adjusting for Wage Effects}
  \label{tab:NTEc}
  \resizebox{\columnwidth}{!}{{
\def\sym#1{\ifmmode^{#1}\else\(^{#1}\)\fi}
\begin{tabular}{l*{4}{d}}
\toprule
                    &\multicolumn{1}{c}{(1)}&\multicolumn{1}{c}{(2)}&\multicolumn{1}{c}{(3)}&\multicolumn{1}{c}{(4)}\\
                    &\multicolumn{1}{c}{Output Value}&\multicolumn{1}{c}{Output Value}&\multicolumn{1}{c}{Output Value}&\multicolumn{1}{c}{Output Value}\\
\midrule
Accepted Hourly Offer&      -0.654\sym{***}&      -0.648\sym{***}&      -0.634\sym{***}&      -0.517\sym{***}\\
                    &     (0.215)         &     (0.208)         &     (0.207)         &     (0.193)         \\
\addlinespace
Number of Previous Tasks/1000&                     &                     &       0.178\sym{***}&       0.178\sym{***}\\
                    &                     &                     &    (0.0337)         &    (0.0324)         \\
\addlinespace
Age                 &                     &                     &                     &     -0.0426\sym{***}\\
                    &                     &                     &                     &   (0.00445)         \\
\addlinespace
Female              &                     &                     &                     &       0.561\sym{***}\\
                    &                     &                     &                     &     (0.113)         \\
\midrule
Task Controls       &\multicolumn{1}{c}{No}         &\multicolumn{1}{c}{Yes}         &\multicolumn{1}{c}{Yes}         &\multicolumn{1}{c}{Yes}         \\
Employment Controls &\multicolumn{1}{c}{No}         &\multicolumn{1}{c}{No}         &\multicolumn{1}{c}{Yes}         &\multicolumn{1}{c}{Yes}         \\
Demographic Controls&\multicolumn{1}{c}{No}         &\multicolumn{1}{c}{No}         &\multicolumn{1}{c}{No}         &\multicolumn{1}{c}{Yes}         \\
R-squared           &\multicolumn{1}{c}{0.042}         &\multicolumn{1}{c}{0.090}         &\multicolumn{1}{c}{0.106}         &\multicolumn{1}{c}{0.223}         \\
\emph{N}            &\multicolumn{1}{c}{3030}         &\multicolumn{1}{c}{3030}         &\multicolumn{1}{c}{3030}         &\multicolumn{1}{c}{3030}         \\
\bottomrule
\end{tabular}
}
}
\begin{tablenotes}
  This table reports estimated coefficients from two-stage least-squares regressions of unresidualized output value against an indicator for accepting an hourly wage offer. I instrument for hourly wage take-up with log wage offer and an indicator variable for being in the no-offer control group. Columns (2)--(4) add control variables for the categories of observable characteristics listed in the bottom panel. Task controls include experimental-wave indicators and start time. Employment controls include employment-status indicators, student enrollment status, and number of previous tasks completed on Prolific. Demographic controls include race, gender, and age. \starnote
\end{tablenotes}
\end{table}

\clearpage
\begin{table}[p]
  \centering
  \caption{2SLS Estimates of Treatment Effects of Hourly Wages on Output Value Based on Completed Sentences}
  \label{tab:TEcnum}
  \resizebox{\columnwidth}{!}{{
\def\sym#1{\ifmmode^{#1}\else\(^{#1}\)\fi}
\begin{tabular}{l*{4}{d}}
\toprule
                    &\multicolumn{1}{c}{(1)}&\multicolumn{1}{c}{(2)}&\multicolumn{1}{c}{(3)}&\multicolumn{1}{c}{(4)}\\
                    &\multicolumn{1}{c}{Output Value}&\multicolumn{1}{c}{Output Value}&\multicolumn{1}{c}{Output Value}&\multicolumn{1}{c}{Output Value}\\
\midrule
Accepted Hourly Offer&      -0.503\sym{**} &      -0.498\sym{**} &      -0.485\sym{**} &      -0.362\sym{**} \\
                    &     (0.206)         &     (0.200)         &     (0.199)         &     (0.185)         \\
\addlinespace
Number of Previous Tasks/1000&                     &                     &       0.164\sym{***}&       0.174\sym{***}\\
                    &                     &                     &    (0.0338)         &    (0.0322)         \\
\addlinespace
Age                 &                     &                     &                     &     -0.0528\sym{***}\\
                    &                     &                     &                     &   (0.00420)         \\
\addlinespace
Female              &                     &                     &                     &       0.365\sym{***}\\
                    &                     &                     &                     &     (0.108)         \\
\midrule
Task Controls       &\multicolumn{1}{c}{No}         &\multicolumn{1}{c}{Yes}         &\multicolumn{1}{c}{Yes}         &\multicolumn{1}{c}{Yes}         \\
Employment Controls &\multicolumn{1}{c}{No}         &\multicolumn{1}{c}{No}         &\multicolumn{1}{c}{Yes}         &\multicolumn{1}{c}{Yes}         \\
Demographic Controls&\multicolumn{1}{c}{No}         &\multicolumn{1}{c}{No}         &\multicolumn{1}{c}{No}         &\multicolumn{1}{c}{Yes}         \\
Dep. Mean           &\multicolumn{1}{c}{7.91}         &\multicolumn{1}{c}{7.91}         &\multicolumn{1}{c}{7.91}         &\multicolumn{1}{c}{7.91}         \\
R-squared           &\multicolumn{1}{c}{0.037}         &\multicolumn{1}{c}{0.080}         &\multicolumn{1}{c}{0.095}         &\multicolumn{1}{c}{0.232}         \\
\emph{N}            &\multicolumn{1}{c}{3030}         &\multicolumn{1}{c}{3030}         &\multicolumn{1}{c}{3030}         &\multicolumn{1}{c}{3030}         \\
\bottomrule
\end{tabular}
}
}
\begin{tablenotes}
  This table reports estimated coefficients from two-stage least-squares regressions of residualized raw output value, where raw output value equals completed sentences per hour times \$0.03, against an indicator for accepting an hourly wage offer. I partial out wage effects by regressing output value against treatment offers and log effective hourly wages among hourly workers, then subtracting the demeaned wage effect implied by the coefficient on log effective hourly wages. I then instrument for hourly wage take-up with log wage offer and an indicator variable for being in the no-offer control group. Columns (2)--(4) add control variables for the categories of observable characteristics listed in the bottom panel. Task controls include experimental-wave indicators and start time. Employment controls include employment-status indicators, student enrollment status, and number of previous tasks completed on Prolific. Demographic controls include race, gender, and age. \starnote
\end{tablenotes}
\end{table}

\clearpage
\begin{table}[p]
  \centering
  \caption{OLS Estimates of Selection on Output Value Based on Completed Sentences by Wage Offer}
  \label{tab:AScnum}
  \resizebox{\columnwidth}{!}{{
\def\sym#1{\ifmmode^{#1}\else\(^{#1}\)\fi}
\begin{tabular}{l*{4}{d}}
\toprule
                    &\multicolumn{1}{c}{(1)}&\multicolumn{1}{c}{(2)}&\multicolumn{1}{c}{(3)}&\multicolumn{1}{c}{(4)}\\
                    &\multicolumn{1}{c}{Output Value}&\multicolumn{1}{c}{Output Value}&\multicolumn{1}{c}{Output Value}&\multicolumn{1}{c}{Output Value}\\
\midrule
Accepted Hourly Offer&      -2.522\sym{***}&      -2.425\sym{***}&      -2.383\sym{***}&      -2.195\sym{***}\\
                    &     (0.325)         &     (0.318)         &     (0.318)         &     (0.297)         \\
\addlinespace
Declined $\times$ Log Hourly Wage Offer&       0.172\sym{*}  &       0.196\sym{**} &       0.213\sym{**} &       0.233\sym{***}\\
                    &    (0.0961)         &    (0.0932)         &    (0.0925)         &    (0.0855)         \\
\addlinespace
Accepted $\times$ Log Hourly Wage Offer&       0.616\sym{***}&       0.565\sym{***}&       0.564\sym{***}&       0.497\sym{***}\\
                    &     (0.116)         &     (0.112)         &     (0.113)         &     (0.104)         \\
\addlinespace
Accepted $\times$ Log Effective Hourly Wage&     -0.0796         &     -0.0583         &     -0.0583         &     -0.0153         \\
                    &     (0.121)         &     (0.117)         &     (0.118)         &     (0.110)         \\
\addlinespace
Control Group (No Hourly Offer)&      -0.330         &      -0.296         &      -0.242         &      -0.228         \\
                    &     (0.226)         &     (0.219)         &     (0.219)         &     (0.201)         \\
\addlinespace
Number of Previous Tasks/1000&                     &                     &       0.166\sym{***}&       0.173\sym{***}\\
                    &                     &                     &    (0.0325)         &    (0.0309)         \\
\addlinespace
Age                 &                     &                     &                     &     -0.0454\sym{***}\\
                    &                     &                     &                     &   (0.00402)         \\
\addlinespace
Female              &                     &                     &                     &       0.435\sym{***}\\
                    &                     &                     &                     &     (0.105)         \\
\midrule
Task Controls       &\multicolumn{1}{c}{No}         &\multicolumn{1}{c}{Yes}         &\multicolumn{1}{c}{Yes}         &\multicolumn{1}{c}{Yes}         \\
Employment Controls &\multicolumn{1}{c}{No}         &\multicolumn{1}{c}{No}         &\multicolumn{1}{c}{Yes}         &\multicolumn{1}{c}{Yes}         \\
Demographic Controls&\multicolumn{1}{c}{No}         &\multicolumn{1}{c}{No}         &\multicolumn{1}{c}{No}         &\multicolumn{1}{c}{Yes}         \\
Dep. Mean           &\multicolumn{1}{c}{7.91}         &\multicolumn{1}{c}{7.91}         &\multicolumn{1}{c}{7.91}         &\multicolumn{1}{c}{7.91}         \\
R-squared           &\multicolumn{1}{c}{0.081}         &\multicolumn{1}{c}{0.122}         &\multicolumn{1}{c}{0.138}         &\multicolumn{1}{c}{0.272}         \\
\emph{N}            &\multicolumn{1}{c}{3030}         &\multicolumn{1}{c}{3030}         &\multicolumn{1}{c}{3030}         &\multicolumn{1}{c}{3030}         \\
\bottomrule
\end{tabular}
}
}
  \begin{tablenotes}
    This table reports estimated coefficients from OLS regressions of raw output value (number of completed sentences per hour times \$0.03) against log hourly wage offers interacted with acceptance status, adjusting for log effective wages among hourly workers (see Equation~\eqref{eq:ols} in the text). The coefficient on ``Declined $\times$ Log Hourly Wage Offer'' captures the change in output value among piece-rate workers for each unit increase in their log hourly wage offer. The coefficient on ``Accepted $\times$ Log Hourly Wage Offer'' captures the change in output value among hourly workers for each unit increase in their log hourly wage offer. The coefficient on ``Accepted $\times$ Log Effective Hourly Wage'' captures the change in output value among hourly workers for each unit increase in the log hourly wage they are \emph{paid}, conditional on the wage they are \emph{offered}. Columns (2)--(4) add control variables for the categories of observable characteristics listed in the bottom panel. Task controls include experimental-wave indicators and start time. Employment controls include employment-status indicators, student enrollment status, and number of previous tasks completed on Prolific. Demographic controls include race, gender, and age. \starnote
\end{tablenotes}
\end{table}

  \clearpage
\begin{table}[p]
  \centering
  \caption{Multiple Hypothesis Tests}
  \label{tab:mht}
  \begin{tabular}{l*{4}{d}}\toprule & \multicolumn{4}{c}{Multiplicity-Adjusted $p$-values}\\ \midrule
Selection on $Y_0$&            &            &            &            \\
$H_0:\beta_0=0$&      0.3483&      0.2303&      0.1527&      0.0767\\
[1em]Selection on $Y_1$&            &            &            &            \\
$H_0:\beta_1=0$&      0.0003&      0.0003&      0.0003&      0.0003\\
[1em]Treatment Effect&            &            &            &            \\
$H_0:\psi=0$&      0.0037&      0.0030&      0.0037&      0.0183\\
[1em]\midrule

Task Controls       &\multicolumn{1}{c}{No}         &\multicolumn{1}{c}{Yes}         &\multicolumn{1}{c}{Yes}         &\multicolumn{1}{c}{Yes}         \\
Employment Controls &\multicolumn{1}{c}{No}         &\multicolumn{1}{c}{No}         &\multicolumn{1}{c}{Yes}         &\multicolumn{1}{c}{Yes}         \\
Demographic Controls&\multicolumn{1}{c}{No}         &\multicolumn{1}{c}{No}         &\multicolumn{1}{c}{No}         &\multicolumn{1}{c}{Yes}         \\
\bottomrule \end{tabular}

  \begin{tablenotes} This table reports $p$-values adjusted for multiple hypothesis testing following Theorem 3.1 from \citet{list2019multiple}. Each cell reports the smallest family-wise error rate (the rate of at least one false rejection) across all three hypotheses that still rejects the null hypothesis listed in that row. The first row corresponds to the null hypothesis of zero selection into higher wage offers on potential output under the piece rate ($H_0:\beta_0=0$ in Equation~\eqref{eq:ols}). The second row corresponds to the null hypothesis of zero selection into higher wage offers on potential output under the hourly wage ($H_0:\beta_1=0$ in Equation~\eqref{eq:ols}). The third row corresponds to the null hypothesis of a zero reduced-form effect of the wage offer on output ($H_0:\psi=0$, where $\psi=\operatorname{Cov}(Y_i,\widetilde{W}_i)/\operatorname{Var}(\widetilde{W}_i)$ and $\widetilde{W}_i$ is log wage offer residualized on controls). Each column's specifications include control variables for the categories of observable characteristics listed in the bottom panel. Task controls include experimental-wave indicators and start time. Employment controls include employment-status indicators, student enrollment status, and number of previous tasks completed on Prolific. Demographic controls include race, gender, and age.
  \end{tablenotes}
\end{table}

\onehalfspacing
  \clearpage
\section[Identification of Experimental Estimands]{Identification of Experimental\\ Estimands}\label{sec:expdetails}
\setcounter{secnumdepth}{2}

This appendix provides additional details and proofs behind the target estimands in my experimental design.

  \subsection{Identifying Average Selection in a Multiple-Offer Experiment}
Consider an example experiment with three offer conditions, $W_i\in \left\{0,L,H\right\}$. As in the previous example, control workers assigned to $W_i=0$ are offered the piece rate with no alternative. But now the remaining workers are randomly separated into two groups---workers assigned to $W_i=L$ are offered the choice between the piece rate and a low hourly wage, while workers assigned to $W_i=H$ are offered the choice between the piece rate and a high hourly wage. Let $D^L_i$ and $D^H_i$ be indicators for individual $i$'s potential take-up of contracts $L$ and $H$, respectively, and assume $D_i^H\geq D_i^L$ for all $i$. Let $\pi^j\equiv\Pr(D_i=1\mid W_i=j)$ denote the take-up rate for offer $j\in\{L,H\}$.\par
As in Equation~\eqref{eq:selection}, comparing decliners of a given wage offer with control workers identifies average selection on $Y_0$ into that offer among all workers. But now I can also compare outcomes between decliners in the high- and low-offer treatment groups to identify selection on $Y_0$ into offer $H$ among those who would reject the less generous offer ($L$):
\begin{align}
\underbrace{E\left[Y_{0i} |D^H_i=1,D^L_i=0\right] - E\left[Y_{0i} |D^H_i=0\right]}_{\text{Average Selection on $Y_0$}}& \nonumber \\=  \frac{1-\pi^L}{\pi^H-\pi^L}(E[Y_i|D_i&=0,W_i=L] - E\left[Y_i |D_i=0,W_i=H\right]).\label{eq:selection0}
\end{align}
At the same time, a comparison between \emph{accepters} in the high- and low-offer treatment groups identifies average selection on $Y_1$ into offer $L$ among those who would accept the more generous offer ($H$):
\begin{align}
\underbrace{E\left[Y_{1i} |D^L_i=1\right] -  E\left[Y_{1i} |D^H_i=1,D^L_i=0\right]}_{\text{Average Selection on $Y_1$}}& \nonumber \\=  \frac{\pi^H}{\pi^H-\pi^L}(E[Y_i|D_i&=1,W_i=L] - E\left[Y_i |D_i=1,W_i=H\right] ).\label{eq:selection1}
\end{align}
In short, because both high- and low-offer treatment arms contain a mix of hourly and piece-rate workers, this multiple treatment design allows me to identify worker selection on \emph{both} potential outcomes---productivity under the piece rate ($Y_0$) and productivity under hourly wages ($Y_1$).

\subsection{Identification of Wage Effects}

So far, I have assumed that a worker's assigned offer condition can only affect their outcome through the choice of hourly versus piece-rate contract. If hourly workers are paid their offered wages, this exclusion restriction could be violated through wage effects---higher pay might induce greater effort through increased motivation or satisfaction, biasing my estimates of both selection and treatment effects.

To separate the potential behavioral response to higher effective wages from the incentive effects of hourly contract structure, my experiment incorporates an additional dimension of randomization in the spirit of \citet{karlanzinman}. Specifically, after workers choose their compensation option, but before they begin the task, I increase hourly wages for a random subset of those accepting lower wage offers, bringing them to parity with higher treatment-offer groups.

To see how randomized wage top-ups separately identify wage effects and selection on $Y_1$, consider an individual $i$ who receives a job offer, $W_i$, at one of two randomized wages: a high offer ($W_i=H$) or a low offer ($W_i=L$). Let $D_{Wi}$ denote the individual's potential acceptance of an offer $W$, so that $D_{Hi}=1$ if $i$ would accept the high offer and $D_{Li}=1$ if $i$ would accept the low offer. Furthermore, let $Y_{Hi}$ and $Y_{Li}$ denote the potential output levels produced by $i$ if they were paid hourly wages of $H$ and $L$, respectively. Note that if realized wages reflected accepted offers, comparing output between those who accept $H$ and those who accept $L$ would yield the following:
\begin{align}
  & E\left[Y_{i} |W_i=H,D_{i}=1\right]  - E\left[Y_{i} |W_i=L,D_{i}=1\right] \nonumber \\
\label{eq:decomp} & =  \underbrace{E\left[Y_{Hi}-Y_{Li} |D_{Li}=1\right]}_{\text{Wage Effect}} + \underbrace{E\left[Y_{Hi} |D_{Hi}=1\right] - E\left[Y_{Hi} |D_{Li}=1\right]}_{\text{Selection}}.
\end{align}
This difference is the sum of both the wage effect and selection of $H$ relative to $L$, which cannot be separated without observing $E\left[Y_{Hi} |D_{Li}=1\right]$.

Now let $R_i$ be an indicator for whether individual $i$ receives a surprise wage increase of $\Delta = H-L$ after accepting their contract. $R_i$ is randomly assigned among those who received low offers ($W_i=L$) and accepted them ($D_{Li}=1$) but is zero for everyone else. With this randomized wage raise, I can estimate wage effects by comparing output between low- and high-wage workers in the low-offer group:
\begin{align}
  \label{eq:TE}
  \text{Wage Effect} &= E\left[Y_{i} |W_i=L,D_{i}=1,R_i=1\right] - E\left[Y_{i} |W_i=L,D_{i}=1,R_i=0\right]\nonumber \\
                          &= E\left[Y_{Hi} - Y_{Li} |D_{Li}=1\right].
\end{align}
And I can estimate selection by comparing output between low- and high-offer groups with high realized wages:
\begin{align}
  \label{eq:AS}
    \text{Selection} &= E\left[Y_{i} |W_i=H,D_{Hi}=1\right] - E\left[Y_{i} |W_i=L,D_{Li}=1,R_i=1\right]\nonumber \\
                   &= E\left[Y_{Hi} |D_{Hi}=1\right] - E\left[Y_{Hi} |D_{Li}=1\right].
\end{align}

\subsection{Identifying Marginal Value in a Linear Model}\label{sec:linearmarg}
Drawing from Equations~\eqref{eq:selection0} and~\eqref{eq:selection1}, consider the average potential outcomes among workers who reject offer $L$ but accept offer $H$.
\begin{align}
 E\left[Y_{1i} |D^H_i=1,D^L_i=0\right]
 &= \frac{\pi^H E\left[Y_i|D_i=1,W_i=H\right] - \pi^L E\left[Y_i |D_i=1,W_i=L\right]}{\pi^H-\pi^L}\label{eq:aselection1}\\
 E\left[Y_{0i} |D^H_i=1,D^L_i=0\right]
 &= \frac{\left(1-\pi^L\right)E\left[Y_i|D_i=0,W_i=L\right]}{\pi^H-\pi^L}\nonumber\\
 &\quad - \frac{\left(1-\pi^H\right)E\left[Y_i |D_i=0,W_i=H\right]}{\pi^H-\pi^L}.\label{eq:aselection0}
\end{align}
Let $\overline{w}_i$ denote the lowest offer individual $i$ is willing to accept. Let $H=w$ and $L=w-\tau$ in Equations~\eqref{eq:aselection1} and~\eqref{eq:aselection0}. The limits of $E\left[Y_{1i} |D^H_i=1,D^L_i=0\right]$ and $E\left[Y_{0i} |D^H_i=1,D^L_i=0\right]$ as $\tau\rightarrow 0$ can be written as
\begin{eqnarray}
  E\left[Y_{1i} |\overline{w}_i=w\right] & = & \frac{\partial\left(E\left[Y_{1i}|\overline{w}_i\leq w\right]S(w)\right)}{\partial S(w)}
\\
 E\left[Y_{0i} |\overline{w}_i=w\right] & = &-\frac{\partial \left(E\left[Y_{0i}|\overline{w}_i>w\right](1-S(w))\right)}{\partial S(w)}.
\end{eqnarray}
Now suppose $E[Y_{1i}|\overline{w}_i\leq w]$, $E[Y_{0i}|\overline{w}_i>w]$, and $S(w) \equiv \Pr\left(\overline{w}_i\leq w \right)$ are linear in the wage offer, $w$:
\begin{eqnarray}
  \label{eq:1}
  S(w) &=& \alpha + \beta w \\
    E[Y_{1i}|\overline{w}_i\leq w]   &=& \gamma_1 + \delta_1 w \\
 E[Y_{0i}|\overline{w}_i>w]      &=& \gamma_0 + \delta_0 w.
\end{eqnarray}
We therefore have
\begin{eqnarray}
  \label{eq:Yw1}
    E\left[Y_{1i} |\overline{w}_i=w\right]&=&\frac{(\gamma_1 + \delta_1w)\beta + (\alpha + \beta w)\delta_1 }{\beta} \\
              &=&\frac{\alpha\delta_1 }{\beta} + \gamma_1 + 2\delta_1w.
\end{eqnarray}
Likewise for $E\left[Y_{0i} |\overline{w}_i=w\right]$:
\begin{eqnarray}
  \label{eq:Yw2}
  E\left[Y_{0i} |\overline{w}_i=w\right]&=&\frac{-(\gamma_0 + \delta_0w)\beta + (1-\alpha - \beta w)\delta_0 }{-\beta} \\
              &=&\frac{(\alpha-1)\delta_0 }{\beta} + \gamma_0 + 2\delta_0w.
\end{eqnarray}
We therefore have
\begin{eqnarray}
  \label{eq:Yw3}
  \frac{\partial E\left[Y_{1i} |\overline{w}_i=w\right]}{\partial w} &=& 2\delta_1 \\
  \frac{\partial E\left[Y_{0i} |\overline{w}_i=w\right]}{\partial w} &=& 2\delta_0.
\end{eqnarray}

\clearpage
\global\long\def\thefigure{\Alph{section}\arabic{figure}}\setcounter{secnumdepth}{1} \setcounter{figure}{0}
\global\long\def\thetable{\Alph{section}\arabic{table}}\setcounter{table}{0}

\section{Experimental Protocol}\label{sec:settingdetails}

\begin{figure}[p]
  \centering \caption{Experiment Timeline}
  \label{fig:timeline}\vspace{-.25cm}
    \includegraphics[scale=0.55,trim=1.8cm 3cm 4.8cm 3cm,clip]{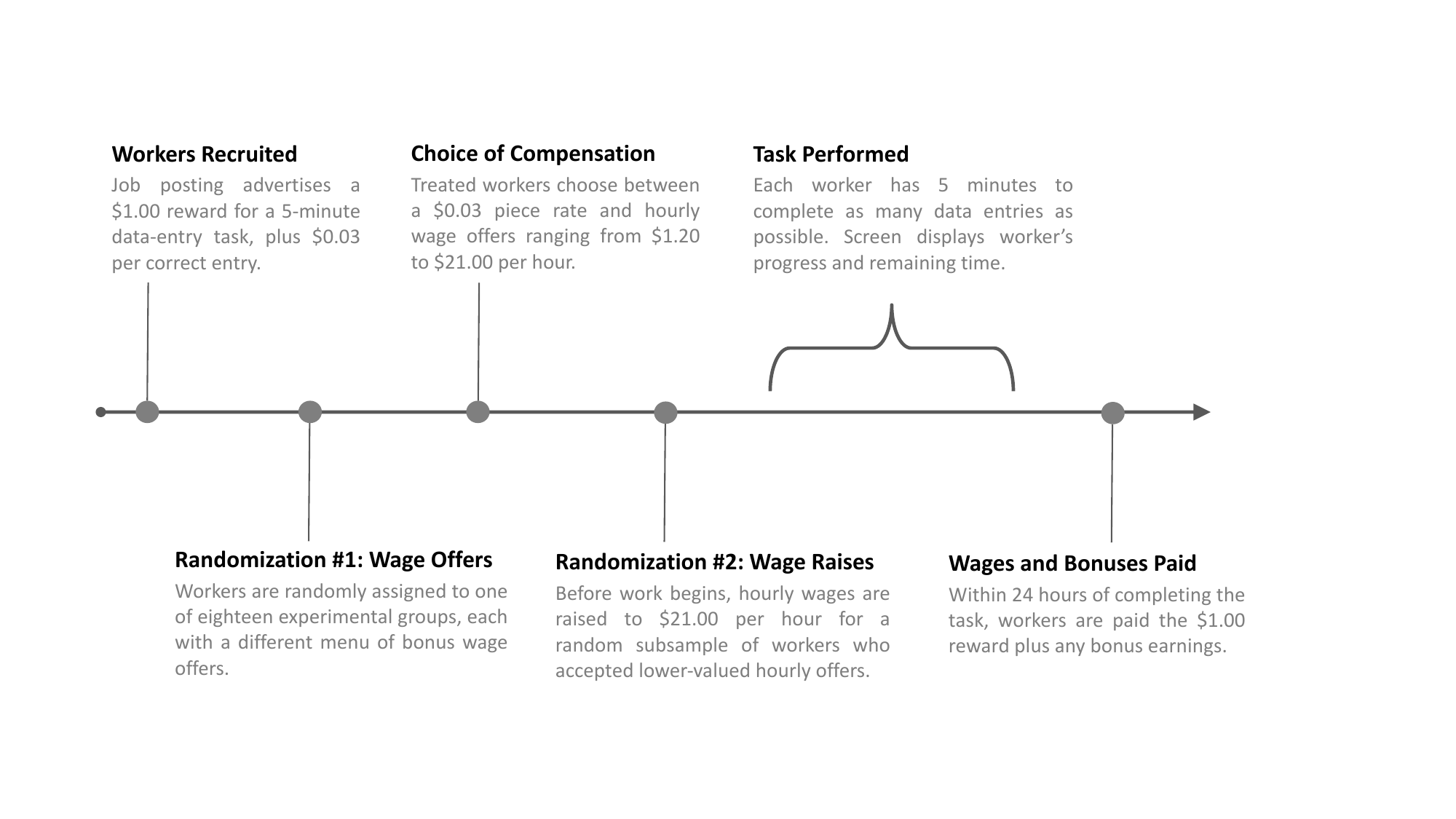}
  \begin{figurenotes}This figure provides a timeline for a single wave of the experiment. \end{figurenotes}
\end{figure}

The design and recruitment details for this experiment were pre-registered on the AEA RCT Registry under ID AEARCTR-13138, titled ``Asymmetric Information in Labor Contracts: Evidence from an Online Experiment'' \citep{registry}. The experiment took place in ten waves of three hundred job postings launched over the course of two weeks beginning August 31, 2024. Figure~\ref{fig:timeline} provides a timeline of the experimental protocol.

Participants in my experiment were recruited on Prolific, an online platform that allows clients to hire online workers for short-term tasks.\footnote{\citet{douglas2023data} find that the Prolific platform compares favorably to Amazon Mechanical Turk (``MTurk'') and other platforms across several dimensions of data quality.}  The experimental job posting offered workers a \$1.00 reward for transcribing handwritten text into typed form for five minutes. Such transcription tasks are commonly requested on Prolific and other online platforms, often for the purpose of training artificial intelligence (AI) algorithms. The posting also informed workers they ``can earn an additional \$0.03 in bonus compensation for each correctly typed sentence.''

Workers could only see my experimental job posting if they met the following screening criteria: (1) were located in the United States, (2) spoke fluent English, (3) successfully completed ten or more previous tasks, and (4) earned an approval rate above 98 percent on previous tasks.\footnote{More than 95 percent of Prolific workers meet the 98-percent approval threshold.} These screening criteria allow me to remove the small number of casual users who may take the tasks less seriously than ``professional'' online workers who regularly perform tasks to earn income. In doing so, they make the sample more representative of the online hiring pool faced by profit-conscious employers.

\begin{figure}[h!]
  \centering \caption{Example Job Posting}
  \label{fig:posting}
  \begin{subfigure}{\textwidth}  \centering    \fbox{\includegraphics[scale=0.3,trim=0cm 6cm 0cm 0cm,clip]{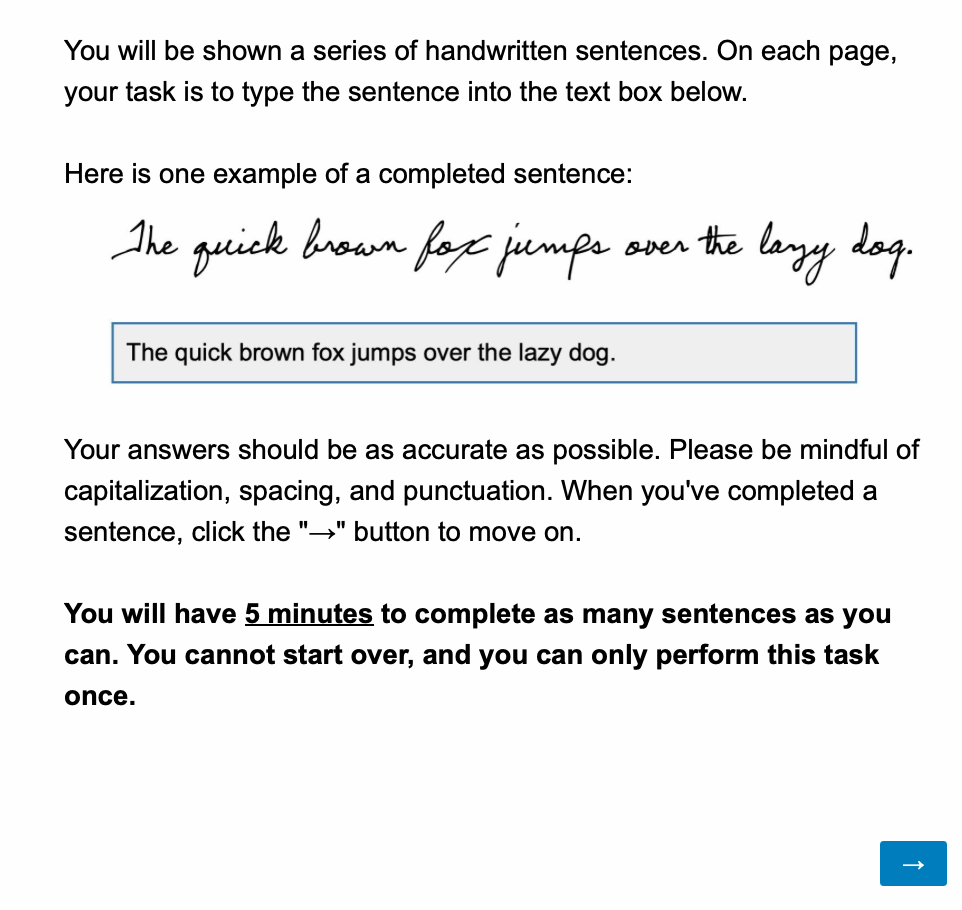}}
    \caption{Task Description}
  \end{subfigure}\\\vspace{.3cm}
  \begin{subfigure}{\textwidth}  \centering
    \fbox{\includegraphics[scale=0.3,trim=0cm 13cm 0cm .5cm,clip]{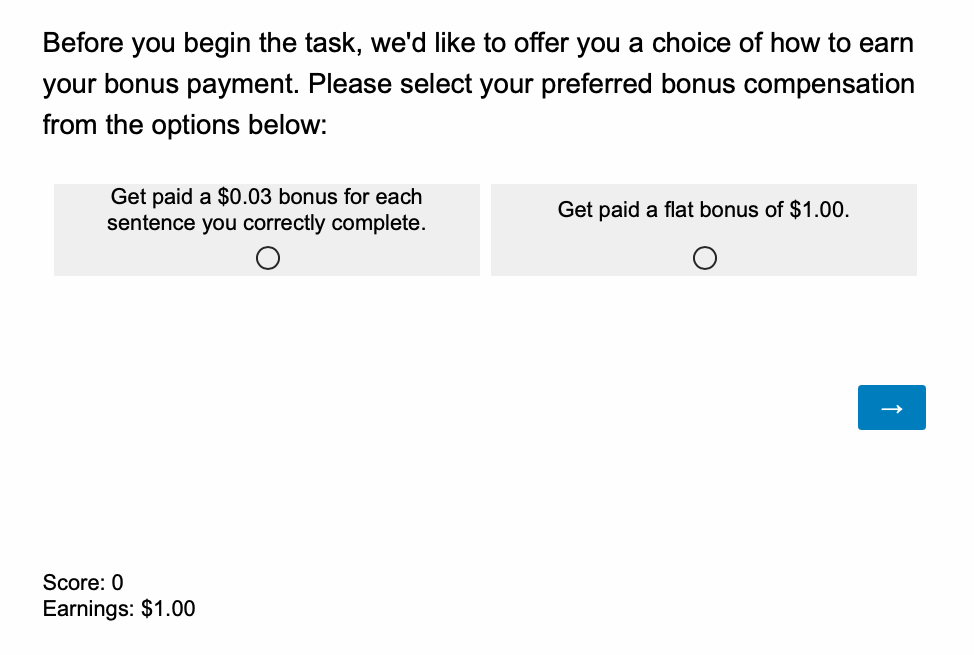}}
    \caption{Example Wage Offer}
  \end{subfigure}\\\vspace{.3cm}
  \begin{subfigure}{\textwidth}  \centering
    \fbox{\includegraphics[scale=0.3]{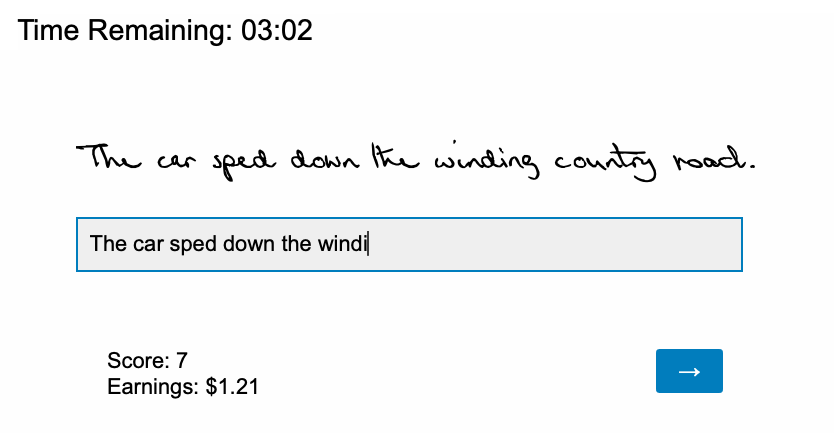}}
    \caption{Typing Task}
    \vspace{.3cm}
  \end{subfigure}
  \begin{figurenotes}[.99]This figure provides screenshots of the experimental intervention. Panel A shows the task description workers see before they see their wage offer. Panel B shows an example wage offer workers see before they begin the task. Panel~C shows the sentence-typing task while it is being performed.
  \end{figurenotes}
\end{figure}

Workers who accept the job posting are provided with a link to the experimental task. Upon clicking this link, workers are shown a screen with a brief task description and example entry.\footnote{The task is hosted on the Qualtrics platform. Readers can view and perform a replication of the task \href{https://uarizona.co1.qualtrics.com/jfe/form/SV_5vf9NKHpxk26Hau}{\color{blue} here}. Screenshots are provided in Figure~\ref{fig:posting}.} After clicking past the description page, workers are randomized into one of eighteen experimental groups. Each group is offered a different menu of compensation options in exchange for completing the five-minute data-entry task. In the first treatment group, workers are offered a choice between a fixed \$0.10 payment (\$1.20 per hour) and a piece rate of \$0.03 per correctly typed sentence.\footnote{A piece rate of \$0.03 per correctly typed sentence was chosen to roughly align with observed market rates for online image-to-text transcription services \citep{TranscriptionUS_TypingRates2024,Upwork_DataEntryRates2024,Upwork_DataEntryGuide2024}.} In the second treatment group, workers are offered a choice between a fixed \$0.15 payment (\$1.80 per hour) and the same \$0.03 piece rate. Additional treatment groups follow the same structure, with each condition offering the \$0.03 piece rate but increasing the flat wage offer by multiples of \$0.05, up to a maximum of \$1.75 (\$21.00 per hour). A control group is offered the \$0.03 piece rate for each correctly typed sentence, with no alternative option. Each of these options is offered as an addition to the \$1.00 reward advertised in the job posting, which all workers receive for agreeing to the task. Experimental conditions are summarized in Table~\ref{tab:groups}.

After receiving detailed instructions for the data-entry task, treated workers are presented with their group's payment options in randomized order, as shown in Figure~\ref{fig:posting}, Panel B. Once workers choose their compensation scheme, they are brought to a new page that states, ``For performing this task, you will receive \$1.00, plus your chosen bonus of [\emph{payment choice}].'' A randomly selected 50 percent of workers who choose lower-valued payment options receive a modified message that increases their base payment by enough to equalize their total compensation with the most generous offer (\$1.00 + \$1.75 = \$2.75). For example, half of those who select the \$0.25 payment are told ``you will receive \$2.50, plus your chosen bonus of \$0.25.''

Once workers are notified of their compensation and click ``Begin Task,'' they are presented with a handwritten sentence and a text box. The worker types a sentence in the box and clicks the ``Next'' button, bringing them to a new page with a different sentence. This process continues for five minutes. Worker output is validated in real time, so workers can see a running tally of their score (the number of correctly typed sentences) and their total earnings in the lower-left corner of each page. Workers also see a countdown timer displaying the number of minutes and seconds remaining in the task.\footnote{Figure~\ref{fig:posting}, Panel C provides a screenshot of the task. The display and submission methods for this task were designed to prevent workers from cheating through automation software or bots. While it is possible that some workers may have tried to make use of such software, performance statistics suggest any such attempts were unsuccessful at increasing output---the maximum score achieved was \fp[0]{\maxscore}.}  When the timer reaches zero, the screen refreshes to an end-of-task page displaying a performance summary and a completion link to redeem their earnings.  Workers are paid the \$1.00 reward plus any bonus earnings within 24 hours of completing the task.

\clearpage
\vspace{-.25cm}

\clearpage

\clearpage
\section{Model Extensions}\label{sec:extensions}

My model is designed so that objects of interest can be mapped to semi-parametric estimands from my experiment. For this reason, it omits features like monitoring costs, multiple contract structures, and dynamic wage-setting, which are absent from my experimental setting and many other short-term labor markets. It also omits wage effects, which are statistically insignificant in my setting (see Section~\ref{sec:results}). In this appendix, I discuss how one might extend the model to incorporate these features.

\subsection{Monitoring Costs}
Existing research shows how relative costs of monitoring worker inputs and outputs can influence worker productivity and equilibrium wage structure in a variety of occupations \citep{lazear1986salaries,lazear2000,goldinpiece,nagin2002,kelley2024monitoring,prendergast2002tenuous}. My paper seeks to complement this literature by identifying the market implications of asymmetric information holding these monitoring costs fixed. As such, worker time and productivity are costlessly observed in both my model and experimental setting. Nonetheless, one could easily extend my framework to incorporate a monitoring cost of measuring worker output, $q_i$. Such output monitoring costs would, all else equal, make fixed-wage contracts more likely than payment schemes that require precise measurement of individual worker productivity.\footnote{Note that fixed wages still require some degree of output or effort monitoring so that firms can credibly threaten low-performing workers with dismissal, rejection, or damaged reputation.}  Likewise, I could allow firms to face an input-monitoring cost of observing workers' time spent on the job, which would make fixed wages less likely than output-based pay. Appendix Figure~\ref{fig:model_mon} shows modified versions of my model that incorporate monitoring costs for inputs and outputs, respectively.

\subsection{Expanding the Contract Space}

The baseline model restricts the contract space to two polar forms: a pure fixed wage and a pure piece rate. This restriction approximates the principal payment margins in my target labor market: Short-term online freelancers are generally paid either a time rate or a task rate \citep{hornuf2022hourly}.\footnote{Upwork permits only hourly and fixed-price contracts \citep{upwork2026contracttypes}, while Prolific, Clickworker, and MTurk generally pay workers per task \citep{prolific2026paymentmodel,clickworker2026crowd,hara2018earnings}.} These patterns do not imply that richer arrangements never occur, but they suggest that the two-contract benchmark captures the main compensation margins in this setting. Nonetheless, my model can be extended to consider alternative contracts that combine fixed-wage and piece-rate components, as in jobs with base wages plus tips or commissions.

\paragraph{Mixed Compensation and Multiple Contracts}
Let a mixed contract be $c=(w,\eta)$, where the firm pays a fixed amount, $w$, in exchange for a share $\eta\in[0,1]$ of the worker's output value. Let $Y_i(\eta)$ denote worker $i$'s potential output value under output share $\eta$, and define the value of the firm's claim as $Y_i^p(\eta)\equiv\eta Y_i(\eta)$. The worker earns $w+(1-\eta)Y_i(\eta)$, while the firm earns $Y_i^p(\eta)-w$. Lowering $\eta$ increases the worker's marginal return to output and, assuming effort responds positively to that return, attenuates moral hazard.

My baseline model compares a pure fixed wage, $c^F=(w,1)$, with a pure piece rate, $c^P=(0,0)$, though the same two-contract framework can accommodate any fixed pair of contracts with output shares in $[0,1]$. Alternatively,  the framework can be extended to $N>2$ contracts, much like in \citet{efc}. In this case, firms can offer any combination of contracts from some menu, $\mathcal{C}=\{c_j=(w_j,\eta_j)\}_{j=0}^{N-1}$. Normalize the pure piece-rate contract, $c_0=c^P=(0,0)$, as the reference contract, and let $\boldsymbol{\eta}=(\eta_1,\ldots,\eta_{N-1})$ denote the exogenously specified output shares and $\mathbf{w}=(w_1,\ldots,w_{N-1})$ the relative fixed payments of the remaining contracts. Holding $\boldsymbol{\eta}$ fixed, let $\mathbf{AV}\left(\mathbf{w}\right)$ denote the vector of average incremental firm-claim values relative to the reference contract among workers choosing each non-reference contract. Under perfect competition and contract-by-contract zero profits, the equilibrium payment vector satisfies
\begin{equation}
  \mathbf{w}^{EQ}=\mathbf{AV}\left(\mathbf{w}^{EQ}\right).\label{eq:multieq}
\end{equation}
Meanwhile, the efficient payment vector satisfies
\begin{equation}  \mathbf{w}^{EF}=\mathbf{MV}\left(\mathbf{w}^{EF}\right),
  \label{eq:multieff}
\end{equation}
where $\mathbf{MV}\left(\mathbf{w}\right)$ corresponds to the vector of firms' claims on output values among workers marginally induced into each contract $j$ given the menu of fixed payments, $\mathbf{w}$. Welfare loss follows from integrating the marginal-value wedges between these allocations. When $N=2$ and $\eta_1=1$, the system reduces to my baseline model. Estimating it for larger $N$ would require contract-specific choices and outcomes, together with exogenous variation across all $N-1$ relative payments. With $N>2$, ensuring existence and uniqueness of solutions to \eqref{eq:multieq} and~\eqref{eq:multieff} requires additional regularity assumptions.

\paragraph{Fully Endogenous Menus}
In the extension above, firms can offer a variety of contracts with different output shares, but those output shares remain restricted to the exogenously specified finite set ($\boldsymbol{\eta}$). A richer extension would relax this restriction by allowing firms to choose output shares from a continuous contract space. This turns the model into a competitive screening problem \citep{rothstig,LaffontMartimort2002}. Depending on worker preferences and production technology, equilibrium could involve pooling, partial separation, or full separation.

Estimating this richer model would likely require imposing substantial structure on worker preferences, contract-specific productivity, feasible menus, and firm competition, placing it beyond the scope of this paper. Nonetheless, the two-contract benchmark I apply in my experiment appears broadly consistent with the coarse contract space observed in the market for online work, which is typically characterized by task- or time-based pay rather than a continuum of mixed contracts. This observation suggests that mixed contracts are either infeasible or unprofitable deviations from competitive equilibrium. In this case, the welfare loss I estimate would provide a weak lower bound on the welfare loss one would expect relative to a full-information benchmark with a richer contract space.

\subsection{Wage Effects}
The theoretical framework in Section~\ref{sec:model} allows workers' expected output to vary between hourly and piece-rate compensation. It does not, however, allow that output to vary with the wage level under an hourly contract. In other words, it ignores any potential wage effects that higher hourly compensation might have on worker output. For example, \citet{gneezy2006putting} find that unexpectedly generous wages can induce short-lived increases in worker effort. While the absence of wage effects in my empirical results would seem to validate this assumption, I can incorporate wage effects into the model by allowing each worker's potential output under the hourly contract to vary with the wage (i.e., $Y_{1i}=Y_{1i}(w)$). With this added dimension to potential outcomes, I rewrite $AV_1(\theta)$ as the average value of output among lower types \emph{at $\theta$'s reservation wage}:
\begin{equation}
  \label{eq:Awage}
  AV^E_1\left(\theta\right)\equiv E\left[Y_{1i}(\overline{w}(\theta))|\theta_i\leq\theta\right].
\end{equation}
Assuming wage effects are weakly positive and non-decreasing in $\theta$, the equilibrium condition is given by $\overline{w}(\theta^{EQ})=AV^E_1\left(\theta^{EQ}\right)$. In this case, firms pay an hourly wage equal to the average value of accepting workers' output \emph{under that wage}, $AV_1^E(\theta^{EQ})$. Relative to the benchmark model, positive wage effects will therefore push the average value curve upwards and increase the share of hourly contracts under asymmetric information.

Note, however, that the full-information equilibrium is also complicated by the presence of wage effects. A fully informed firm may benefit from paying a worker above their reservation wage if their expected increase in output exceeds the wage premium (i.e., if $E\left[Y_{1i}(w) - Y_{1i}(\overline{w}_i)|\theta_i=\theta\right] > w - \overline{w}(\theta)$ for some $w$).\footnote{I avoid the term ``efficiency wages,'' which refers to a class of models explaining unemployment as a general-equilibrium consequence of firms' strategic wage-setting behavior \citep{weiss2014efficiency,krueger1988efficiency,yellen1984efficiency}. In many efficiency-wage models, above-market wages are driven not by causal effects of wages on productivity, but by worker selection, firms' monitoring ability, or turnover costs \citep{salop1979model,weiss1980job}.
} I thus rewrite $MV_1(\theta)$ as the marginal value of type $\theta$'s output \emph{at their profit-maximizing wage}, so
\begin{equation}
  \label{eq:MVwage}
  MV^E_1\left(\theta\right)\equiv E\left[Y_{1i}(w^*(\theta))|\theta_i=\theta\right],
\end{equation}
where
\begin{equation}
  \label{eq:weff}
w^*(\theta)\equiv \argmax_{w\geq \overline{w}(\theta)}E\left[Y_{1i}(w)-w|\theta_i=\theta\right].
\end{equation}
Note that allowing for wage effects means I can no longer interpret Equation~\eqref{eq:MHtheta} as the marginal treatment effect of hourly-contract take-up---if the wage level influences worker output independently of the hourly compensation structure, the wage-offer instrument no longer satisfies the exclusion restriction. The randomized wage raises in my experiment are designed to address this concern. By equalizing the paid wages of low-offer accepters with those of high-offer accepters, these surprise wage increases isolate variation in \emph{offered} wages conditional on a given \emph{effective} wage. I discuss this instrument validity and estimation of wage effects in Section~\ref{sec:estwage}.

\subsection{Dynamic Contracts}
In dynamic labor markets, workers tend to reveal information about their productivity over time. In long-term employment relationships, employers might use repeated observations of worker output to inform promotions, wage cuts, or dismissals, effectively updating wage contracts to better align each worker's compensation with their latent productivity and effort \citep{FarberGibbons1996}. In short-term gig markets like my experimental setting, this dynamic screening process operates primarily through work histories, ratings, and reviews, which have been shown to narrow the informational gap between workers and firms \citep{pallais2014inefficient,pallais2016referential}.\footnote{While informative to firms, these measures of worker experience and reputation are likely less predictive than the repeated instances of past worker performance observed in long-term employment relationships. These relatively limited information sets make short-term employers more vulnerable to adverse selection and moral hazard, perhaps explaining the predominance of output-based pay observed among gig workers.} Rather than explicitly incorporate this dynamic evolution of wage contracts, my model considers contracts for a single realization of labor output. Recall, however, that workers in the model are assumed to be pre-screened on observable characteristics. Contracts can be placed in a dynamic context by simply defining this screenable information set to include prior realizations of output or observable measures of reputation. Indeed, my empirical analysis conditions on the information available to firms in my experimental setting, including prior work history (see Footnote~\ref{fn:screen}).

\clearpage
\section{Value Curve Estimation Details}\label{sec:curvedetails}

This appendix provides details for estimating the hourly supply and value curves described in Section~\ref{sec:estmodel}.

For positive hourly wage offers, the baseline first stage is
\begin{equation}
  \Pr(D_i=1\mid W_i=w,X_i=x)=\logit^{-1}\left(a(x)+\beta\ln w\right),
\end{equation}
where $a(x)$ contains the intercept and the index of first-stage covariates; the no-offer control group is assigned a propensity score of zero by design. The corresponding conditional inverse supply curve is
\begin{equation}
  \label{eq:conditionalinverse}
  \overline{w}(\theta;x)=\exp\left\{\frac{\operatorname{logit}(\theta)-a(x)}{\beta}\right\}.
\end{equation}
Thus, $\theta$ indexes resistance ranks conditional on $X=x$: the same rank can imply different reservation wages at different values of $x$.

I use the local polynomial approach from \citet{carneiro2011estimating} to estimate average and marginal values. First, I residualize $Y_i$ with respect to covariates separately for hourly and piece-rate workers using double-residual regression methods \citep{robinson1988root}, imposing
\begin{equation}
  E\left[Y_{Ji}\mid X_i=x,\theta_i=\theta\right]
  =MV_J(\theta;\overline X)+\xi_J'\widetilde x,\qquad J\in\{0,1\},
\end{equation}
where $\widetilde x$ is normalized to have sample mean zero. This additive-separability assumption allows $X$ to shift the levels, but not the slopes, of the value curves. The screenable covariates include controls for number of previous tasks, task start time, and employment status.\footnote{Race, gender, and age are excluded because employers cannot legally use these characteristics in employment or wage-setting decisions. Estimates of linear selection effects from Table~\ref{tab:ASc} suggest including these demographic controls would have minimal effect on the slopes of value curves.} For hourly workers, I also residualize $Y_i$ with respect to the effective wage paid after randomized top-up payments, preventing potential wage effects from violating the exclusion restriction. This effective-wage control is held fixed across the profiles used for welfare aggregation because it is an exclusion control rather than an employer-screenable characteristic. I then estimate marginal and average values using local polynomial regressions of residualized $Y_i$ on $S(W_i;X_i)$ with a bandwidth of 0.2.\footnote{The no-offer control group allows for identification of $AV_0(\theta)$ at infinitely high reservation wages ($\theta \rightarrow 1$). However, because take-up rates fall below one for even the most generous fixed-wage offers, estimating right tails of the value-curve distributions requires some extrapolation of local polynomials. This extrapolation has no meaningful effect on welfare estimates because equilibrium and efficient allocations fall in the middle of the $\theta$ distribution.} For each observed profile, additive separability yields
\begin{equation}
  \begin{aligned}
    MV_J(\theta;X_i)&=MV_J(\theta;\overline X)+\xi_J'\widetilde X_i,\\
    AV_J(\theta;X_i)&=AV_J(\theta;\overline X)+\xi_J'\widetilde X_i.
  \end{aligned}
\end{equation}

I combine these profile-specific value curves with $\overline w(\theta;X_i)$ from Equation~\eqref{eq:conditionalinverse}. For each sample member, the equilibrium and efficient cutoffs are the highest ranks satisfying
\begin{equation}
  \begin{aligned}
    \theta_i^{EQ}&=\sup\left\{\theta:\overline w(\theta;X_i)\leq AV_1(\theta;X_i)\right\},\\
    \theta_i^{EF}&=\sup\left\{\theta:\overline w(\theta;X_i)\leq MV_1(\theta;X_i)\right\},
  \end{aligned}
\end{equation}
and profile-specific welfare loss is
\begin{equation}
  DWL_i=\int_{\theta_i^{EQ}}^{\theta_i^{EF}}
  \left[MV_1(\theta;X_i)-\overline w(\theta;X_i)\right]d\theta.
\end{equation}
The primary estimates report $N^{-1}\sum_i\theta_i^{EQ}$, $N^{-1}\sum_i\theta_i^{EF}$, and $N^{-1}\sum_iDWL_i$, using equal weights over the estimation sample and a rank grid spaced by 0.001. Figure~\ref{fig:curves} instead displays the curves and intersections at $X=\overline X$. To estimate welfare loss without moral hazard, I repeat the profile-specific calculation using $MV_0$ and $AV_0$. Each of the 1,000 bootstrap replications re-estimates the curves and repeats the complete profile-specific solve-and-average procedure.

\subsection{Alternative Specification with First-Stage Monotone Spline}\label{sec:fullsemi}

The baseline semi-parametric specification estimates the first-stage propensity score using a logit model. As a robustness check, I also estimate a more flexible first stage that allows the wage-offer component of the propensity score to vary semi-parametrically, subject to the monotonicity restriction that fixed-wage take-up is weakly increasing in the offered wage.

Let $D_i$ indicate take-up of the hourly contract, let $w_i\equiv \ln(W_i)$ denote the log hourly wage offer, and let $X_i$ denote the first-stage covariates. Among workers who receive a positive hourly wage offer, the monotone-spline first stage is
\begin{equation}
  \Pr(D_i=1\mid w_i,X_i)=\logit^{-1}\left(\alpha_0+X_i'\alpha+g(w_i)\right),
\end{equation}
where $g(\cdot)$ is a continuous piecewise-linear function of the log wage offer. The no-offer control group is assigned a propensity score of zero by design. Let $\kappa_0<\kappa_1<\cdots<\kappa_R$ denote knots spanning the support of $w_i$. I write the wage-offer component as
\begin{equation}
  g(x)=\sum_{r=1}^{R} \exp(\delta_r) h_r(x),
\end{equation}
where $h_1(x)=\min\{x,\kappa_1\}-\kappa_0$, $h_R(x)=[x-\kappa_{R-1}]_+$, and, for $1<r<R$, $h_r(x)=\left[\min\{x,\kappa_r\}-\kappa_{r-1}\right]_+$, with $[a]_+\equiv\max\{a,0\}$. Thus, the first and last segments extend linearly beyond the boundary knots, as in the estimator. Equivalently, $g(x)$ is linear within each knot interval, with slope $\exp(\delta_r)$ on interval $r$. Since $\exp(\delta_r)>0$, the fitted first-stage index is monotone increasing in the log wage offer holding $X_i$ fixed. Because the logit link is also monotone, the fitted propensity score is likewise monotone increasing in the offered wage. The profile-specific inverse supply curve is therefore
\begin{equation}
  \overline w(\theta;X_i)=\exp\left\{g^{-1}\left[\operatorname{logit}(\theta)-\alpha_0-X_i'\alpha\right]\right\}.
\end{equation}
I use this inverse in the same profile-specific solve-and-average procedure described above; Appendix Figure~\ref{fig:curvesfullsemi} displays the spline curves at the mean covariate profile.

The parameters are estimated by penalized logit maximum likelihood:
\begin{equation}
  \left(\widehat{\alpha}_0,\widehat{\alpha},\widehat{\delta}\right)=\argmin_{\alpha_0,\alpha,\delta}
  \sum_i \omega_i\left[\ln\left\{1+\exp(\eta_i)\right\}-D_i\eta_i\right]
  +\frac{\lambda}{2}\alpha'\alpha,
\end{equation}
\enlargethispage{2\baselineskip}
where $\eta_i=\alpha_0+X_i'\alpha+g(w_i)$ and $\omega_i$ denotes any estimation weight. Before estimation, each nonconstant first-stage covariate is centered and scaled by its estimation-sample mean and standard deviation. The ridge penalty is applied only to the linear covariate coefficients in the first stage, not to the wage-offer spline. This regularization stabilizes specifications with richer control sets while preserving full monotonicity and flexibility in the wage-offer dimension. The implementation uses five interior knots, placed at equally spaced quantiles of log wage offers unless otherwise specified, and sets $\lambda=25$. The second-stage MTE estimation is otherwise held fixed at the baseline semi-parametric specification, including the second-degree local polynomial and bandwidth of 0.2.

\clearpage
\section{MVPF and Optimal-Policy Derivations}\label{sec:mvpfdetails}

In this appendix, I derive optimal tax and subsidy policies from Section~\ref{sec:policy}.

\subsection*{Tax}
Consider an ad-valorem piece-rate tax, $\rho$, assessed on the value of labor product produced under the piece rate. For each dollar of labor product compensated under the piece rate, the government collects an additional $\rho$ dollars in tax revenue. Under this wedge, procuring labor product through the piece rate becomes $1+\rho$ times as expensive as in the baseline. For simplicity, I assume interior equilibria before and after the tax, $0<\theta^{EQ}\leq \theta^{\rho}<1$, so that both fixed-wage and piece-rate workers exist with or without the piece-rate tax. The equilibrium condition under tax $\rho$ is given by
\begin{equation}
\overline{w}(\theta^{\rho})=(1+\rho)AV_1(\theta^{\rho}),
\end{equation}
where $\theta^{\rho}$ is the equilibrium share of hourly workers under the tax.

To calculate the fiscal externality associated with piece-rate tax $\rho$, let $\mathcal{T}(\rho)$ denote aggregate receipts from preexisting taxes on worker and firm income, calculated before applying any deduction for piece-rate-tax payments.\footnote{As mentioned in Section~\ref{sec:policy}, I treat direct piece-rate-tax payments as deductible from the preexisting tax base. I include the associated reduction in income-tax receipts in the policy’s direct net transfer, $(1-\tau)\rho\int_{\theta^\rho}^{1}MV_0(\theta)\,d\theta$, rather than its fiscal externality $FE(\rho)\equiv\mathcal T(\rho)-\mathcal T(0)$.} Let $TI_F(0)$ denote taxable income accruing to firms or their owners in the untaxed equilibrium. Holding the worker-received piece rate fixed, I assume that the tax-induced gross-up in fixed wages reduces taxable firm income dollar-for-dollar. Total tax receipts are therefore
\begin{equation}
  \begin{aligned}
    \mathcal{T}(\rho)
    ={}&\underbrace{\tau_W\left[(1+\rho)\int_0^{\theta^\rho}MV_1(\theta)d\theta
      +\int_{\theta^\rho}^1MV_0(\theta)d\theta\right]}_{\text{Taxes Paid by Workers}} \\
    &+\underbrace{\tau_F\left[TI_F(0)-\rho\int_0^{\theta^\rho}MV_1(\theta)d\theta\right]}_{\text{Taxes Paid by Firms or Their Owners}}.
  \end{aligned}
\end{equation}
The two $\rho\int_0^{\theta^\rho}MV_1(\theta)d\theta$ terms reflect a transfer from firms to workers---the mechanical increase in wages arising from the higher cost of labor in the piece-rate sector. Assuming worker and firm income face a common effective tax rate, $\tau_W=\tau_F=\tau$, the fiscal externality is the resulting change in receipts from preexisting taxes:
\begin{equation}
  \begin{aligned}
    FE(\rho)\equiv\mathcal{T}(\rho)-\mathcal{T}(0)
    ={}&\tau\int_{\theta^{EQ}}^{\theta^\rho}MH(\theta)d\theta \\
       &+\underbrace{\tau\rho\int_0^{\theta^\rho}MV_1(\theta)d\theta}_{\text{Higher Worker Tax Payments}}
       -\underbrace{\tau\rho\int_0^{\theta^\rho}MV_1(\theta)d\theta}_{\text{Lower Firm Tax Payments}} \\
    ={}&\tau\int_{\theta^{EQ}}^{\theta^\rho}MH(\theta)d\theta.
  \end{aligned}
\end{equation}

The welfare-maximizing level of tax is given by
\begin{equation}
  \max_\rho \left\{ \eta NR(\rho) - WTP(\rho) \right\},
\end{equation}
where $NR(\rho)$ is the government's revenue from the piece-rate tax net of any fiscal externalities, and $WTP(\rho)$ is aggregate private willingness-to-pay to \emph{avoid} the tax. $\eta$ reflects the highest-MVPF policy for which revenue might be used. Under a balanced budget, $\eta$ would represent the MCPF of the least efficient revenue source one could replace with the piece-rate tax.

\begin{equation}
  \label{eq:rhofoc}
  \max_\rho \left\{
  \begin{aligned}
    &\eta \underbrace{\left((1-\tau)\int^1_{\theta^\rho}\rho MV_0(\theta)d\theta + \int_{\theta^{EQ}}^{\theta^\rho}\tau MH(\theta)d\theta \right)}_{NR(\rho)} \\
    &-\underbrace{(1-\tau)\left(\int^1_{\theta^\rho}\rho MV_0(\theta)d\theta - \int_{\theta^{EQ}}^{\theta^\rho}\left(MV_1(\theta) - \overline{w}(\theta)\right)d\theta\right)}_{WTP(\rho)}
  \end{aligned}
  \right\}.
\end{equation}
The first-order condition for \eqref{eq:rhofoc} implies

\begin{eqnarray}  MCPF_{d\text{Tax}}({\rho^*})\equiv\frac{WTP'(\rho^*)}{NR'(\rho^*)}&=&\eta \\
  \frac{\frac{d\rho}{d\theta^{\rho}} \int^1_{\theta^{\rho^*}}MV_0(\theta)d\theta - \rho^* MV_0(\theta^{\rho^*}) - \left(MV_1(\theta^{\rho^*}) - \overline{w}(\theta^{\rho^*}) \right)}{\frac{d\rho}{d\theta^{\rho}} \int^1_{\theta^{\rho^*}}MV_0(\theta)d\theta - \rho^* MV_0(\theta^{\rho^*}) + \frac{\tau}{1-\tau} MH(\theta^{\rho^*})}&=&\eta.\label{eq:rhostarapp}
\end{eqnarray}
$MCPF_{d\text{Tax}}({\rho})$ is the MCPF for a \emph{marginal increase} in the piece-rate tax.

To calculate $\frac{d\rho}{d\theta^{\rho}}$, consider the equilibrium condition from Equation~\eqref{eq:equilibrium} under the tax wedge defined in Section~\ref{sec:policy}:
\begin{equation}
\overline{w}(\theta^{\rho})=(1+\rho)AV_1(\theta^{\rho}).
\end{equation}
Differentiating with respect to $\theta^\rho$ yields
\begin{eqnarray}
  \frac{d\rho}{d\theta^\rho} & = & \frac{\frac{d\overline{w}}{d\theta^\rho}}{AV_1(\theta^\rho)} - \frac{\overline{w}(\theta^\rho)}{AV_1(\theta^\rho)^2}\frac{dAV_1(\theta^\rho)}{d\theta^\rho} \\
                             & = & \left(\frac{dS}{d\overline{w}}\right)^{-1}\frac{1}{AV_1(\theta^\rho)} - \frac{\overline{w}(\theta^\rho)}{AV_1(\theta^\rho)^2}\frac{MV_1(\theta^\rho) - AV_1(\theta^\rho)}{\theta^\rho} \\
                             & = &\frac{\overline{w}(\theta^\rho)}{\beta \theta^\rho(1-\theta^\rho)AV_1(\theta^\rho)} - \frac{\overline{w}(\theta^\rho)\left(MV_1(\theta^\rho) - AV_1(\theta^\rho)\right)}{\theta^\rho AV_1(\theta^\rho)^2},
\end{eqnarray}
where $\beta$ is the coefficient on $\ln w$ in a logistic model of hourly labor supply.

\subsection*{Subsidy}

Consider an hourly-wage subsidy of \$$\delta$ per hour worked. In my model, the subsidy lowers nominal reservation wages by \$$\delta$, yielding a new equilibrium share of hourly workers, $\theta^\delta$, such that
\begin{equation}
\overline{w}(\theta^{\delta})=AV_1(\theta^{\delta})+\delta.
\end{equation}
Assuming the subsidy is treated as taxable income, the aggregate willingness-to-pay for the subsidy is
\begin{equation}
  \label{eq:wtp}
  WTP(\delta) = \underbrace{(1-\tau)\delta\theta^\delta}_{\text{Transfer}} + \underbrace{(1-\tau)\int_{\theta^{EQ}}^{\theta^\delta}\left(MV_1(\theta) - \overline{w}(\theta)\right)d\theta}_{\text{Insurance Benefit}},
\end{equation}
and the government's net cost is
\begin{equation}
  \label{eq:cost}
NC(\delta) = \underbrace{(1-\tau)\delta\theta^\delta}_{\text{Direct Cost of Transfer}} - \underbrace{\int_{\theta^{EQ}}^{\theta^\delta} \tau MH(\theta)d\theta}_{\text{Fiscal Externality from Moral Hazard}}.
\end{equation}
The corresponding MVPF is
\begin{equation}
  \label{eq:mvpf}
  MVPF_{\text{Sub}}(\delta) \equiv \frac{WTP(\delta)}{NC(\delta)}=  \frac{\delta\theta^\delta  + \int_{\theta^{EQ}}^{\theta^\delta}\left(MV_1(\theta) - \overline{w}(\theta)\right)d\theta }{\delta\theta^\delta - \int_{\theta^{EQ}}^{\theta^\delta}\frac{\tau}{1-\tau} MH(\theta)d\theta }.
\end{equation}

The welfare-maximizing level of subsidy solves
\begin{equation}
  \label{eq:optimalsub}
  \max_\delta \left\{\begin{aligned}
    &\underbrace{(1-\tau)\left(\delta\theta^\delta + \int_{\theta^{EQ}}^{\theta^\delta}\left(MV_1(\theta) - \overline{w}(\theta)\right)d\theta\right)}_{WTP(\delta)}\\
    &-\lambda \underbrace{\left((1-\tau)\delta\theta^\delta - \int_{\theta^{EQ}}^{\theta^\delta}{\tau}MH(\theta)d\theta\right)}_{NC(\delta)}
  \end{aligned}\right\},
\end{equation}
where $\lambda$ reflects the marginal cost of public financing---the cost of raising one dollar of revenue through taxation, or the MVPF of some alternative policy from which funds are redirected. The first-order condition for \eqref{eq:optimalsub} implies
\begin{equation}
  \label{eq:deltastar}
  MVPF_{d\text{Sub}}({\delta^*})\equiv\frac{WTP'(\delta^*)}{NC'(\delta^*)} =
  \frac{\frac{d\delta}{d\theta^{\delta}}\theta^{\delta^*} + \delta^* + MV_1(\theta^{\delta^*}) - \overline{w}(\theta^{\delta^*})}{\frac{d\delta}{d\theta^{\delta}}\theta^{\delta^*} + \delta^* - \frac{\tau}{1-\tau} MH(\theta^{\delta^*})} =\lambda.
\end{equation}
$MVPF_{d\text{Sub}}({\delta})$ is the MVPF for a \emph{marginal increase} in the hourly-wage subsidy.

To calculate $\frac{d\delta}{d\theta^{\delta}}$, consider the equilibrium condition from Equation~\eqref{eq:equilibrium} in the presence of an hourly-wage subsidy, $\delta$:
\begin{equation}
\overline{w}(\theta^{\delta})=AV_1(\theta^{\delta})+\delta.
\end{equation}
Differentiating with respect to $\theta^\delta$ yields
\begin{eqnarray}
  \frac{d\delta}{d\theta^\delta} & = & \frac{d\overline{w}}{d\theta^\delta} - \frac{dAV_1(\theta^\delta)}{d\theta^\delta}  \\
                                 & = & \left(\frac{dS}{d\overline{w}}\right)^{-1} - \frac{MV_1(\theta^\delta) - AV_1(\theta^\delta)}{\theta^\delta}  \\
                                 & = & \frac{\overline{w}(\theta^\delta)}{\beta \theta^\delta(1-\theta^\delta)} - \frac{MV_1(\theta^\delta) - AV_1(\theta^\delta)}{\theta^\delta},
\end{eqnarray}
where $\beta$ is the coefficient on $\ln w$ in a logistic model of hourly labor supply.

\subsection*{Worker Pass-Through and General Equilibrium Effects}
The policy counterfactuals above hold the worker-received piece rate, $p$, fixed, placing the full piece-rate tax wedge on firms' procurement costs. If $p$ is instead determined in equilibrium, some of the tax may be passed through to piece-rate workers through a lower net piece rate. The effect of the  policy would still be to encourage fixed wages over piece rates, though that effect would now operate partly through a smaller upward shift in the average value curve and partly through a downward shift in hourly reservation wages, $\overline{w}(\theta)$. An hourly-wage subsidy could similarly lower $p$ by reducing the demand for piece-rate labor, shifting both hourly reservation wage and average value curves downward. Incorporating this margin would require estimates of firms' demand for labor product, $q$, and the responsiveness of workers' reservation wages to $p$.

Relatedly, the policy counterfactuals also abstract from effects on the aggregate demand for labor product, $q$. Incorporating these effects would introduce conventional quantity distortions beyond those captured by moral hazard: piece-rate taxes would tend to reduce $q$, whereas hourly-wage subsidies would tend to increase it. Because these distortions operate in opposite directions, an appropriate combination of the two policies could, in principle, offset their effects on aggregate quantity and welfare.

\subsection*{Population Aggregation and Numerical Implementation}
The analytical derivations fix $X=x$ and suppress this conditioning in the notation. The empirical implementation restores the conditioning and applies a single policy rate to every observed profile of employer-screenable characteristics. At each tax $\rho$, I solve
\begin{equation}
  \overline w(\theta_i^\rho;X_i)=(1+\rho)AV_1(\theta_i^\rho;X_i)
\end{equation}
separately for each $i$. The resulting cutoff determines the profile-specific willingness-to-pay and net-revenue components:
\begin{equation}
  \label{eq:conditionalmcpfcomponents}
  \begin{aligned}
    WTP(\rho;X_i)
    &=(1-\tau)\left[{}\rho\int_{\theta_i^\rho}^1 MV_0(\theta;X_i)d\theta
      -\int_{\theta_i^{EQ}}^{\theta_i^\rho}
      \left[MV_1(\theta;X_i)-\overline w(\theta;X_i)\right]d\theta\right],\\
    NR(\rho;X_i)
    &=(1-\tau)\rho\int_{\theta_i^\rho}^1 MV_0(\theta;X_i)d\theta
      +\tau\int_{\theta_i^{EQ}}^{\theta_i^\rho}MH(\theta;X_i)d\theta,
  \end{aligned}
\end{equation}
where $MH(\theta;X_i)=MV_1(\theta;X_i)-MV_0(\theta;X_i)$. I use unqualified notation for the resulting averages over the empirical distribution of $X_i$:
\begin{equation}
  \label{eq:mcpfaggregation}
  \begin{aligned}
    WTP(\rho)&=N^{-1}\sum_i WTP(\rho;X_i),\\
    NR(\rho)&=N^{-1}\sum_i NR(\rho;X_i),\\
    MCPF_{\mathrm{Tax}}(\rho)
    &\equiv\frac{WTP(\rho)}{NR(\rho)}
    =\frac{N^{-1}\sum_i WTP(\rho;X_i)}{N^{-1}\sum_i NR(\rho;X_i)}.
  \end{aligned}
\end{equation}
Thus, the reported MCPF is a ratio of averages, not an average of profile-specific MCPF ratios. The subsidy calculations analogously solve $\overline w(\theta_i^\delta;X_i)=AV_1(\theta_i^\delta;X_i)+\delta$. They then average $WTP(\delta;X_i)$ and $NC(\delta;X_i)$ and take their ratio, $MVPF_{\mathrm{Sub}}(\delta)$.

I evaluate taxes on a $0(.001)0.5$ grid and subsidies on a $0(.01)2$ grid. The conditional rank grid is spaced by 0.001; because its final point lies just below one, I value the residual tail of the direct tax base using the terminal fitted value of $MV_0(\theta;X_i)$. The marginal schedules numerically differentiate the aggregated willingness-to-pay and fiscal components separately, then form their ratio. Specifically, the reported derivatives use symmetric secants with five grid cells on either side---half-width 0.005 for taxes and 0.05 for subsidies---clipped at the endpoints. The optimal tax is the direct grid solution to
\begin{equation}
  \rho^*(\eta)=\argmin_\rho\left\{WTP(\rho)-\eta NR(\rho)\right\}.
\end{equation}
This optimization is repeated within each bootstrap draw, with exact ties assigned to the smaller tax.

\clearpage
\section{External Validity}\label{sec:external}

My experiment was designed to demonstrate how moral hazard and adverse selection can lead to an underprovision of fixed-wage jobs. In the parlance of \citet{list2020trial}, it aims to provide ``WAVE1'' insights validating the theory presented in Section~\ref{sec:model}. While point estimates directly speak to the importance of information asymmetries in some settings, they are not intended to measure welfare losses across the myriad of labor markets characterized by risky forms of compensation. Nonetheless, it is important to consider the settings to which my estimates can be directly applied. I therefore assess the external validity of my results
using the SANS conditions from \citet{list2020trial}.

\paragraph{Selection}
The experimental sample was drawn from the population of workers on Prolific, a widely used and well-established freelancing platform with over 100,000 workers. Among online workers, Prolific is widely considered the most desirable micro-task platform due to its ease of use and high pay. Consequently, the platform has a waitlist, and workers rarely turn down a task for which they are eligible \citep{douglas2023data}.
Workers in this population were eligible for my experimental job posting if they met four screening criteria: (1) were located in the United States, (2) spoke fluent English, (3) successfully completed ten or more previous tasks, and (4) earned an approval rate above 98 percent on previous tasks. More than 95 percent of Prolific workers meet the 98-percent approval threshold. Importantly, my experimental job posting advertised generous compensation for a five-minute task---a \$1.00 flat fee in addition to the \$0.03 piece rate for each correctly typed sentence---to avoid a restricted sample of workers with low reservation wages or limited outside options. Experimental jobs were posted across a broad range of days and times to ensure the sample was not restricted to workers with a particular schedule.

\paragraph{Attrition}
There was virtually no attrition from the sample. One worker exited the task before observing their experimental wage offer and was dropped from the experimental sample. All other workers remained in the sample, even if they failed to enter sentences or click the submit button after the five-minute timer expired.

\paragraph{Naturalness}
Each aspect of the experimental intervention was designed to place participants in a naturally occurring setting. Contract options are presented in a simple and straightforward manner, offering workers a choice between a randomized hourly wage offer and a standardized \$0.03 piece rate for each correctly typed sentence. This \$0.03 piece rate was set to approximate observed market rates for online image-to-text transcription services \citep{TranscriptionUS_TypingRates2024,Upwork_DataEntryRates2024,Upwork_DataEntryGuide2024}.
Like most online employment, hired workers faced a threat of job dismissal and reputational damage for unapproved tasks, so they had an incentive to maintain a minimum standard of performance under either compensation scheme. The transcription task, described in Appendix~\ref{sec:settingdetails}, closely resembles those commonly requested on online platforms, which typically take no more than a few minutes \citep{hornuf2022hourly,Fiverr_Bilalkhan651,Fiverr_HaxeebAhmad}. More generally, ``traditional keyboarding'' is a job requirement for 66 percent of American workers \citep{blsocc}. Finally, participants were not aware that they were part of an experiment until after performing the task, so estimates are not biased by any desire to generate a particular result.\par

\paragraph{Scaling}
Apart from monetary costs, there would be few constraints on running an experiment like the one in this paper at a larger scale. There is also little reason to doubt that this paper's prescribed policy intervention---a piece-rate tax on data labeling---could be implemented at scale. The Internal Revenue Code already conditions tax treatment on both industry and compensation form. For example, Section~45B provides food-and-beverage businesses with an income-tax credit for employer FICA taxes paid on certain tipped compensation \citep{rane2024consequences}, while the 2025 reconciliation law created separate deductions for qualified tips and qualified overtime compensation \citep{irsOBBBA2025}. There would therefore be clear administrative precedent for a specific tax on firms' task-based payments to workers in the data-labeling industry.

A separate but related question is how well my results might extrapolate to contexts outside my target setting. My experimental task is more predictable, lower stakes, and shorter duration than many of those found in other labor markets. As a result, workers might exhibit less fatigue, more inattention, and a greater tolerance for risk than those in other work settings. I would expect these attributes to attenuate my estimates towards zero: Workers facing higher stakes or more uncertainty over their labor products would pay a higher premium (lower reservation wage) for the implicit insurance offered by fixed wages.\footnote{Workers' uncertainty over their labor product depends on their prior knowledge of the task. To be conservative, I intentionally design the task and instructions to maximize this task-related knowledge while maintaining realism. If instructions were less informative, each worker would face a higher subjective variance in earnings and a larger benefit from insurance.} Under plausible assumptions, this increased risk premium would lead to a greater welfare loss than the one I estimate.\footnote{Assuming both equilibrium and efficient allocations remain interior, uniformly higher willingness-to-pay for insurance increases the welfare loss from adverse selection when the increase in the value of missing contracts exceeds the reduction in missing contracts generated by the equilibrium response. The implied restriction on the curve shapes can be satisfied under a broad range of functional forms (e.g., linear), or by assuming $MV'(\theta^{EQ})>AV'(\theta^{EQ})$ and $\overline w(\theta)-MV(\theta)$ is weakly concave between efficient and equilibrium allocations.}Likewise, less inattention would make hourly supply more elastic and more correlated with workers' latent productivity, exacerbating adverse selection problems. Finally, more fatigue would likely lead to larger moral hazard effects---if the cost of effort increases with the duration of the task, so would the benefits of shirking. In short, my analysis would likely underestimate the welfare loss one would expect to find in settings with longer, higher-stakes tasks.\par

While my welfare estimates are likely attenuated relative to many short-term labor markets, this lower-bound relationship is unlikely to hold in every setting, especially those with dynamically adjusting contracts or costly output monitoring.\footnote{I discuss how my framework can be adapted to accommodate labor-market features like monitoring costs and dynamic contracts in Appendix~\ref{sec:extensions}.} Policies aimed at addressing information asymmetries in these markets warrant further evidence targeted to those settings.

\clearpage
\section{Rideshare Selection: Calibrations from  \texorpdfstring{\citet{angrist2021uber}}{Angrist et al. (2021)}}\label{sec:uberapp}

In \citet{angrist2021uber}, experimental ``taxi'' offers consist of a reduced proportional fee on fares charged to drivers, $t_1$, and a weekly lease payment, $L$. For a driver who earns $Y_i$ per week in ride fares, accepting an offer would increase their take-home pay from $(1-t_0)Y_i$ to $(1-t_1)Y_i$, where $t_0$ is the pre-existing Uber fee of 20 or 25 percent, in exchange for paying Uber \$$L$ each week. For a given lease offer $(t_1,L)$, the ``breakeven value,'' $w_B$, is defined as the weekly fare revenue required for a driver's increase in take-home earnings, $(t_0-t_1)Y_i$, to cover the cost of their weekly lease payment, $L$:
    \begin{equation}
      \label{eq:uberbreakeven}
    w_B\equiv\frac{L}{\Delta_t},
  \end{equation}
  where $\Delta_t\equiv t_0-t_1$. So, a lease offer $(L,t_1)$ effectively gives the driver an opportunity to buy back a $\Delta_t$-share of their own earnings at a valuation or ``share price'' of $w_B$. Viewed through this lens, opting into a taxi-style lease is the same as opting \emph{out} of a (partial) fixed wage contract---a driver who accepts a lease offer gives up \$$L$ in fixed weekly income in favor of a $\Delta_t$-higher share of their labor product. This equivalence means I can calibrate my model using estimates of selection \emph{out of} taxi leases from \citet{angrist2021uber} to approximate welfare losses from inefficient compensation in the market for rideshare drivers.

  To calibrate my model to this setting, I estimate a supply curve for lease-free driving by fitting a logit model to the aggregated statistics from Table~4 of \citet{angrist2021uber}, which reports lease take-up rates and observation counts by week, experimental strata, and treatment condition. Specifically, I estimate the following:
  \begin{equation}\label{eq:ubersupply}
 \Pr[\mathrm{Lease}_i=0\mid {w_{B}}_i=w_B,X_i]= \logit^{-1}\left(\alpha + \beta \ln w_B + X_i'\gamma\right),
  \end{equation}
  where $\mathrm{Lease}_i$ is a binary indicator for accepting a lease offer and $X_i$ is a vector of controls for experimental strata and week. I weight each cell by its driver count, reconstructed from reported group sizes, offer rates, and take-up rates and rounded to the nearest integer. I then derive the lease-free supply curve by averaging over observed $X_i$ in the full estimation sample using the same weights, so $S(w_B)= E_X\left[\Pr\left[{Lease}_i=0|{w_{B}}_i=w_B, X_i=X\right]\right]$.

  To derive the average value curve, I combine the information on take-up rates with reported gains across accepters and decliners of leases. First, I use values in Table~4 to compute average experimental breakeven offers, residualized by strata- and week-fixed effects and recentered at the overall mean, separately for accepters and decliners of lease offers. I then combine these estimates with information from Table~6, which reports average dollar gains from leases and the share driving more than zero hours separately by accepters and decliners of experimental offers. To express gains in fare-revenue units, I divide them by each offer's proportional fee reduction, $\Delta_t$, before adding them to breakeven fare revenue. I average the resulting values across offers using driver counts within each take-up group. Specifically, letting $j$ index offers, $d$ denote lease take-up, $n_{jd}$ denote driver counts, $g_d$ denote the reported gain conditional on driving, and $q_d$ denote the share driving, I compute
  \begin{equation*}
    \overline{Y}_d=q_d\left(\overline{w}_{B,d}+g_d\frac{\sum_j n_{jd}/\Delta_{t,j}}{\sum_j n_{jd}}\right),
  \end{equation*}
  where $\overline{w}_{B,d}$ is the mean residualized and recentered breakeven offer in group $d$.

  From Table~4, the average residualized breakeven value among decliners is \$\fp[0]{\meanbreakevendecline}. From Table~6, the share of decliners who drove more than zero hours is ${\obsconditionaldecline} / {\obsunconditionaldecline} = \fp[0]{100*\obsconditionaldecline/\obsunconditionaldecline}$ percent. Among those who drove, decliners' expected gain from a lease averaged ${\$\gainconditionaldecline}$. After dividing gains by offer-specific fee reductions and averaging across offers, estimated mean fare revenue among decliners is $\overline{Y}_0=\$\fp[0]{\uberalpha+\uberbeta*\thetaav}$.
 Because this value is computed among decliners in all experimental groups,\footnote{Unfortunately, Table~6 does not report gains by take-up status at the strata, week, or breakeven level, so I apply each group's pooled gain and driving share to all offers before averaging. This imposes a common conditional dollar gain across offers with different fee reductions. The average value curve is then calibrated using two points, assigning the pooled decliner mean to the supply share at the average log offer and the implied population mean to $\theta=1$.} I assign it to the supply share at the driver-weighted mean log breakeven offer, $\overline{\theta}=S\left(\exp\{E[\ln w_B]\}\right)=\fp[3]{\thetaav}$, evaluated over the full estimation sample, so
 \begin{equation}
   \label{eq:uberav}
 AV(\fp[2]{\thetaav})\equiv E[Y_i|\theta_i\leq\fp[2]{(\thetaav)}]=\$\fp[0]{\uberalpha+\uberbeta*\thetaav}.
 \end{equation}

  Meanwhile, the average residualized breakeven value among accepters in Table~4 is \$\fp[0]{\meanbreakevenaccept}. From Table~6, the share of  accepters who drove more than zero hours is
  ${\obsconditionalaccept} / {\obsunconditionalaccept} = \fp[0]{100*\obsconditionalaccept/\obsunconditionalaccept}$ percent. Among those who drove, accepters gained an average of ${\$\gainconditionalaccept}$. Applying the same fee conversion and averaging across offers gives estimated mean fare revenue $\overline{Y}_1=\$\fp[0]{\uberalpha+\uberbeta*(1+\thetaav)}$. I assign this value to the remaining \fp[0]{100*(1-\thetaav)} percent of drivers, so $E[Y_i|\theta_i>\fp[2]{(\thetaav)}]=\$\fp[0]{\uberalpha+\uberbeta*(1+\thetaav)}$.\footnote{I use observed treatment-week gains for accepters and expected treatment-week gains for decliners, as reported in Table~6. The calibration abstracts from moral hazard, treating the resulting values as potential output under the partial fixed wage contract.} I combine this value with $ AV(\fp[2]{\thetaav})$ from Equation~\eqref{eq:uberav} to take the total expectation,
  \begin{eqnarray}
    AV(1)&\equiv& E[Y_i|\theta_i\leq 1]=E[Y_i] \nonumber\\
& = & \fp[2]{(\thetaav)}\times E[Y_i|\theta_i\leq\fp[2]{(\thetaav)}] + (1-\fp[2]{(\thetaav)})\times E[Y_i|\theta_i>\fp[2]{(\thetaav)}] \nonumber\\
& = &     \fp[2]{(\thetaav)}\times\$\fp[0]{\uberalpha+\uberbeta*\thetaav} + (1-\fp[2]{(\thetaav)})\times\$\fp[0]{\uberalpha+\uberbeta*(1+\thetaav)} \nonumber\\
    AV(1) &=& \$\fp[0]{\uberalpha+\uberbeta}.\label{eq:av1}
  \end{eqnarray}
I therefore calibrate the average value curve by fitting a line between two points:  $AV(\fp[2]{\thetaav})=\$\fp[0]{\uberalpha+\uberbeta*\thetaav}$ and $AV(1)= \$\fp[0]{\uberalpha+\uberbeta}$, yielding the following average value curve:
  \begin{equation}
    AV(\theta) = \fp[1]{\uberalpha} + \fp[1]{\uberbeta}\theta.
  \end{equation}
The marginal value curve follows by differentiation:
  \begin{eqnarray}
    MV(\theta) & = & \frac{\partial}{\partial \theta}\left(\theta AV(\theta)\right)\nonumber\\
               & = &\frac{\partial }{\partial \theta}\left(\fp[1]{\uberalpha}\theta + \fp[1]{\uberbeta}\theta^2 \right) \nonumber\\
               MV(\theta) & = & \fp[1]{\uberalpha} + \fp[1]{2*\uberbeta}\theta.
  \end{eqnarray}

  With average value, marginal value, and supply curves in hand, I can calculate equilibrium and efficient shares of lease-free driving. The equilibrium share of lease-free contracts is $\theta^{EQ}=\fp[1]{100*\ubereq}$ percent. The associated ``breakeven price'' is $w_B(\theta^{EQ})=\$\fp[0]{\uberalpha + \uberbeta*\ubereq}$. Scaling this price by $\Delta_t$ represents the hypothetical lease that drivers would need to pay in lieu of the proportional fee on rider fares to generate the same net revenue for Uber. Meanwhile, the efficient share of lease-free contracts one would expect in a full-information equilibrium is $\theta^{EF}=\fp[1]{100*\ubereff}$ percent. The difference of $\theta^{EF} - \theta^{EQ} = \fp[1]{100*(\ubereff-\ubereq)}$ percentage points represents potential drivers who would generate profitable levels of fare revenue if they were charged a lower proportional fee or paid some additional fixed weekly compensation that implicitly valued their labor product somewhere between \$\fp[0]{\uberalpha + \uberbeta*\ubereq} and \$\fp[0]{\uberalpha + 2*\uberbeta*\ubereff} per week.\footnote{One way to frame this comparison is to imagine that all drivers must pay a fixed weekly ``lease,'' $\bar{L}$, to operate a taxi or Uber. In equilibrium, the $\theta^{EQ}$-share of drivers who choose Uber forfeit $\Delta_t$ of their earnings in exchange for a fixed weekly wage of $w^{EQ}_{B}\Delta_t=\bar{L}$, exactly offsetting the lease. A higher breakeven value could be accomplished with a lower proportional fee (as in the \citet{angrist2021uber} experiment), or with a positive net-of-lease fixed wage (i.e., for some $w_B>w_B^{EQ}$, pay drivers $w^+ \equiv w_B \Delta_t - \bar{L} > 0$ per week).} However, the cost of adverse selection among less productive drivers would make offering this compensation unprofitable. The corresponding welfare loss, shaded in yellow, is $\$\fp[2]{\uberdwl}$ per week. For an approximate hourly conversion, I use \fp[2]{\hoursperweek} driving hours per week, the average over the four weeks before sample selection among drivers assigned taxi-lease offers, calculated from the group means and sample sizes in Tables~A3--A4 of \citet{angrist2021uber}. Dividing by these hours gives $\$\fp[2]{\uberdwl/\hoursperweek}$ per hour in fare-revenue units. Scaling by the average fee reduction of $\fp[1]{100*\feereduction}$ percentage points yields an approximate welfare loss of $\$\fp[2]{\feereduction*\uberdwl/\hoursperweek}$ per hour worked, comparable to my estimates for online data-entry.\footnote{Note that, because all drivers retain at least 75 percent of rider fares, I exclude moral hazard from the welfare analysis. In effect, I am identifying the insurance value drivers place on contracts with marginally lower lease payments ($L$) and marginally higher Uber fees ($t$). As such, my estimates represent the welfare losses from adverse selection only.}

  There are several limitations to this calibration exercise. Calibrated curves, particularly the average value curve, rely on transformations across a small number of point estimates. Linear extrapolation of market-wide selection patterns from these points comes with strong functional-form assumptions, and the absence of standard errors prevents any meaningful discussion of statistical significance.
  Moreover, the experimental sample in \citet{angrist2021uber} consists of \emph{existing} Uber drivers. Observed selection into taxi-style lease offers likely excludes drivers with the strongest preference for this type of compensation because they are already working as taxi drivers! More generally, studies of workers under existing employment contracts are likely to understate the effects of information asymmetries, which depend upon marginal treatment and selection across all workers in a hypothetical market.
  Finally, while drivers in \citet{angrist2021uber} are stratified into high- and low-hour groups with separate experimental offers, this coarse conditioning likely falls short of the kind of dynamic contract that's feasible on modern rideshare platforms. For example, the ``Uber Pro'' program gives drivers with favorable histories access to more favorable rates and amenities, potentially mitigating welfare loss from adverse selection \citep{uberpro2025}. Despite these limitations, however, the calibration exercise provides an example of how my framework can be applied to a variety of labor markets where workers' selection on potential productivity can prevent efficient contracting.

\end{appendices}

\end{document}